\documentclass[twocolumn,showpacs,aps,floatfix,prd]{revtex4}

\usepackage{graphicx}
\usepackage{dcolumn}
\usepackage{amsmath}
\usepackage{epsfig}
\usepackage{hhline}
\usepackage{soul}
\usepackage{bm}
\usepackage{tabularx,pifont,multirow}
\usepackage{longtable}
\usepackage{psfrag}

\RequirePackage{xspace}

\newcommand{\singlewidth}{3in}
\newcommand{\thirdofsinglewidth}{1in}

\newcommand{\BABARPubYear}    {04}
\newcommand{\BABARPubNumber}  {030}
\newcommand{\SLACPubNumber} {10808}

\RequirePackage{xspace}

\usepackage{relsize}
\def\babar{\mbox{\slshape B\kern-0.1em{\smaller A}\kern-0.1em
    B\kern-0.1em{\smaller A\kern-0.2em R}}}

\def\epem       {\ensuremath{e^+e^-}\xspace}
\def\ee         {\ensuremath{e^-e^-}\xspace}

\def\mumu       {\ensuremath{\mu^+\mu^-}\xspace}

\def\ellm       {\ensuremath{\ell^-}\xspace}
\def\ellp       {\ensuremath{\ell^+}\xspace}

\def\piz   {\ensuremath{\pi^0}\xspace}

\def\pip   {\ensuremath{\pi^+}\xspace}
\def\pim   {\ensuremath{\pi^-}\xspace}

\def\pipm  {\ensuremath{\pi^\pm}\xspace}
\def\pimp  {\ensuremath{\pi^\mp}\xspace}

\def\Kbar  {\kern 0.2em\overline{\kern -0.2em K}{}\xspace}

\def\Kz    {\ensuremath{K^0}\xspace}
\def\Kzb   {\ensuremath{\Kbar^0}\xspace}
\def\KzKzb {\ensuremath{\Kz \kern -0.16em \Kzb}\xspace}
\def\Kp    {\ensuremath{K^+}\xspace}
\def\Km    {\ensuremath{K^-}\xspace}
\def\Kpm   {\ensuremath{K^\pm}\xspace}

\def\KpKm  {\ensuremath{\Kp \kern -0.16em \Km}\xspace}
\def\KS    {\ensuremath{K^0_{\scriptscriptstyle S}}\xspace} 
 
\def\Kstarz  {\ensuremath{K^{*0}}\xspace}

\def\Kstar   {\ensuremath{K^*}\xspace}

\def\Kstarp  {\ensuremath{K^{*+}}\xspace}

\def\Kstarpm {\ensuremath{K^{*\pm}}\xspace}

\def\Dbar    {\kern 0.2em\overline{\kern -0.2em D}{}\xspace}

\def\Dz      {\ensuremath{D^0}\xspace}
\def\Dzb     {\ensuremath{\Dbar^0}\xspace}
\def\DzDzb   {\ensuremath{\Dz {\kern -0.16em \Dzb}}\xspace}
\def\Dp      {\ensuremath{D^+}\xspace}
\def\Dm      {\ensuremath{D^-}\xspace}

\def\DpDm    {\ensuremath{\Dp {\kern -0.16em \Dm}}\xspace}

\def\B       {\ensuremath{B}\xspace}
\def\Bbar    {\kern 0.18em\overline{\kern -0.18em B}{}\xspace}

\def\BB      {\ensuremath{B\Bbar}\xspace} 
\def\Bz      {\ensuremath{B^0}\xspace}
\def\Bzb     {\ensuremath{\Bbar^0}\xspace}
\def\BzBzb   {\ensuremath{\Bz {\kern -0.16em \Bzb}}\xspace}
\def\Bu      {\ensuremath{B^+}\xspace}
\def\Bub     {\ensuremath{B^-}\xspace}
\def\Bp      {\ensuremath{\Bu}\xspace}
\def\Bm      {\ensuremath{\Bub}\xspace}
\def\Bpm     {\ensuremath{B^\pm}\xspace}

\def\BpBm    {\ensuremath{\Bu {\kern -0.16em \Bub}}\xspace}

\def\BorBbar    {\kern 0.18em\optbar{\kern -0.18em B}{}\xspace}
\def\DorDbar    {\kern 0.18em\optbar{\kern -0.18em D}{}\xspace}
\def\KorKbar    {\kern 0.18em\optbar{\kern -0.18em K}{}\xspace}

\def\jpsi     {\ensuremath{{J\mskip -3mu/\mskip -2mu\psi\mskip 2mu}}\xspace}

\mathchardef\Upsilon="7107
\def\Y#1S{\ensuremath{\Upsilon{(#1S)}}\xspace}

\def\FourS {\Y4S}

\def\chic#1{\ensuremath{\chi_{c#1}}\xspace} 

\mathchardef\Deltares="7101
\mathchardef\Xi="7104
\mathchardef\Lambda="7103
\mathchardef\Sigma="7106
\mathchardef\Omega="710A

\def\Deltabar{\kern 0.25em\overline{\kern -0.25em \Deltares}{}\xspace}
\def\Lbar{\kern 0.2em\overline{\kern -0.2em\Lambda\kern 0.05em}\kern-0.05em{}\xspace}
\def\Sigbar{\kern 0.2em\overline{\kern -0.2em \Sigma}{}\xspace}
\def\Xibar{\kern 0.2em\overline{\kern -0.2em \Xi}{}\xspace}
\def\Obar{\kern 0.2em\overline{\kern -0.2em \Omega}{}\xspace}
\def\Nbar{\kern 0.2em\overline{\kern -0.2em N}{}\xspace}
\def\Xb{\kern 0.2em\overline{\kern -0.2em X}{}\xspace}

\def\mes        {\mbox{$m_{\rm ES}$}\xspace}

\def\DeltaE     {\mbox{$\Delta E$}\xspace}

\newcommand{\tev}{\ensuremath{\mathrm{\,Te\kern -0.1em V}}\xspace}
\newcommand{\gev}{\ensuremath{\mathrm{\,Ge\kern -0.1em V}}\xspace}
\newcommand{\mev}{\ensuremath{\mathrm{\,Me\kern -0.1em V}}\xspace}
\newcommand{\kev}{\ensuremath{\mathrm{\,ke\kern -0.1em V}}\xspace}
\newcommand{\ev}{\ensuremath{\mathrm{\,e\kern -0.1em V}}\xspace}
\newcommand{\gevc}{\ensuremath{{\mathrm{\,Ge\kern -0.1em V\!/}c}}\xspace}
\newcommand{\mevc}{\ensuremath{{\mathrm{\,Me\kern -0.1em V\!/}c}}\xspace}
\newcommand{\gevcc}{\ensuremath{{\mathrm{\,Ge\kern -0.1em V\!/}c^2}}\xspace}
\newcommand{\mevcc}{\ensuremath{{\mathrm{\,Me\kern -0.1em V\!/}c^2}}\xspace}

\def\cm   {\ensuremath{{\rm \,cm}}\xspace}

\def\mm   {\ensuremath{{\rm \,mm}}\xspace}

\def\mum  {\ensuremath{{\,\mu\rm m}}\xspace}

\def\invfb   {\ensuremath{\mbox{\,fb}^{-1}}\xspace}
\def\invab   {\ensuremath{\mbox{\,ab}^{-1}}\xspace}

\def\mus  {\ensuremath{\rm \,\mus}\xspace}

\def\ps   {\ensuremath{\rm \,ps}\xspace}

\def\mus        {\ensuremath{\,\mu{\rm s}}\xspace}    
\def\ps         {\ensuremath{{\rm \,ps}}\xspace}  

\def\mrad{\ensuremath{\rm \,mrad}\xspace}               
\def\rad{\ensuremath{\rm \,rad}\xspace}

\def\to                 {\ensuremath{\rightarrow}\xspace}

\newcommand{\stat}{\ensuremath{\mathrm{(stat)}}\xspace}
\newcommand{\syst}{\ensuremath{\mathrm{(syst)}}\xspace}

\def\pep2{PEP-II}

\newcommand{\dedx}{\ensuremath{\mathrm{d}\hspace{-0.1em}E/\mathrm{d}x}\xspace}

\def\gsim{{~\raise.15em\hbox{$>$}\kern-.85em
          \lower.35em\hbox{$\sim$}~}\xspace}
\def\lsim{{~\raise.15em\hbox{$<$}\kern-.85em
          \lower.35em\hbox{$\sim$}~}\xspace}

\def\CP                {\ensuremath{C\!P}\xspace}

\def\stwob{\ensuremath{\sin\! 2 \beta   }\xspace}

\def\deltaz{\ensuremath{{\rm \Delta}z}\xspace}
\def\deltat{\ensuremath{{\rm \Delta}t}\xspace}

\xspace

\def\jetset74   {\mbox{\tt Jetset \hspace{-0.5em}7.\hspace{-0.2em}4}\xspace}

\def\minuit     {\mbox{\tt Minuit}\xspace}

\def\az    {\ensuremath{A_{0}}}
\def\ap    {\ensuremath{A_{\parallel}}}
\def\at    {\ensuremath{A_{\perp}}}

\def\azd   {\ensuremath{|\az|^{2}}}
\def\apd   {\ensuremath{|\ap|^{2}}}
\def\atd   {\ensuremath{|\at|^{2}}}

\def\Imm       {\ensuremath{\Im m}}
\def\Ree       {\ensuremath{\Re e}}

\def\thetakstar {\ensuremath{\theta_{\Kstar}}}

\def\cthetakstar{\ensuremath{\cos{\thetakstar}}}

\def\pipt       {\ensuremath{\Imm{(\ap^{*}\at)}}}
\def\przp       {\ensuremath{\Ree{(\az^{*}\ap)}}}
\def\pizt       {\ensuremath{\Imm{(\az^{*}\at)}}}

\newcommand{\dd}{\text{d}}

\newcommand{\vomega}{\boldsymbol{\omega}}

\def\beq{\begin{equation}}
\def\eeq{\end{equation}}
\def\bea{\begin{eqnarray}}
\def\eea{\end{eqnarray}}
\def\bq{\begin{quote}}
\def\eq{\end{quote}}
\def\ben{\begin{enumerate}}
\def\een{\end{enumerate}}

\def\ctwob{\ensuremath{\cos\! 2 \beta   }\xspace}
\def\ctwobz{\ensuremath{\cos\! 2 \beta_{0}   }\xspace}
\def\stwobz{\ensuremath{\sin\! 2 \beta_{0}   }\xspace}
\newcommand{\SP}{$S$--$P$\xspace}
\newcommand{\jchan}{{b}}
\newcommand{\ichan}{{a}}
\newcommand{\phm}{\hphantom{-}}
\newcommand{\ER}{e_{\cal R}}
\newcommand{\CR}{c_{\cal R}}
\newcommand{\SR}{s_{\cal R}}

\newcommand{\WAVE}{}
\newcommand{\remark}[1]{\relax}

\long\def\inst#1{\par\nobreak\kern 4pt\nobreak
  {\it #1}\par\vskip 10pt plus 3pt minus 3pt}

\begin{document}

\begin{flushright}
~\\
~\\
~\\
~\\
~\\
~\\
~\\
\babar-PUB-\BABARPubYear/\BABARPubNumber \\
SLAC-PUB-\SLACPubNumber \\
hep-ex 0411016 \\
{\it Submitted to Physical Review D}\\
November, 2004\\
\end{flushright}

\title{\large \bf \boldmath \Large Ambiguity-Free Measurement of $\cos 2\beta$:  Time-Integrated and Time-Dependent Angular Analyses   of $\B\to\jpsi K\pi$
\begin{center}  \vskip 5mm The \babar\ Collaboration \end{center} } 

%
\author{B.~Aubert}
\author{R.~Barate}
\author{D.~Boutigny}
\author{F.~Couderc}
\author{J.-M.~Gaillard}
\author{Y.~Karyotakis}
\author{J.~P.~Lees}
\author{V.~Poireau}
\author{V.~Tisserand}
\author{A.~Zghiche}
\affiliation{Laboratoire de Physique des Particules, F-74941 Annecy-le-Vieux, France }
% ...
\collaboration{The \babar\ Collaboration}
\noaffiliation

\date{\today}

\begin{abstract}
We present results on $\B\to \jpsi K\pi$ decays using \epem
annihilation data collected with the \babar\ detector at the \Y4S
resonance. The detector is located at the \pep2\
asymmetric-energy storage ring facility at the
Stanford Linear Accelerator Center.
Using approximately 88 million \BB pairs, we measure the decay
amplitudes for the flavor eigenmodes and observe strong-phase
differences indicative of final-state interactions with a significance
of 7.6 standard deviations.
We use the interference between the $K\pi$ $S$-wave and $P$-wave amplitudes in
the region of the $\Kstar(892)$ to resolve the ambiguity in the
determination of these strong phases.  
We then perform an ambiguity-free measurement of $\cos2\beta$ using
the angular and time-dependent asymmetry in $B\to \jpsi
K^{*0}(\KS\piz$) decays.  
With \stwob fixed at its measured value and \ctwob treated as an independent parameter,
we find $\cos 2\beta=2.72_{-0.79}^{+0.50}\stat \pm 0.27\syst$, determining the
sign of $\cos 2\beta$ to be positive at 86\% CL.
\end{abstract}


\maketitle

\newpage

\setcounter{footnote}{0}

The Standard Model of electroweak interactions describes \CP\ violation 
in weak interactions of quarks by the presence of a nonzero phase in
the three-generation Cabibbo-Kobayashi-Maskawa (CKM) quark-mixing
matrix~\cite{KobayashiMaskawa}. In this framework, the \CP-violation parameter
\stwob\ can be measured by examining the proper-time distributions
of neutral $B$-meson
decays to final states containing a charmonium
meson  and a neutral kaon.  The Belle~\cite{BelleSin2BetaPLB2002} and 
\babar~\cite{BaBarSin2BetaPRL2002} experiments have recently performed
precise measurements  of  $\sin 2\beta$, leading to a world average of 
$0.731 \pm 0.056$~\cite{PDG2004}.
  These measurements determine $\beta$  up to a four-fold ambiguity, corresponding
to the two different signs of $\ctwob$ and the transformation $\beta \rightarrow \pi + \beta$.

One of the possible values of  $\beta$  is compatible with  measurements
of other quantities that constrain the Unitarity Triangle~\cite{PDG2004}.  However,
it is still possible that, because of contributions from new physics,
the actual value of $\beta$ is one of the three other values consistent
with the measurement of $\sin2\beta$~\cite{GronauLondonNPPRD97,Kayser:1999bt,Fleischer:2003xx}.
A measurement of the sign of $\ctwob$ would either
agree with the standard interpretation $\beta\approx 0.41$ and with its
indistinguishable nonstandard alternative $\beta\approx 0.41 +\pi$, or would exclude
these and instead imply the nonstandard solutions $\beta\approx 1.16$ and $\beta\approx 1.16+\pi$.

Several strategies to determine $\cos2\beta$ have been proposed 
 \cite{Kayser:1999bt}, 
\cite{Azimov:1990xv,Grossman:1997gd,Charles:1998vf,Browder:1999ng,Quinn:2000jy,Dunietz:2000cr}.
In particular, $\cos2\beta$ appears as a factor in the  interference between the
$CP$-odd and the two $CP$-even amplitudes in the time- and angle-dependent  distribution
describing the decay $B\to J/\psi K^{*0}$ ($K^{*0}\to K_S^0\pi^0$, $\jpsi\to\ellp\ellm$)
\cite{DunietzPRD1990,CharlesLeYaouancOliverPenePRD1998,DigheDunietzFleischer,ChiangPRD2000}.
However, neither this distribution nor the time-integrated angular
distributions of the companion channels $B^0\to J/\psi
K^{*0}(K^{*0}\to K^+\pi^-)$ and $B^+\to J/\psi K^{*+}$ (and related
charge-conjugate decays) can resolve a two-fold ambiguity in the
relative strong phases among the
three  
$\B\to\jpsi\Kstar$ decay
amplitudes.
This  leaves an overall sign ambiguity in  $\cos 2\beta$ 
\cite{DigheDunietzFleischerResolving,BaBarFullAngularAnalysisPRL2001}.
Resolving the ambiguity from these partial waves alone would require the measurement of the polarization of the leptons from the $J/\psi$ decay
\cite{ChiangWolfensteinPRD2000}.
This could be done in principle using $\jpsi\to\mu^+\mu^-$ decays or with
$\psi(2S)\to\tau^+\tau^-$ decays by measuring the lepton polarizations.
Such measurements are not feasible today.
Theoretical arguments, based on the analysis of $s$-quark helicity
conservation, suggest a ``preferred'' set of strong phases~\cite{Suzuki:2001za},
but cannot guarantee the validity of this set.

In this analysis, we use the known dependence on $K\pi$ invariant mass of the
relative phase between the $S$-wave and $P$-wave $K\pi$ $I=1/2$ scattering amplitudes
in the vicinity of the $\Kstar(892)$ to resolve the two-fold ambiguity in the
relative strong phases among the three amplitudes for $\B\to\jpsi\Kstar$.
The dominant $P$-wave has the canonical Breit-Wigner form
with a phase $\delta_P$ that increases rapidly with $K\pi$ invariant mass $m_{K\pi}$, while the
$S$-wave phase $\delta_S$ increases slowly with $m_{K\pi}$~\cite{Aston:1987ir}. Accordingly,
$\delta_S - \delta_P$, where $\delta_P$ is assumed to be common to the
three \WAVE{$B\to\jpsi K\pi$} $P$-wave amplitudes, is expected to decrease rapidly
as $m_{K\pi}$ increases from below to above the \Kstar resonance.
We find that one solution for \WAVE{$\delta_S - \delta_P$} yields this expected behaviour while the other has precisely the opposite behavior.
In this way, the ambiguity is
resolved, so that the subsequent time-dependent analysis yields a value
of $\ctwob$ that is free of the associated sign ambiguity.

\begin{table*}
\caption{\label{tab:fitted-values-comp}
The $\B\to\jpsi K^*(892)$ amplitude parameters (described in
Sec. \ref{sec:AngularTimeDepDecayRateProb}) as measured by several
experiments.  
The results in Ref. \protect\cite{BaBarFullAngularAnalysisPRL2001} are
superseded by this work.  Note that the phases are subject to a
two-fold ambiguity, as described by
Eq.~(\protect\ref{eqn:phase_ambiguity}).
}
\begin{ruledtabular}
\begin{tabular}{lcccc}
 & $|A_0|^2$ & $|A_\perp|^2$ & $\delta_\parallel-\delta_0$ (\rad) & $\delta_\perp-\delta_0$ (\rad) \\ \hline
\text{CLEO}
\protect\cite{CLEOFullAngularAnalysisPRL97} & $0.52 \pm 0.07 \pm 0.04$ & $0.16 \pm 0.08 \pm 0.04$ & $3.00 \pm 0.37 \pm 0.04$ & $-0.11 \pm 0.46 \pm 0.03 $\\
\text{CDF }
\protect\cite{CDFFullAngularAnalysisPRL2000} & $0.59 \pm 0.06 \pm 0.01$ & $0.13^{+0.12}_{-0.09} \pm 0.06$ & $2.2 \pm 0.5 \pm 0.1$ & $-0.6 \pm 0.5 \pm 0.1$ \\
\babar\ 
\protect\cite{BaBarFullAngularAnalysisPRL2001} & $0.597 \pm 0.028 \pm 0.024$ & $0.160 \pm 0.032 \pm 0.014$ & $2.50 \pm 0.20 \pm 0.08$ & $-0.17 \pm 0.16 \pm 0.07$ \\
\text{Belle}
\protect\cite{BelleFullAngularAnalysisICHEP2002} & $0.617 \pm 0.020 \pm 0.027$ & $0.192 \pm 0.023 \pm 0.026$ & $2.83 \pm 0.19 \pm 0.08$ & $-0.09 \pm 0.13 \pm 0.06$ \\
\end{tabular}
\end{ruledtabular}
\end{table*}

We perform an angular analysis of the $\B \to \jpsi \Kstar$ decay mode and measure \ctwob on a sample
of $ (88 \pm 1) \times 10^{6}$ \BB pairs
collected with the  \babar\ detector at the
\pep2\ asymmetric-energy \B Factory.
These data
correspond to an integrated luminosity of 81.9 \invfb
recorded at the \FourS resonance. The analysis is  
performed in three distinct stages.

     In the first stage the time-integrated angular distributions
   describing the decay channels $\Bz\to\jpsi\Kstarz, \Kstarz\to\Kp\pim$ and
   $\Bp\to\jpsi\Kstarp, \Kstarp\to\KS\pip$ and $\Kstarp\to\Kp\piz$,
   together with those of the related charge-conjugate modes, are analyzed
   assuming that the $K\pi$ system may be described entirely in terms of
   $P$-wave amplitudes.
   The weak process  $b\to c \overline{c} s$ is a $\Delta I=0$ interaction, so the amplitudes for \Bz and \Bp decay should be equal, as should be those for \Bzb and \Bm.
   A convenient description of the decays is provided in the transversity basis~\cite{DunietzPRD1990}
   since the related amplitudes have well-defined  $CP$ parities, which is of particular relevance for the $\ctwob$ measurement.
   The formalism is described in 
   Sec.~\ref{TimeIntegratedAngDist} and the results of its application to the data are
   presented in Sec.~\ref{sec:AngularAnalysis} in the form of transversity-amplitude
   magnitudes and relative phases. There is an
   intrinsic mathematical ambiguity associated with the phases; the
   relevant transformation expressed in Eq.~(\ref{eqn:phase_ambiguity}) below
   leaves the angular distribution unchanged.

     This ambiguity can be resolved by extending the
   formalism to include a $K\pi$ $S$-wave amplitude and then measuring
   the $K\pi$ mass-dependence of its phase difference with respect to the
   $P$-waves.
   \WAVE{We will show that including a $K\pi$ $S$-wave with a significant \SP interference is required to describe the
   data (see Sec.~\ref{sec:probingTheSPInterference}).}
   The extended angular distribution
   is presented in Sec.~\ref{subsec:ProbingthepresenceofaKpiSwave}, and its use in
   resolving the phase ambiguity is described in Sec.~\ref{sec:SolvingtheStrongPhasesAmbiguity}. This is the
   second stage in the analysis procedure.

     The $P$-wave parameters extracted in Sec.~\ref{sec:AngularAnalysis} are
   only slightly affected by the presence of an $S$-wave amplitude since  in performing the analysis
   the data are integrated over a broad $(\pm 100 \mevcc)$ $K\pi$ mass
   interval centered on the $K^*(892)$. The
   \SP interference contributions essentially average out over this
   region, and since the $S$-wave intensity proves to be only a few 
   percent of that of the $P$-wave, the presence of the $S$-wave can be accounted for by 
   a small additional source of systematic uncertainty (Table~\ref{tab:syst-ch234},
   line 7).

The third stage of the analysis is the application of
the time-dependent formalism to the $\Bz\to \jpsi \Kstarz(\KS\piz)$ decay
channel, as described in Sec.~\ref{sec:time_dependent_ang_dist}. There, the $K\pi$ $S$-wave is omitted
and the $P$-wave parameters are fixed to those obtained during the
first stage of the analysis.
   The phase ambiguity discussed
   in Sec.~\ref{TimeIntegratedAngDist} translates into a sign ambiguity for
   $\ctwob$. The $K\pi$ \SP interference analysis of Sec.~\ref{sec:AngularAnalysis}
   resolves the ambiguity and results in the assignment of a
   unique sign to the term in the time-dependent angular distribution
   that is proportional to $\ctwob$. The time-dependent analysis of the $\Bz\to \jpsi \Kstarz(\KS\piz)$
   data sample, which is statistically independent of that
   used for the measurement of the amplitudes, is presented in Sec.~\ref{sec:Measurementofcos2beta}.
   We summarize the results
   of the paper in Sec.~\ref{sec:Conclusion}.

Several experiments have determined the decay amplitudes
in $\B \to \jpsi \Kstar$ decay.  These results
are summarized in Table~\ref{tab:fitted-values-comp}.
The measurements presented here supersede previous \babar\ results
\cite{BaBarFullAngularAnalysisPRL2001}, which are based on  subsets of the data used for this analysis.
The data reconstruction and Monte Carlo simulation procedures have undergone significant improvement since
our previous analysis;
the reconstruction of $\jpsi K^*$ channels involving a \piz has been improved (Sec.~\ref{sec:EventSelection})
leading to a better purity; a dedicated background subtraction method has been developed
(Sec.~\ref{subsec:BackgroundSubtraction} and Appendix~\ref{appendx:pseudoLK}).

\section{Angular- and Time-Dependent Differential Decay Rates}
\label{sec:AngularTimeDepDecayRateProb}

The \B decay amplitudes are measured from the time-integrated
differential decay distribution, expressed in the transversity basis.
\begin{figure}
\includegraphics[width=\singlewidth]{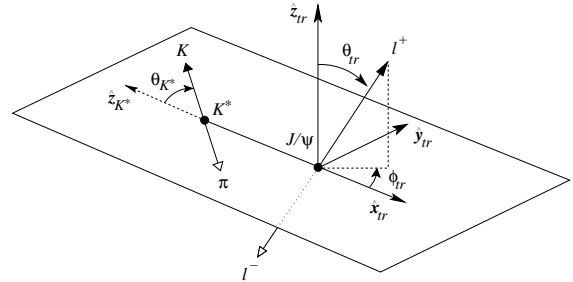}
\caption{\label{fig:trans-frame}Definition of the transversity angles $(\theta_{K^*},\theta_{tr},\phi_{tr})$ and 
coordinate system $(\hat{\bm{ x}}_{tr},\hat{\bm{y}}_{tr},\hat{\bm{z}}_{tr})$. 
The direction opposite to the \B meson momentum  in the $\jpsi$ rest frame is $\hat{\bm{x}}_{tr}$; 
$\hat{\bm{y}}_{tr}$ is perpendicular to $\hat{\bm{x}}_{tr}$ in the
plane that contains
$\hat{\bm{x}}_{tr}$ and $\vec{\bm{p}}_{K}$,  chosen so
$\vec{\bm{p}}_{K}\cdot \hat{\bm{y}}_{tr} > 0$;
$\hat{\bm{z}}_{tr} = \hat{\bm{x}}_{tr}\times\hat{\bm{y}}_{tr}$.
The helicity angle $\theta_{K^*}$ of the \Kstar decay is the 
 angle between the direction opposite to the \B meson flight
direction and the kaon momentum, in the \Kstar rest frame.
Finally, $\theta_{tr}$ and $\phi_{tr}$
are the polar and azimuthal angle of the positive lepton defined in the \jpsi rest frame.
}
\end{figure}
The definitions of the transversity frame and  the related
transversity angles $(\theta_{K^*},\theta_{tr},\phi_{tr})$  are shown in Fig.~\ref{fig:trans-frame}.

\subsection{The Time-Integrated $\jpsi K^*$ Angular Distribution}
\label{TimeIntegratedAngDist}

We first consider only the $K^*(892)$ $K\pi$ mass region.
The amplitude for  longitudinal polarization of the two
vector mesons is $A_0$. 
There are two amplitudes for polarizations of the vector mesons
transverse to the decay axis: $A_\parallel $ for parallel polarization
of the two vector mesons and $A_\perp$ for their perpendicular polarization.
The three independent amplitudes are assumed to have a common dependence on $m_{K\pi}$.
This common dependence is irrelevant to the angular distribution and each of the three amplitudes is thus represented by a complex constant.

In terms of  the angular variables
$\vomega \equiv (\theta_{K^*},\theta_{tr},\phi_{tr})$,
the  time-integrated differential decay rate for the decay of the $B$ meson to 
the $\jpsi\left(\Kp\pim\right)_{P\text{-wave}}$, $\jpsi\left(\Kp\piz\right)_{P\text{-wave}}$, or $\jpsi\left(\KS\pip\right)_{P\text{-wave}}$ final state is
\begin{eqnarray}
 \label{eqn:g_definition}
 g(\vomega;\boldsymbol{A}) &\equiv& \frac{1}{\Gamma}\frac{{\rm d}^3\Gamma}{{\rm d}\cos\theta_{K^*}{\rm d}\cos\theta_{tr}{\rm d}\phi_{tr}} \nonumber \\
 &=& f_1(\vomega) |A_0|^2 + f_2(\vomega) |A_\parallel |^2 + f_3(\vomega) |A_\perp|^2+\nonumber\\
 &~& f_4(\vomega) \Imm (A_\parallel ^*A_\perp)+ f_5(\vomega) \Ree (A_\parallel A_0^*)+  \nonumber\\
&~&f_6(\vomega) \Imm (A_\perp A_0^*),
\end{eqnarray}
where the functions  $f_i(\vomega)$ are
\begin{eqnarray}
f_1(\vomega) &\equiv& \;\;\,\frac{9}{32\pi} 2\cos^2\theta_{K^*}\left[1-\sin^2\theta_{tr}\cos^2\phi_{tr}\right], \nonumber \\
f_2(\vomega) &\equiv& \;\;\,\frac{9}{32\pi} \sin^2\theta_{K^*}\left[1-\sin^2\theta_{tr}\sin^2\phi_{tr}\right], \nonumber \\
f_3(\vomega) &\equiv& \;\;\,\frac{9}{32\pi} \sin^2\theta_{K^*}\sin^2\theta_{tr}, \nonumber \\
f_4(\vomega) &\equiv& \;\;\,\frac{9}{32\pi} \sin^2\theta_{K^*}\sin2\theta_{tr}\sin\phi_{tr}, \nonumber \\
f_5(\vomega) &\equiv& -\frac{9}{32\pi} \frac{1}{\sqrt{2}}\sin2\theta_{K^*}\sin^2\theta_{tr}\sin2\phi_{tr}, \nonumber \\
f_6(\vomega) &\equiv& \;\;\,\frac{9}{32\pi} \frac{1}{\sqrt{2}}\sin2\theta_{K^*}\sin2\theta_{tr}\cos\phi_{tr}.
\label{eq:theAngularFunctions}
\end{eqnarray}
Equations~(\ref{eqn:g_definition},\ref{eq:theAngularFunctions})
have been obtained by summing  over the unobserved lepton polarizations~\cite{Dighe:1995pd,ChiangWolfensteinPRD2000,StephThese}.
 
The symbol $\boldsymbol{A}$ denotes the transversity amplitudes for the decay of the \B\ meson:
$\boldsymbol{A} \equiv (A_0,A_\parallel ,A_\perp)$.
 We set $|A_0|^2 + |A_\parallel |^2 + |A_\perp|^2 = 1$, so that
$g(\vomega;\boldsymbol{A})$~(Eq.~(\ref{eqn:g_definition})) is a probability density function (PDF).
We denote by $\boldsymbol{\overline{A}}$ the amplitudes for the
$\Bbar$ meson decay. In the absence of direct \CP violation, we
can choose a phase convention so that these amplitudes are related by
$\overline{A}_0 = +A_0 $,
$\overline{A}_\parallel = +A_\parallel $,
$\overline{A}_\perp = -A_\perp $,
so that $A_\perp $ is \CP-odd and $A_0$ and ${A}_\parallel $ are \CP-even.
Fixing this phase convention also fixes the phase of the amplitude
for \Bz--\Bzb mixing.

The phases $\delta_i$ of the amplitudes, where $i= 0, \parallel , \perp$, are
defined by $A_i = |A_i| e^{i \delta_i}$.
Obviously, only differences of phases appear in the differential decay rate  through
the observables 
\begin{eqnarray}
\label{eqn:phase_observable}
\Imm (A_\parallel ^*A_\perp) &=& |A_\parallel ||A_\perp| \sin(\delta_\perp - \delta_\parallel ), \nonumber\\
\Ree (A_\parallel A_0^*)     &=& |A_\parallel ||A_0|     \,\cos(\delta_\parallel -\delta_0), \nonumber\\
\Imm (A_\perp A_0^*)         &=& |A_\perp|     |A_0|     \sin(\delta_\perp-\delta_0),
\end{eqnarray}
so that the differential decay rate (Eq.(\ref{eqn:g_definition})) is invariant under the tranformation
\begin{eqnarray}
\label{eqn:phase_ambiguity}
(\delta_\parallel-\delta_0 , \delta_\perp-\delta_0) \longleftrightarrow (\delta_0-\delta_\parallel , \pi+\delta_0 - \delta_\perp).\\ \nonumber
\end{eqnarray}
This is the above-mentioned ambiguity.

The three terms that would allow to resolve the ambiguity 
 ($\Ree (A_\parallel ^*A_\perp), \Imm (A_\parallel A_0^*)$
and $\Ree (A_\perp A_0^*)$)
vanish after summation over the unobserved final lepton
polarizations.

We ensure $|A_0|^2 + |A_\parallel |^2 + |A_\perp|^2 = 1$ by parametrizing the magnitudes
of the three \B-decay amplitudes by
\begin{eqnarray}
\label{eqn:polar_description_of_amplitudes}
\cos\theta_A &\equiv& |A_0|, \nonumber\\
\sin\theta_A \cos\phi_A &\equiv& |A_\parallel |, \nonumber\\
\sin\theta_A \sin\phi_A &\equiv& |A_\perp|. 
\end{eqnarray}
with $0\leq \theta_A \leq \pi/2$, $0\leq\phi_A\leq\pi/2$.

\subsection{Angular  Distributions Including a $\boldsymbol{K\pi}$ $\boldsymbol{S}$-Wave Contribution}
\label{subsec:ProbingthepresenceofaKpiSwave}
The $K\pi$ system originating from $B\to\jpsi(K\pi)$ can, in principle, have any integer spin.
The experiment with the largest $K\pi$ sample, LASS~\cite{Aston:1987ir}, showed however that below 1.3 \gevcc, the $S$ and $P$ waves dominate.
We have previously observed a broad structure \cite{BaBarFullAngularAnalysisPRL2001} in the 1.1 -- 1.3 \gevcc range of the $m_{K \pi}$ spectrum and found it to be
compatible with a significant $S$-wave amplitude contribution.
When a $K\pi$ $S$ wave  in the \B decay amplitude is included in addition to the
$K\pi$ $P$ wave, the differential decay rate (Eq.~(\ref{eqn:g_definition})) becomes \cite{StephThese}
\begin{widetext}
\begin{eqnarray}
G_{S+P}(\vomega,m_{K\pi};\boldsymbol{A},A_P,A_S) &\equiv& \frac{1}{\Gamma}
\frac{{\rm d}^4 \Gamma}{{\rm d} m_{K\pi}{\rm d}\cos\theta_{K^*}{\rm d}\cos\theta_{tr}{\rm d}\phi_{tr}} \nonumber\\
&\propto& pq \times \Bigg{[} A_P^2 g(\vomega;\boldsymbol{A}) + |A_S|^2 f_7(\vomega) + \nonumber \\
 &~& \hspace{1.0cm}A_P \left[
 f_8(\vomega) \Ree \left( A_\parallel A_S^* \right) + 
 f_9(\vomega) \Imm \left(A_\perp A_S^* \right) + 
 f_{10}(\vomega) \Ree \left(A_0 A_S^* \right) \right]
\Bigg{]},
\label{eqn:g_S+P_definition}
\end{eqnarray}
\end{widetext}
where we have kept the notation $\theta_{K^*}$ for the $(K\pi)$
helicity angle;
$p$ is the $K\pi$-system momentum in the \B rest frame and
$q$ is the kaon momentum in the $K\pi$ rest frame;
we chose $A_P$ to a be real and positive function of $m_{K\pi}$. Its square is
indicative of the overall strength of the $P$-wave amplitudes.
 We represent the $m_{K\pi}$-dependent $S$-wave amplitude as $A_S=|A_S|e^{i\delta_S}$.
The phases of the $P$-wave amplitudes reside in $A_0$, $A_\|$, and $A_\perp$.

Using the same phase convention as for the $P$-wave amplitudes,  $\overline{A}_S = A_S$.
The angular functions $f_{7\cdots10}$ are
\begin{eqnarray}
\label{eqn:f7-10_definition}
f_7(\vomega) &\equiv& \;\;\,\frac{3}{32\pi} 2\left[1-\sin^2\theta_{tr}\cos^2\phi_{tr}\right], \nonumber\\
f_8(\vomega) &\equiv& -\frac{3}{32\pi} \sqrt{6}\sin\theta_{K^*}\sin^2\theta_{tr}\sin2\phi_{tr},\nonumber\\
f_9(\vomega) &\equiv& \;\;\,\frac{3}{32\pi} \sqrt{6}\sin\theta_{K^*}\sin2\theta_{tr}\cos\phi_{tr},\nonumber\\
f_{10}(\vomega) &\equiv& \;\;\,\frac{3}{32\pi} 4\sqrt{3}\cos\theta_{K^*}\left[1-\sin^2\theta_{tr}\cos^2\phi_{tr}\right].\nonumber\\
\end{eqnarray}

At a given $m_{K\pi}$, the normalization is obtained by introducing the parametrization
\begin{eqnarray}
\label{eqn:lambda_definition}
\cos\lambda &\equiv& \frac{A_P}{\sqrt{A_P^2 + |A_S|^2}}, \nonumber \\
\sin\lambda &\equiv& \frac{|A_S|}{\sqrt{A_P^2 + |A_S|^2}},
\end{eqnarray}
where $\lambda$ is in the range $[0,\pi/2]$. The term
$\cos^2\lambda$ ($\sin^2\lambda$) represents the fraction of the $P$-wave ($S$-wave) intensity at that value of $m_{K\pi}$.
The distribution~(Eq.~(\ref{eqn:g_S+P_definition})), normalized so that at any fixed $m_{K\pi}$ the integral over the angular variables yields unity, is given by
\begin{eqnarray}
\label{eqn:g_S+P_reduite}
g_{S+P}(\vomega;m_{K\pi},\boldsymbol{A},\lambda) &\equiv& \cos^2\lambda\, g(\vomega;\boldsymbol{A})
 + \sin^2\lambda\, f_7(\vomega)  \nonumber \\ &+& \frac{1}{2}\sin 2\lambda\Big{[} \nonumber\\
&~&\;\;\, f_8(\vomega) \cos(\delta_\parallel -\delta_S) |A_\parallel | \nonumber\\
&~&+ f_9(\vomega) \sin(\delta_\perp-\delta_S) |A_\perp| \nonumber \\
&~&+ f_{10}(\vomega) \cos(\delta_S-\delta_0) |A_0| \,\,\,\Big{]}.\nonumber \\
\end{eqnarray}
In Eq.~(\ref{eqn:g_S+P_reduite}), the dependence of $g_{S+P}$ on $m_{K\pi}$ follows from that
of $\lambda$ and of the strong phases $\delta_i$ $(i=\perp,\parallel,0,S)$.
We see that at a given value of $m_{K\pi}$ the equations are invariant under the transformation
\begin{eqnarray}
\label{eqn:phase_S+P_ambiguity}
&&(\delta_\parallel-\delta_0 , \delta_\perp-\delta_0, \delta_S-\delta_0) 
\longleftrightarrow \nonumber
\\
&& \qquad(\delta_0-\delta_\parallel , \pi+\delta_0-\delta_\perp, \delta_0-\delta_S).
\end{eqnarray}
We will use the change of the \SP relative phase in the region of the $K^*(892)$
to resolve this ambiguity.

The phase of a weak decay amplitude is determined by phases introduced
through the weak interaction itself, that is from the CKM matrix, and
by strong final-state interactions.
If in the decay $B\to \jpsi K\pi$ the $\jpsi$ were known not to
interact with the $K\pi$ system, Watson's final-state interaction
theorem~\cite{watson} would guarantee that the phases for the $P$-wave
and $S$-wave final states would be simply the corresponding phase
shifts in $P$-wave and $S$-wave $K\pi$ scattering at the appropriate
invariant mass, taking $K\pi$ scattering to be elastic in this range.
However, we know this is not exactly the case, for if it were, the
three individual $P$-wave amplitudes would be relatively real (relative phases $0$ or $\pi$).  This
is not the experimental result, as we shall show.
Nonetheless, we will provisionally adopt the assumption that the
interactions with the $\jpsi$ are small, and in particular that they do not 
change significantly with $m_{K\pi}$.
We then anticipate that the difference $\delta_S-\delta_0$ will behave much like the difference $\delta(K\pi({L=0})) -
\delta(K\pi({L=1}))$, where we restrict ourselves to the $I=1/2$ channel, which is produced in the $B\to \jpsi K\pi$ decay.
According to Wigner's causality principle~\cite{Wigner}, the phase of a resonant amplitude increases with increasing invariant mass.
Since the $K\pi$, $I=1/2$ $P$-wave phase shift increases rapidly in
the vicinity of the $K^*(892)$, while the corresponding $S$-wave
increases only very gradually, we expect $\delta_S-\delta_0$,
$\delta_S-\delta_\perp$, and $\delta_S-\delta_\parallel$ to fall
rapidly with increasing $m_{K\pi}$ in this region.

\subsection{Time-Dependent  Angular Distribution}
\label{sec:time_dependent_ang_dist}

The time-dependent angular distribution for a \Bz meson produced at
time $t=0$ decaying as \Bz\to \jpsi \Kstarz (\Kstarz\to \KS \piz)
at proper time $t$ has the same form
as in Eq.~(\ref{eqn:g_definition}) but with time-dependent amplitudes $\boldsymbol{A}(t)$:
\begin{eqnarray}
\label{eqn:time_dependent_g}
g(\vomega;\boldsymbol{A}(t),\sin2\beta,\cos 2\beta) &\equiv& \frac{1}{\Gamma}\frac{{\rm d}^4 \Gamma}{{\rm d}t\; {\rm d}\cos\theta_{K^*}{\rm d}\cos\theta_{tr}{\rm d}\phi_{tr}}.
\nonumber\\
\end{eqnarray}
Under the hypothesis of no direct \CP violation in the decay, i.e.
 $|A_i(0)| = |\overline{A}_i(0)|,\; i = 0,\parallel ,\perp$,
the corresponding terms that enter Eq.~(\ref{eqn:time_dependent_g}) are
\cite{DigheDunietzFleischer,ChiangPRD2000,StephThese}
\begin{widetext}
\begin{eqnarray}
\label{eqn:amplitude_de_t}
\left|\stackrel{(-)}{A_0}\!(t)\right|^2 &\equiv& e^{-\Gamma_0 t}|A_0|^2 \left[1\stackrel{(-)}{+} \sin 2\beta\sin\Delta m\, t\right], \nonumber\\
\left|\stackrel{(-)}{A_\parallel} \!(t)\right|^2 &\equiv& e^{-\Gamma_0 t}|A_\parallel |^2 \left[1\stackrel{(-)}{+} \sin 2\beta\sin\Delta m\, t\,\right], \nonumber\\
\left|\stackrel{(-)}{A_\perp}\!(t)\right|^2 &\equiv& e^{-\Gamma_0 t}|A_\perp|^2 \, \left[1\stackrel{(+)}{-} \sin 2\beta\sin\Delta m\, t\,\right], \nonumber\\
\Imm \left( \stackrel{(-)}{A_\parallel}^*\!\!\!\!(t)\stackrel{(-)}{A_\perp}\!\!(t)\right) &\equiv& \stackrel{(-)}{+} e^{-\Gamma_0 t}|A_\parallel ||A_\perp| \left[\sin(\delta_\perp-\delta_\parallel )\cos\Delta m\, t\,-
 \cos(\delta_\perp-\delta_\parallel )\cos 2\beta\sin\Delta m\, t\,\right], \nonumber\\
\Ree \left(\stackrel{(-)}{A_\parallel}\!(t)\stackrel{(-)}{A_0}^*\!\!\!\!(t)\right) &\equiv& e^{-\Gamma_0 t}|A_\parallel ||A_0|\,\, \cos(\delta_\parallel-\delta_0) \left[1\stackrel{(-)}{+} \sin 2\beta\sin\Delta m\, t\,\right], \nonumber\\
\Imm \left(  \stackrel{(-)}{A_\perp}\!\!(t) \stackrel{(-)}{A_0}^*\!\!\!\!(t)\right) &\equiv& \stackrel{(-)}{+} e^{-\Gamma_0 t}|A_\perp||A_0| \left[\sin(\delta_\perp-\delta_0)\cos\Delta m\, t\,-
 \cos(\delta_\perp-\delta_0)\cos 2\beta\sin\Delta m\, t\,\right],
\end{eqnarray}
for an initial $\B^0$ ($\overline{B}^0$) meson.
The mass difference between the two neutral \B mass eigenstates is $\Delta m$,
and $\Gamma_0$ is the common neutral \B-meson decay rate,
neglecting the lifetime difference between these mass eigenstates.
The expression for the differential decay rate can be recast in the following form \cite{StephThese}:
\begin{eqnarray}
g_{\eta}(\vomega,t;\boldsymbol{A},\sin 2\beta,\cos 2\beta) &=& \frac{\Gamma_0}{2} e^{-\Gamma_0 t}{\cal A}(\vomega;\boldsymbol{A})\times \nonumber\\
&~&\hspace{-2.0cm}\left\{1 + \eta \left[
 \cos\Delta m\, t\, \frac{{\cal P}(\vomega;\boldsymbol{A})}{{\cal A}(\vomega;\boldsymbol{A})}+
 \sin\Delta m\, t\,\left(
 \frac{{\cal S}(\vomega;\boldsymbol{A})}{{\cal A}(\vomega;\boldsymbol{A})} \sin 2\beta + 
 \frac{{\cal C}(\vomega;\boldsymbol{A})}{{\cal A}(\vomega;\boldsymbol{A})} \cos 2\beta
 \right)
 \right]\right\},
\label{eqn:asymm_CP_1}
\end{eqnarray}
with $\eta=+1$ ($\eta=-1$) for an initial $\B^0$ ($\overline{B}^0$) meson. The angular terms $\cal{A}$, $\cal{P}$, $\cal{S}$, and ${\cal C}$ are
\begin{eqnarray}
\label{eqn:def_ABCD}
{\cal A}(\vomega;\boldsymbol{A}) &\equiv& \phm f_1(\vomega)|A_0|^2+f_2(\vomega)|A_\parallel |^2+f_3(\vomega)|A_\perp|^2+f_5(\vomega)|A_0||A_\parallel |\cos(\delta_\parallel-\delta_0) , \nonumber \\
{\cal P}(\vomega;\boldsymbol{A}) &\equiv& \phm f_4(\vomega)|A_\parallel ||A_\perp|\sin(\delta_\perp-\delta_\parallel )+f_6(\vomega)|A_0||A_\perp|\sin(\delta_\perp-\delta_0), \nonumber \\
{\cal S}(\vomega;\boldsymbol{A}) &\equiv& \phm f_1(\vomega)|A_0|^2+f_2(\vomega)|A_\parallel |^2-f_3(\vomega)|A_\perp|^2+f_5(\vomega)|A_0||A_\parallel |\cos(\delta_\parallel-\delta_0) , \nonumber \\
{\cal C}(\vomega;\boldsymbol{A}) &\equiv& -f_4(\vomega)|A_\parallel ||A_\perp|\cos(\delta_\perp-\delta_\parallel )-f_6(\vomega)|A_0||A_\perp|\cos(\delta_\perp-\delta_0).
\end{eqnarray}
The time-dependent asymmetry in the decay then reads
\begin{eqnarray}
\label{eqn:asymm_CP}
a(\vomega,t;\boldsymbol{A},\sin 2\beta,\cos 2\beta) &\equiv&
\frac{
g_{+1}(\vomega,t;\boldsymbol{A},\sin 2\beta,\cos 2\beta) -
g_{-1}(\vomega,t;\boldsymbol{A},\sin 2\beta,\cos 2\beta) 
}
{
g_{+1}(\vomega,t;\boldsymbol{A},\sin 2\beta,\cos 2\beta) +
g_{-1}(\vomega,t;\boldsymbol{A},\sin 2\beta,\cos 2\beta) 
} \nonumber \\
 &=& 
\cos\Delta m\, t\, \frac{{\cal P}(\vomega;\boldsymbol{A})}{{\cal A}(\vomega;\boldsymbol{A})}+
 \sin\Delta m\, t\,\left(
 \frac{{\cal S}(\vomega;\boldsymbol{A})}{{\cal A}(\vomega;\boldsymbol{A})} \sin 2\beta + 
 \frac{{\cal C}(\vomega;\boldsymbol{A})}{{\cal A}(\vomega;\boldsymbol{A})} \cos 2\beta
 \right).
\end{eqnarray}
\end{widetext}
This reduces to the usual expression for decays to \CP eigenstates 
when only the \CP-even ($A_0$, $A_\parallel$) amplitudes are nonzero
or when only the \CP-odd ($A_\perp$) amplitude is nonzero.
We now examine the terms on the right hand side of Eq.~(\ref{eqn:asymm_CP}):
\begin{figure}[htpb]
\psfrag{PSURA}{\ \ ${\cal P}/{\cal A}$}
\psfrag{SSURA}{\ \ ${\cal S}/{\cal A}$}
\psfrag{CSURA}{\ \ ${\cal C}/{\cal A}$}
\begin{center}
\includegraphics[width=\singlewidth]{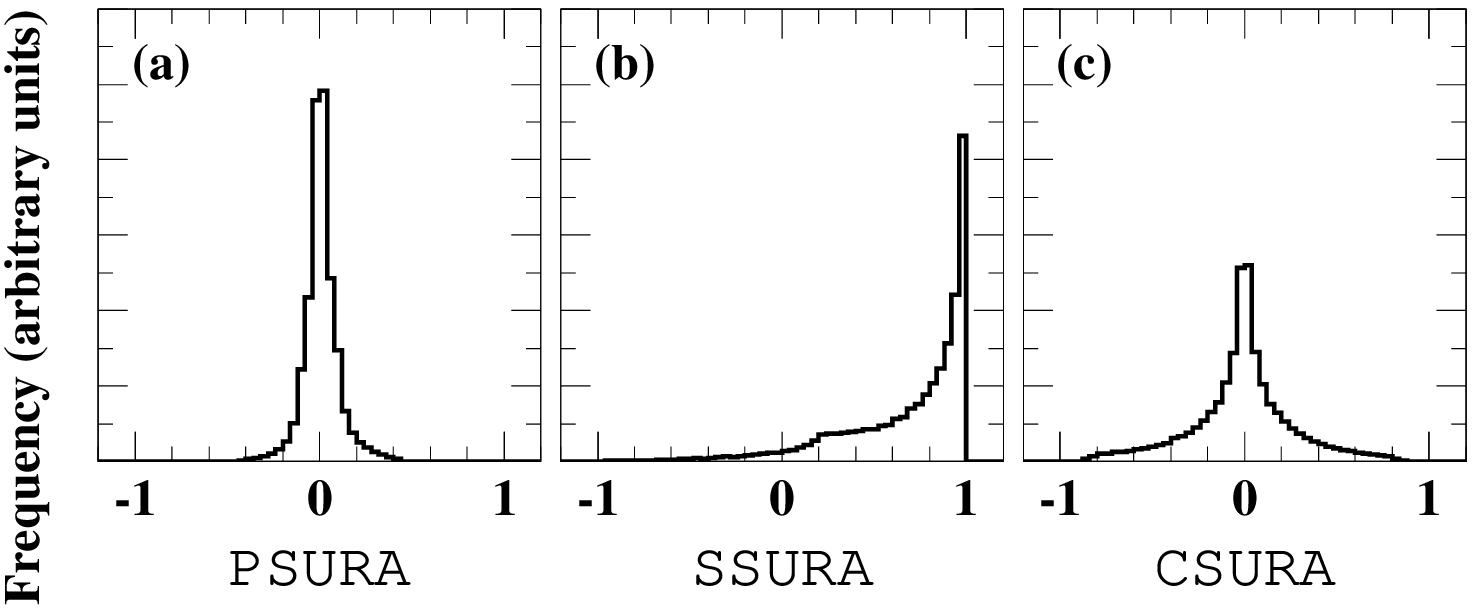}
\end{center}
\caption{\label{fig:ABCD}(a) The distribution of 
${{\cal P}(\vomega;\boldsymbol{A})}/{{\cal A}(\vomega;\boldsymbol{A})}$,
(b) ${{\cal S}(\vomega;\boldsymbol{A})}/{{\cal A}(\vomega;\boldsymbol{A})}$ and
(c) ${{\cal C}(\vomega;\boldsymbol{A})}/{{\cal A}(\vomega;\boldsymbol{A})}$,
where ${\cal A}$,  ${\cal P}$,  ${\cal S}$ and ${\cal C}$ are defined by Eq.~(\ref{eqn:def_ABCD}),
for a set of events generated according to the amplitudes $\boldsymbol{A}$ corresponding to the \babar\ values in Table~\ref{tab:fitted-values-comp}.}
\end{figure}
\begin{itemize}
\item 
The $\cos\Delta m\, t\,$ term makes the smallest contribution to $g_{\eta}(\vomega,t;\boldsymbol{A},\sin 2\beta,\cos 2\beta)$ because
the distribution of values taken by ${{\cal P}(\vomega;\boldsymbol{A})}/{{\cal A}(\vomega;\boldsymbol{A})}$, as shown in Fig.~\ref{fig:ABCD}(a), peaks at zero.
\item 
The $\sin\Delta m\, t\,$ term has explicit dependence on both
$\sin 2\beta$ and $\cos 2\beta$:
\begin{itemize}
\item The usual $\sin(\Delta m t) \sin 2\beta$ factor  is weighted by the angular term 
${{\cal S}(\vomega;\boldsymbol{A})}/{{\cal A}(\vomega;\boldsymbol{A})}$,
which can take values between $-1$ and $+1$, and whose distribution is shown in Fig.~\ref{fig:ABCD}(b).
This distribution reduces to one peak at $+1$ or $-1$ for a pure $CP$-even ($\vert A_\perp\vert = 0$) or $CP$-odd ($\vert A_\perp\vert = 1$) decay, respectively.
\item 
The $\cos 2\beta$ contribution is characteristic of a vector-vector
channel. This contribution appears only {\em via} the interference
terms involving the \CP-odd amplitude $A_\perp$ and the 
\CP-even amplitudes $A_0$ and $A_\parallel $ (Eq.~(\ref{eqn:def_ABCD})).
The angular term ${{\cal C}(\vomega;\boldsymbol{A})}/{{\cal A}(\vomega;\boldsymbol{A})}$
takes values in a range smaller than $[-1,+1]$  (Fig.~\ref{fig:ABCD}(c)), whose bounds depend
on the amplitudes and phases. The distribution of this angular term tends
to peak at zero (Fig.~\ref{fig:ABCD}(c)), inducing some loss in sensitivity
to $\ctwob$ compared to that to $\stwob$.
\item 
The $\sin 2\beta$ and $\cos 2\beta$ contributions are distinguished by the angular information only.
\item 
From the orthogonality of the angular functions ${\cal S}$ and  ${\cal C}$  (Eq.~(\ref{eqn:def_ABCD}))
and the angular symmetry of $g_\eta$ (Eq.~(\ref{eqn:asymm_CP_1})),
the $\sin 2\beta$ and $\cos 2\beta$ parameters, if regarded as independent quantities, 
are uncorrelated in a fit of the differential decay rate (Eq.~(\ref{eqn:asymm_CP_1})), in the limit of infinite statistics and in the absence of experimental effects.
\end{itemize}
\end{itemize}
Under the transformation
$(\delta_\parallel-\delta_0 ,\delta_\perp-\delta_0)\to (\delta_0-\delta_\parallel ,\pi+\delta_0-\delta_\perp)$, 
 $\cal{A}$, ${\cal P}$, and $\cal{S}$ are unchanged, while $\cal{C}$ changes sign, 
showing that the ambiguity in the strong phases  translates into an 
 ambiguity in the sign of $\cos 2\beta$:
\begin{eqnarray}
\label{eqn:ambiguity-strong-phase-cos2beta}
&&\left(\delta_\parallel-\delta_0  ,\delta_\perp-\delta_0 , \cos 2\beta\right)
\longleftrightarrow  \nonumber \\ 
&&  \qquad\left(\delta_0-\delta_\parallel ,\pi+\delta_0-\delta_\perp, -\cos 2\beta\right).
\end{eqnarray}

\section{The \babar\ Detector}
\label{sec:BaBarDetector}

A detailed description of the \babar\ detector is presented in
Ref.~\cite{ref:babar}. 
Charged particles are detected with a five-layer, double-sided
silicon vertex tracker (SVT) and a 40-layer drift chamber (DCH) with a
helium-isobutane gas mixture, placed in a 1.5-T solenoidal field produced
by a superconducting magnet. The  charged-particle momentum resolution 
is approximately $(\delta p_T/p_T)^2 = (0.0013 \, p_T)^2 +
(0.0045)^2$, where $p_T$ is the transverse momentum   in \gevc. 
The SVT, with a typical
single-hit resolution of 10\mum,  measures the impact
parameters of charged-particle tracks in both the plane transverse to
the beam direction and along the beam.
Charged-particle types are identified from the ionization energy loss
(\dedx) measured in the DCH and SVT, and from the Cherenkov radiation
detected in a ring-imaging Cherenkov device. Photons are
detected by a CsI(Tl) electromagnetic calorimeter (EMC) with an
energy resolution $\sigma(E)/E = 0.023\cdot(E/\gev)^{-1/4}\oplus
0.019$. 
The return yoke of the superconducting coil is instrumented with
resistive plate chambers (IFR) for the identification and muons and the
detection of neutral hadrons.

\section{Event Reconstruction and Selection}
\label{sec:EventSelection}

The event selection is similar to that used in our previous analysis
\cite{BaBarFullAngularAnalysisPRL2001}.
Multihadron  events are selected 
by demanding a minimum of three reconstructed charged tracks in the
polar angle range $0.41 < \theta_{lab} < 2.54$\rad.
A charged track must be reconstructed in the DCH, and,
if it does not result from a \KS decay, it must
originate  at the nominal interaction point within 1.5\cm in the plane transverse to the beam and 10\cm along the beam.
Events are required to have a primary vertex within 0.5\cm of the
average position of the interaction point in the plane transverse to
the beamline, and  within 6\cm longitudinally.
Electromagnetic  depositions in the calorimeter in the polar angle range
$0.410 < \theta_{lab} < 2.409~\rad$ that are not associated with
charged tracks, that have an energy greater than 30\mev and that have a shower shape
consistent with a photon interaction are taken as neutral clusters.
We require the total energy for charged tracks and photon
candidates in the fiducial region
to be greater than 4.5\gev.
To reduce continuum
background, we require the normalized second Fox-Wolfram
moment $R_2$~\cite{FoxWolfram} of the event, calculated with both charged
tracks and neutral clusters, to be less than 0.5.

Charged tracks are required to be in regions of polar angle for which the particle identification (PID) efficiency is well-measured.
For electrons, muons, and kaons the acceptable ranges are  0.40 to 2.40, 0.30 to 2.70, and
0.45 to 2.50 rad, respectively.

Candidates for \jpsi mesons are reconstructed in  the \epem and
\mumu decay modes, from a pair of identified leptons that form a good vertex.
A {\tt Loose} \cite{BaBarTheBigPRD_CP_AsymFlavorOsc} identification condition is
required for each muon.
(The number of interaction lengths it traverses in the EMC and IFR must be consistent with the expectation,
as must be the average number of hits in each layer of the IFR; the IFR hits and the track extrapolation must match;
the energy deposition in the EMC must be small.)
A {\tt Tight} condition
\cite{BaBarTheBigPRD_CP_AsymFlavorOsc} is required for each electron.
(The measured \dedx must be consistent with expectations;
the energy deposition in the calorimeter must be consistent with the momentum measured in the drift chamber,
and the lateral shower shape must be consistent with an electromagnetic shower.)
Electrons that have no EMC information are selected on the basis of \dedx
information alone.
For \jpsi\to\epem decays, where an electron may have radiated one or several
Bremsstrahlung photons, the missing energy is recovered by identifying EMC
clusters with energy greater than 30\mev lying within 35\mrad in polar angle
and 50\mrad in azimuth of the electron direction projected onto the
EMC.
The  lepton-pair invariant mass must be between 3.06 and 3.14\gevcc
for muons, and between 2.95 and 3.14\gevcc for electrons.
This corresponds to a $\pm3\sigma$ interval for muons, and accounts for
the partially recovered radiative tail due to Bremsstrahlung for electrons.

A candidate \KS\  consists of a vertexed pair of oppositely-charged tracks
with invariant mass between 489 and 507\mevcc, when interpreted as pions.
The \KS flight length must be greater than 1\mm, and its direction
must form an angle with the \KS momentum vector in the plane
perpendicular to the beam line that is less than 0.2\rad.

Neutral clusters, as defined above, are used as 
photon candidates for the reconstruction of  $\piz \to \gamma \gamma$ decays.
A \piz candidate consists of a pair of photons with invariant mass in
the interval 106 to 153\mevcc, and a total energy greater than 200 \mev.

The \jpsi, \KS, and \piz candidates are constrained to their 
corresponding nominal masses.  Except   in the
analysis  that includes an $S$-wave  contribution, 
\Kstar candidates must have a $K\pi$ invariant mass within 100\mevcc
of the nominal $\Kstar(892)$ mass.

The \jpsi and \Kstar candidates are combined to form $\B \to \jpsi \Kstar$ candidates.
It may happen that a genuine $\jpsi\Kstar$ event is reconstructed
incorrectly, most often with the true \jpsi, but with a wrongly reconstructed \Kstar.
This happens mainly for $B$ candidates with a daughter \piz, with cross-feed (CF) from the companion channel with a \pipm,
or self cross-feed (SCF) when the genuine \piz is incorrectly reconstructed with at least one wrong photon candidate.
The (S)CF is reduced by demanding, for channels with a \piz in the final state, that
$\cthetakstar <0.7$, where $\theta_{K^*}$ is the $K^*$-decay helicity angle (see Fig.~\ref{fig:trans-frame}).
In addition, as was done in Ref.~\cite{StephThese}, if a single event can be reconstructed in two different $K^*$ modes
and if one reconstruction uses a $\piz$ and the other does not, the reconstruction without a $\piz$ is retained.
This  reduces the cross-feed by 75\% for  a 1\% relative loss in signal efficiency.
In modes with a \pipm, no \cthetakstar\ cut is applied.

Two kinematic variables are used to further discriminate against incorrect \B candidates.
The first is the difference $\Delta E = E^*_B - E^*_{\rm beam}$
between the candidate-\B energy and the beam energy in the \FourS\ rest frame. In the absence of
experimental effects, 
reconstructed signal candidates have $\Delta E = 0$. The  second
is the beam-energy-substituted mass
$\mes = (E^2_{\rm exp} - {\vec p}^{\,2}_B)^{1/2}$
where, in the laboratory frame $E_{\rm exp} = \left(s/2 + {\vec p}_B.{\vec p}_i\right)/E_i$ is the
\B-candidate expected energy, ${\vec p}_B$, its measured momentum,   $(E_i, {\vec p}_i)$, the
$\epem$ initial-state four-momentum, and  $\sqrt{s}$ is the center-of-mass energy.
For the signal region, $\Delta E$ is required to be between $-70$\mev and $+50$\mev for
channels involving a \piz, and within $30$\mev of zero otherwise.
If several \B\ candidates are found in an event, the one with the
smallest $|\DeltaE|$ is retained. 

\begin{figure}[tpb]
\begin{center}
\includegraphics[width=\singlewidth]{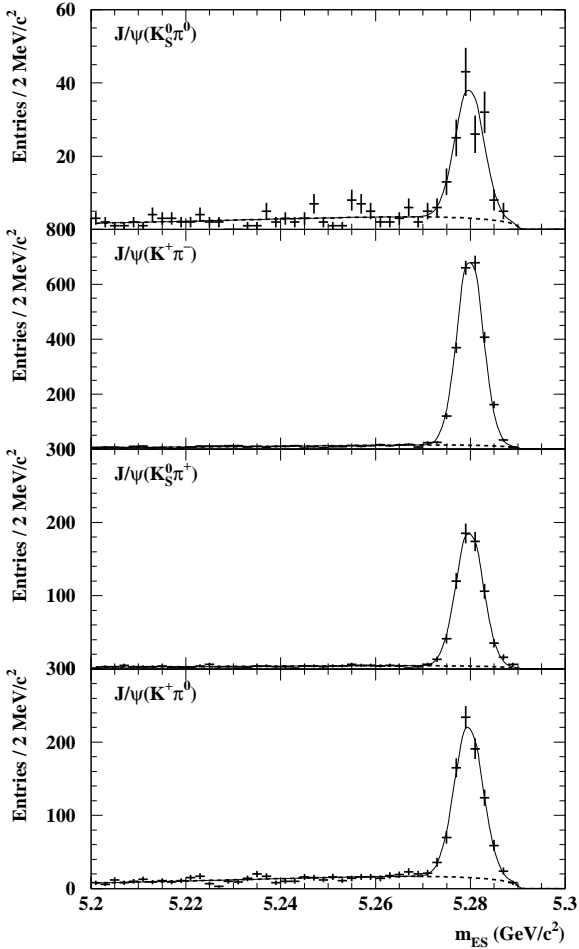}
\end{center}
\caption{The \mes\ distributions for the $\Delta E$ intervals described in the text with overlaid  Gaussian and ARGUS fit functions, for $\jpsi K\pi$
candidates in data. \label{fig:resmesde}}
\end{figure}

The \mes distributions for the $\B \to \jpsi K\pi$ candidates are shown in
Fig.~\ref{fig:resmesde}. Corresponding signal yields and purities are given
in Table~\ref{tab:resmes}.
These results are obtained from fits to the \mes\ distributions using 
a Gaussian distribution for the signal and an ARGUS shape~\cite{argus} for the combinatorial background of the form
\begin{eqnarray}
a( \mes  ) &=& a_{0} \mes 
 \sqrt{ 1 - (\mes/m_{0})^2 } \times \nonumber \\
&~&{\rm e}^{\displaystyle \xi \left( 1 - (\mes/m_{0})^2 \right)},
\label{eqn:argus}
\end{eqnarray}
for $\mes < m_{0}$, where $m_0$ represents the kinematic upper limit
and is  fixed at the center-of-mass beam energy
$E^{*}_{\rm beam}=5.291\gev$. The parameter $\xi$ determines the shape of the spectrum.

With the signal region defined by $\mes>5.27\gevcc$ and the above \DeltaE ranges,
the \B reconstruction efficiencies, summed over $\jpsi\to\ee$ and $\jpsi\to\mumu$, are 
$(9.6\pm0.1)\%$, 
$(24.5\pm0.1)\%$,
$(19.7\pm0.2)\%$,  and 
$(12.5\pm0.2)\%$
for the modes \KS\piz, \Kpm\pimp, \KS\pipm, and \Kpm\piz, respectively.
\begin{table}
\caption{Event yield and purity, estimated from a fit to the \mes distribution (Fig.~\ref{fig:resmesde}),
with a  Gaussian signal distribution and an ARGUS threshold function
\protect\cite{argus} describing the combinatorial background.
The spectra are integrated  over the  range \mes $>$ 5.27 \gevcc.  
No correction for cross-feed is made since these numbers are not used in
the actual analysis; rather they provide an indication of the purity of the data sample.
\label{tab:resmes}}
\begin{center}
\begin{ruledtabular}
\begin{tabular}{ccc}
Channel & Yield & Purity (\%) \\ \hline
$\jpsi (\KS\piz)$    & $~131 \pm 14$ & $81.6$ \\ 
$\jpsi (\Kpm\pimp)$  & $2376 \pm 51$ & $95.8$ \\ 
$\jpsi (\KS\pipm)$ & $~670 \pm 27$ & $95.7$ \\ 
$\jpsi (\Kpm\piz)$ & $~791 \pm 33$ & $85.0$ \\
\end{tabular} \end{ruledtabular}
\end{center}
\end{table}
The composition of the remaining background events is given in Table
\ref{tab:backgrounds}.  The contribution of $B$ candidates with a 
fake $\jpsi$ candidate is less than 2\%.
\begin{table*}[htbp]
\caption{\label{tab:backgrounds}
The expected number of background events for each decay mode in the signal region, in an
on-peak 81.9 \invfb sample.
The contribution from continuum is estimated using a 9.6 \invfb
off-peak data sample.
The \BB contribution is estimated using a fully-simulated sample of
generic \BB decays equivalent to 72 \invfb (with the inclusive \jpsi events
removed from the sample).
The inclusive \jpsi contribution is estimated using a fully-simulated
sample equivalent to 91 \invfb ($\B \to \jpsi \Kstar(892)$ events
removed). }
\begin{ruledtabular}
\begin{tabular}{lcccc}
Background &\multicolumn{4}{c}{\B Decay Mode} \\
Source & $\jpsi (\KS\piz)$ & $\jpsi (\Kpm\pimp)$ & $\jpsi(\KS\pipm)$ & $\jpsi (\Kpm\piz)$ \\ \hline
Continuum   & $0.2\pm 0.1$ & $1.2 \pm 0.4$ & $0.1\pm 0.1$ & $0.7 \pm 0.4$ \\
Generic \BB & $0.2\pm 0.1$ & $1.2 \pm 0.2$ & $0.3\pm 0.1$ & $1.2 \pm 0.3$ \\
Inclusive \jpsi & $22\pm 5 $ & $126\pm 12 $ & $38\pm 7 $ & $135\pm 12$ \\
\end{tabular} \end{ruledtabular}
\end{table*}

\section{Angular Analysis}
\label{sec:AngularAnalysis}

The parameters $\theta_A$, $\phi_A$ (Eq.~(\ref{eqn:polar_description_of_amplitudes})), $\delta_\|-\delta_0$ and $\delta_\perp-\delta_0$
of the angular-dependent  time-integrated decay rates
are determined using a simultaneous unbinned likelihood fit to the three
flavor-eigenstate $\B \to \jpsi \Kstar$ channels: 
$\jpsi (\Kpm\pimp)$, $\jpsi(\KS\pipm)$, and $\jpsi (\Kpm\piz)$.
The PDF, before accounting for the experimental 
effects described below, is given by Eq.~(\ref{eqn:g_definition}).
We first consider only the $P$-wave amplitudes;
the effect of the $S$-wave amplitude is discussed in
Sec.~\ref{subsec:Systematics}.
The \B flavor is taken into account in the fit through the relations
$(\overline{A}_0=A_0,\ \overline{A}_\|=A_\|,\ \overline{A}_\perp = -A_\perp)$ as explained
in Sec.~\ref{sec:AngularTimeDepDecayRateProb}.

\subsection{Acceptance Correction}
\label{subsec:AcceptanceCorrection}

The acceptance correction is applied as in our previous measurement 
\cite{BaBarFullAngularAnalysisPRL2001}.
We perform an unbinned likelihood fit of the 
PDF $g^{obs}$ to the observed events, where
\begin{equation}
\label{eq:gObsDef}
g^{obs}(\vomega;\boldsymbol{A}) = g(\vomega;\boldsymbol{A}) 
 \frac{\varepsilon({\vomega})}{\langle \varepsilon\rangle(\boldsymbol{A})},
\end{equation}
where $g(\vomega;\boldsymbol{A})$ is given by Eq.~(\ref{eqn:g_definition}), $\varepsilon(\vomega)$ is the angle-dependent acceptance, and 
\begin{equation}
\langle \varepsilon\rangle(\boldsymbol{A}) \equiv \int{g(\vomega;\boldsymbol{A})\varepsilon(\vomega){\rm d}\vomega}
\end{equation}
is the average  acceptance 
over the event-weighted phase space, which depends
on the amplitudes $\boldsymbol{A}$, and which ensures the normalization of $g^{obs}$.

In the case of the $\jpsi\Kstar$ channels studied here, the presence of
cross-feed from the companion channels, which have,  as a consequence of isospin
symmetry, the same $\boldsymbol{A}$ dependence as that of the signal, must be
taken into account.
The observed PDF for channel $\jchan$ $(\jchan=\Kpm\pimp, \KS\pipm, \Kpm\piz)$ is then
\begin{eqnarray}
\label{eq:gobs}
g^{\jchan,obs}(\vomega;\boldsymbol{A}) &=&
g(\vomega;\boldsymbol{A})
\frac{\varepsilon^{\jchan}(\vomega)}
{\sum_{k}{\cal A}_k(\boldsymbol{A}) \Phi^{\jchan}_k},
\end{eqnarray}
with
\begin{eqnarray}
\label{eqn:effectiveEff}
\varepsilon^{\jchan}(\vomega) &\equiv& \sum_\ichan F_\ichan\, \varepsilon^{\ichan\to \jchan}(\vomega), \\
\Phi^{\jchan}_k &\equiv&
 \sum_\ichan F_\ichan \int{ f_k(\vomega) \varepsilon^{\ichan\to \jchan}(\vomega)
\dd \vomega}
\label{eqn:coefs:Eff}
\end{eqnarray}
and $\ichan=\KS\piz,\Kpm\pimp, \KS\pipm, \Kpm\piz$.
In the above expressions, the ${\cal A}_{1\dots 6}$ terms are (see~Eq.(\ref{eqn:g_definition}))
\begin{eqnarray}
{\cal A}_1&=&\azd, \nonumber\\
{\cal A}_2&=&\apd, \nonumber\\
{\cal A}_3&=&\atd, \nonumber\\
{\cal A}_4&=&\pipt, \nonumber\\
{\cal A}_5&=&\przp, \nonumber\\
{\cal A}_6&=&\pizt,                \label{eqn:numerics}
\end{eqnarray}
and 
$F_\ichan$
is the fraction
of mode $\ichan$ in $\B \to \jpsi \Kstar$ decays (with
$\sum_\ichan F_\ichan =1$). 
We assume that ${\cal B} ( \FourS \to \Bz \Bzb ) =
{\cal B} ( \FourS \to \Bp \Bm)$,
$\Gamma(\Kstarz \to \Kp \pim) = 2 \times \Gamma(\Kstarz \to \Kz \piz)$,
and
$\Gamma(\Kstarp \to \Kz \pip) = 2 \times \Gamma(\Kstarp \to \Kp \piz)$.
The measured values \cite{ref:pdg} of  the branching fractions for the decays 
$\Bz \to \jpsi \Kstarz$ and   $\Bp \to \jpsi \Kstarp$ are used.
The angular functions $f_k(\vomega)$ ($k=1\dots 6$) have been
defined in Eq.~(\ref{eq:theAngularFunctions})
and $\varepsilon^{\ichan\to \jchan}(\vomega)$ is the probability for an event
generated in channel $\ichan$ and with  angles $\vomega$ to be detected as an
event in channel $\jchan$.
Finally, 
$\varepsilon^{\jchan}(\vomega)$ is the efficiency for
reconstructed channel $\jchan$ considering $\B\to\jpsi\Kstar$ channels as a
whole, that is counting cross-feed events as signal.
The   $\Phi^{\jchan}_k$ are the $f_k(\vomega)$ moments of the
``whole'' efficiency $\varepsilon^{\jchan}$.

The angular resolution has been neglected, even for (self)cross-feed events.
Also the possibility of  doubly misidentifying the daughters of
the $\Kstarz \to \Kp \pi^-$ candidate ($K$--$\pi$ swap)
is not taken into account.
The induced biases have been estimated with   Monte Carlo
(MC) based studies, and found to be negligible.
Under these two approximations, the  acceptance
$\varepsilon^{\jchan}(\vomega)$ can be factorized as in
Eq. (\ref{eq:gobs}), and only the coefficients
$\Phi^{\jchan}_k$ are needed.

The coefficients $\Phi^{\jchan}_k$ are computed with exclusive
signal MC samples obtained using a full simulation of the experiment
\cite{Evtgen,Geant4BaBar,Geant4ToolKit}.
Particle identification efficiencies measured with data control samples
are used to adjust the MC simulation to represent the actual behavior of the detector.
Separate coefficients are used for 
different charges of the final state mesons, in particular to take into account
the charge dependence of the interaction of charged kaons
with matter, and any other possible charge asymmetry of the detector.
Writing the log-likelihood function for a pure signal sample we have, for each channel $\jchan$,
\begin{eqnarray}
\label{eqn:likelihoodDef}
L^{\jchan}(\boldsymbol{A}) &\equiv& \sum_{i=1}^{N_{evt}}\ln\left(g^{\jchan,obs}(\vomega_i;\boldsymbol{A})\right) \nonumber \\
&=&
\sum_{i=1}^{N_{evt}}\ln\left( g(\vomega_i;\boldsymbol{A}) \right) - N_{evt} \ln\left(\sum_{k}{\cal A}_k(\boldsymbol{A}) \Phi^{\jchan}_k\right) \nonumber \\
&+& \sum_{i=1}^{N_{evt}}\ln\left(\varepsilon^{\jchan}(\vomega_i)\right).
\end{eqnarray}
where $\vomega_i$ represents the measured angular variables for event $i$, and $N_{evt}$ is the total number of signal candidates.
When maximizing $L^{\jchan}(\boldsymbol{A})$, the third term, which does not
depend on the amplitudes, can be ignored
\cite{BaBarFullAngularAnalysisPRL2001}.

\subsection{Background Subtraction}
\label{subsec:BackgroundSubtraction}

In our previous measurement \cite{BaBarFullAngularAnalysisPRL2001} of
the decay amplitudes,  it was assumed that the combinatorial
background could be taken into account with an expansion in the same basis functions as the signal.  The systematic bias due to
neglecting the missing  components of the background angular
distribution was checked with MC-based studies.

Here, we use an improved background correction method in which events from 
the \mes sideband are added to the log-likelihood
that is maximized, but with a negative weight.

The sample of $N_{evt}$ events selected in the signal region contains 
 $n_S$ signal events and $n_B$ background events, so that 
$N_{evt} = n_S + n_B$.
The values of $n_S$ and $n_B$ are unknown {\em a priori}.
The quantity we would like to maximize  is
$\displaystyle
\sum_{i=1}^{n_S}\ln\left(g^{\jchan,obs}(\vomega_i;\boldsymbol{A})\right)
$,
while we have
\begin{eqnarray}
\label{eqn:log-likelihood-def}
L^{\jchan}(\boldsymbol{A}) &\equiv& 
\sum_{i=1}^{N_{evt}}\ln\left(g^{\jchan,obs}(\vomega_i;\boldsymbol{A})\right) 
\nonumber
\\
&=&
\sum_{i=1}^{n_S}\ln\left(g^{\jchan,obs}(\vomega_i;\boldsymbol{A})\right)
\nonumber \\ &+&
\sum_{j=1}^{n_B}\ln\left(g^{\jchan,obs}(\vomega_j;\boldsymbol{A})\right).
\end{eqnarray}
Note that the same PDF appears for both the signal events and the background events:
the PDF $g^{\jchan,obs}$ of the signal.
We use a pure sample of background
events to obtain an estimate of
the second term. This sample is from
the \mes sideband region
 $5.20 < \mes < 5.27\gevcc$,
which contains $N_{B}$ events.
It can be shown that maximizing the modified expression ${L^{\jchan\prime}}$, where
\begin{eqnarray}
\label{eqn:pseudo-log}
 L^{\jchan\prime}(\boldsymbol{A}) &\equiv& \sum_{i = 1}^{N_{evt}}\ln(g^{\jchan,obs}(\vomega_i;\boldsymbol{A})) 
 \nonumber \\ &-&
 \frac{\tilde{n}_B}{N_{B}}\sum_{k = 1}^{N_{B}}\ln(g^{\jchan,obs}(\vomega_k;\boldsymbol{A})) \nonumber \\ 
&=&
 \sum_{i = 1}^{n_S}\ln(g^{\jchan,obs}(\vomega_i;\boldsymbol{A})) +
 \sum_{j = 1}^{n_B}\ln(g^{\jchan,obs}(\vomega_j;\boldsymbol{A})) \nonumber \\ &-& 
 \frac{\tilde{n}_B}{N_{B}}\sum_{k = 1}^{N_{B}}\ln(g^{\jchan,obs}(\vomega_k;\boldsymbol{A})),
\end{eqnarray}
yields an unbiased estimator of the true parameters if
$\tilde{n}_B$ is an unbiased estimator of $n_B$.
The quantity $\tilde{n}_B$ is obtained by fitting
the data from the \mes sideband and signal regions
with a combination of an ARGUS and a Gaussian
function. Since there is
no peaking background contribution in the signal
region, we take for $\tilde{n}_B$ the portion
of the ARGUS fit that falls in
the signal region.

As $L^{\jchan\prime}$ is not a log-likelihood, the uncertainties yielded by the minimization 
program \minuit~\cite{minuit} are biased estimates of the actual uncertainties.
An unbiased estimation of the uncertainties is described and  validated in Appendix~\ref{appendx:pseudoLK}.

With this pseudo-log-likelihood technique, we avoid parametrizing the
acceptance as well as the background angular distributions.
This technique and the combined
(\mes, angular) likelihood fit used in Ref.~\cite{BaBarFullAngularAnalysisPRL2001} rely on the assumption that the
angular behavior of the combinatorial background is the same in
the \mes signal region and sideband.
The possible bias related to this assumption is discussed in the next section.

\subsection{Validation}
\label{subsec:Validation:Ang} 

The complete fit scheme, including acceptance and background corrections
as described above, has been validated with a \BB Monte Carlo sample equivalent to an integrated luminosity of 590 \invfb,
produced with a full simulation of the \babar\ detector (based on {\tt GEANT}4 \cite{Evtgen,Geant4BaBar,Geant4ToolKit}).
In this sample only events with a true $\jpsi \to \ell \ell$ decay with 
center-of-mass momentum $p^*_{\jpsi}$ greater than $1.3$~\gevc are simulated.
This momentum cut is not applied in the analysis. It
does not affect the signal region $(\mes > 5.27 \gevcc)$, but means
that only a subset of the events in the \mes sideband region 
is included.

An additional study has been performed with a larger sample generated with a parametrized
simulation from the same event generator \cite{Evtgen}  with resolution effects and efficiencies incorporated.
The equivalent integrated luminosity of this  sample  is 16 \invab.

The results of the two simulations
are found to be compatible with each
other. No statistically significant bias is observed with the 
full simulation.  However,
the high-statistics fast simulation 
shows small biases in the fitted parameters (Table~\ref{tab:bias-comparison}).
A contribution to the systematic uncertainty is derived from these biases in  Sec.~\ref{subsec:Systematics}.
\begin{table}
\begin{center} 
\caption{\label{tab:bias-comparison}
Bias (in units of $10^{-3}$) observed in fits for the individual $K^*$
channels and the combined channel, based on parametrized Monte Carlo, taking as input
 the values of the amplitudes  from
Ref.~\protect\cite{BaBarFullAngularAnalysisPRL2001}.
The upper part of the table presents results for the fitted
quantities $\theta_A, \phi_A, \delta_\parallel-\delta_0$, and
$\delta_\perp-\delta_0$, all expressed in radians.
The lower part presents  results for the amplitude moduli squared, which are computed
from $\theta_A$ and $\phi_A$.}
\[
{
\begin{array}{rrrrrr}\hline\hline
 & & \multicolumn{2}{c}{{\rm Bias\ } (10^{-3})}\\
 & \Kpm\pimp &\KS\pipm &\Kpm\piz & {\rm all}\ K^*\\ \hline
\theta_A & 
 2.9\pm 1.0 & 
 -1.3\pm 1.9 &
 -8.7\pm 1.9 &
 -0.2\pm0.8\\ \phi_A &
 3.3\pm 3.5 &
 13.0\pm 6.6 &
 5.5\pm 6.4 &
 5.5\pm 2.8\\ \delta_\parallel-\delta_0 &
 -34.9\pm 7.8 &
 -19.2\pm 14.7&
 -54.5\pm 13.9&
 -36.2\pm 6.2\\ \delta_\perp-\delta_0 &
 -29.3\pm 6.4 &
 -7.8\pm 11.8&
 -29.0\pm 11.4 &
 -25.2\pm 5.0\\ \hline
 |A_0|^2 &
 -2.9\pm 1.0 &
 1.3\pm 1.9 &
 8.5\pm 1.9 &
 0.2\pm 0.8 \\ |A_\parallel |^2 &
 0.4\pm 1.5 &
 -5.9\pm 2.8 &
 -7.2\pm 2.7 &
 -2.2\pm 1.2\\ |A_\perp|^2 &
 2.5\pm 1.4 &
 4.6\pm 2.7 &
 -1.3\pm2.6&
 2.1\pm 1.1
\\ \hline\hline
\end{array}
}
\]
\end{center}
\end{table}

\subsection{Systematic Uncertainties}
\label{subsec:Systematics}

\begin{table}
\begin{center}
\caption{\label{tab:syst-ch234}Systematic uncertainties in the relative phases (\rad) and in the amplitude moduli squared, for the three \Kstar channels combined.}
\[
\begin{array}{clccccccc}\hline\hline
&{\rm Source} & \delta_\parallel-\delta_0 & \delta_\perp-\delta_0 &
|A_0|^2 & |A_\parallel |^2 & |A_\perp|^2 \\ \hline 1.& {\rm c.m.\
energy} & 0.003 & 0.005 & 0.0018 & 0.0002 & 0.0016 \\ 2.& {\rm Bkg.\
shape} & 0.002 & 0.003 & 0.0012 & 0.0001 & 0.0011 \\ 3.& {\rm BR} &
0.000 & 0.001 & 0.0001 & 0.0001 & 0.0001 \\ 4.& {\rm MC\ stat.}  &
0.014 & 0.008 & 0.0027 & 0.0023 & 0.0024 \\ 5.& {\rm Fit~bias} & 0.036
& 0.025 & 0.0002 & 0.0022 & 0.0021 \\ 6.& {\rm PID} & 0.005 & 0.004 &
0.0013 & 0.0023 & 0.0019 \\ 7.& S\ {\rm wave} & 0.033 & 0.038 &
0.0029 & 0.0025 & 0.0005 \\\hline &{\rm Total} & 0.052 & 0.046 &
0.0048 & 0.0046 & 0.0042 \\\hline\hline
\end{array}
\]
\end{center}
\end{table}

Table~\ref{tab:syst-ch234} summarizes the systematic
uncertainties for the measurement of the amplitudes.
The sources of uncertainty we have considered are described here.
\begin{enumerate}
\item 
``c.m. energy'': The center-of-mass energy, which defines the \mes
endpoint spectrum, enters as the parameter $m_0$ of the  ARGUS
function (Eq.~(\ref{eqn:argus})). The value (5.291 \gevcc) is changed by $\pm$ 2 \mevcc (uncertainty on the beam energy in the c.m. frame)
and the largest deviation from the nominal fit result
is taken as the estimate of the systematic
uncertainty.
\item 
 ``Backgrounds shape'': 
The ARGUS function shape parameter $\xi$  (Eq.~(\ref{eqn:argus})), fitted to
the \mes spectrum, is changed by $\pm 1$ standard deviation
and the largest deviation from the nominal fit result
is taken as the estimate of the systematic
uncertainty.
\item 
``BR'': The relative branching fractions of neutral and charged \B 
mesons to \jpsi \Kstar affects the amount of cross-feed. The
branching ratios are changed independently by $\pm 1$ standard
deviation \cite{ref:pdg} and the largest difference is taken as the
systematic uncertainty.
\item 
``MC stat.'': The finite  Monte Carlo sample size induces a limited
knowledge of the coefficients $\Phi^{\jchan}_k$. This effect is
evaluated by splitting the original   Monte Carlo sample into ten equal-sized
subsamples, each of which is used to compute the $\Phi^{\jchan}_k$
coefficients. These coefficients are then used for  ten angular fits on
the data, all differences being thus due to differences of the
$\Phi^{\jchan}_k$ coefficients. For each fitted parameter, the standard
deviation is computed, and divided by $\sqrt{10}$ to estimate the ``MC
stat.'' effect due to the original Monte Carlo  finite size.
\item 
``Fit bias'': Biases are observed in validation studies
(Table~\ref{tab:bias-comparison}).
The  observed bias is used as an estimate of the systematic uncertainty.
\item 
``PID'': The efficiency of the particle identification has
angular dependence.
The induced effect on the fitted parameters is corrected 
by the acceptance-correction scheme.
Imperfect knowledge of the particle-identification efficiency will result in a bias.
A conservative estimate of the systematic uncertainty is obtained by using
acceptance-correction factors for different beam conditions,
corresponding to the years 2000, 2001, and 2002, and using the largest
differences as estimates of the systematic uncertainties.
\item 
``$S$ wave'': An additional fit is performed with the $g_{S+P}$ PDF (see
next Section).  The full $g_{S+P}$-to-$g_{P}$ shift is used as a
conservative estimate of the contribution to the systematic uncertainty, as was
done in Ref.~\cite{BaBarFullAngularAnalysisPRL2001}.
\end{enumerate}

\subsection{Results of the Angular Analysis}
\label{subsec:Results}

\begin{table*}[tbp]
\begin{center}
\caption{\label{tab:fitted-values}
Values of 
$\vert A_0 \vert^2$, 
$\vert A_\parallel \vert^2$,
 $\vert A_\perp\vert^2$,
$\delta_\parallel-\delta_0$, and $\delta_\perp-\delta_0$,
for subsamples of the data divided according to decay channel.
 The first uncertainty is statistical; the second, when given, systematic. 
Note that the phases are subject to a two-fold ambiguity (Eq.~(\protect\ref{eqn:phase_ambiguity})).
}
\begin{ruledtabular}
\begin{tabular}{cccccc} 
 {\rm Sample} & $\vert A_0 \vert^2$ & $\vert A_\parallel \vert^2$ & $\vert A_\perp\vert^2$ 
 & $\delta_\parallel-\delta_0$ (\rad) & $\delta_\perp-\delta_0$ (\rad) \\ \hline
\Kpm\pimp & $0.560 \pm 0.015 \pm 0.005$ & $0.208 \pm 0.019 \pm 0.004$ & $0.232 \pm 0.020 \pm 0.005$ 
 & $2.673 \pm 0.121 \pm 0.052$ & $0.159 \pm 0.084 \pm 0.048$ \\
\KS\pipm  & $0.560 \pm 0.028 \pm 0.006$ & $0.232 \pm 0.034 \pm 0.010$ & $0.208 \pm 0.034 \pm 0.007$ 
 & $2.747 \pm 0.220 \pm 0.052$ & $0.124 \pm 0.174 \pm 0.050$ \\
\Kpm\piz  & $0.592 \pm 0.028 \pm 0.013$ & $0.165 \pm 0.032 \pm 0.011$ & $0.243 \pm 0.036 \pm 0.009$ 
 & $2.904 \pm 0.287 \pm 0.090$ & $0.329 \pm 0.176 \pm 0.066$ \\ \hline
{\rm Total} & $0.566 \pm 0.012 \pm 0.005$ & $0.204 \pm 0.015 \pm 0.005$ & $0.230 \pm 0.015 \pm 0.004$ 
 & $2.729 \pm 0.101 \pm 0.052$ & $0.184 \pm 0.070 \pm 0.046$ \\
\end{tabular} \end{ruledtabular}
\end{center}
\end{table*}

Table~\ref{tab:fitted-values} summarizes the
results of the fit to the angular distribution.
The fitted values for each channel and for each year
of data collection (with statistical uncertainties only) are shown.
Keeping in mind the two-fold phase ambiguity~(Eq.~(\protect\ref{eqn:phase_ambiguity})), we obtain
\begin{eqnarray}
\delta_\parallel-\delta_0 &=& (2.73 \pm 0.10 \pm 0.05) \rad, \nonumber \\ 
\delta_\perp-\delta_0 &=&  (0.18 \pm 0.07 \pm 0.05) \rad,  \nonumber \\ 
\vert A_0 \vert^2 &=&  0.566 \pm 0.012 \pm 0.005,  \nonumber \\ 
\vert A_\parallel \vert^2 &=&  0.204 \pm 0.015 \pm 0.005,  \nonumber \\ 
\vert A_\perp\vert^2 &=&  0.230 \pm 0.015 \pm 0.004,
\label{eqn:angFitNum}
\end{eqnarray}
where the correlation matrix of the fitted parameters $(\theta_A, \phi_A, \delta_\|-\delta_0, \delta_\perp-\delta_0)$ (Eq.~(\ref{eqn:polar_description_of_amplitudes})) is
\[
\left(
\begin{array}{rrrr}
  1.00    &    0.00      &     -0.04    &  +0.04      \\
          &    1.00      &     -0.23    &  -0.09      \\
          &              &      1.00    &  +0.65      \\
          &              &              &   1.00      
\end{array}
\right)\begin{array}{r}\\ \\ \\ \!\!\!\!\!\!.\end{array}
\]

\begin{figure}[tbp]
\begin{center}
\includegraphics[width=\thirdofsinglewidth]{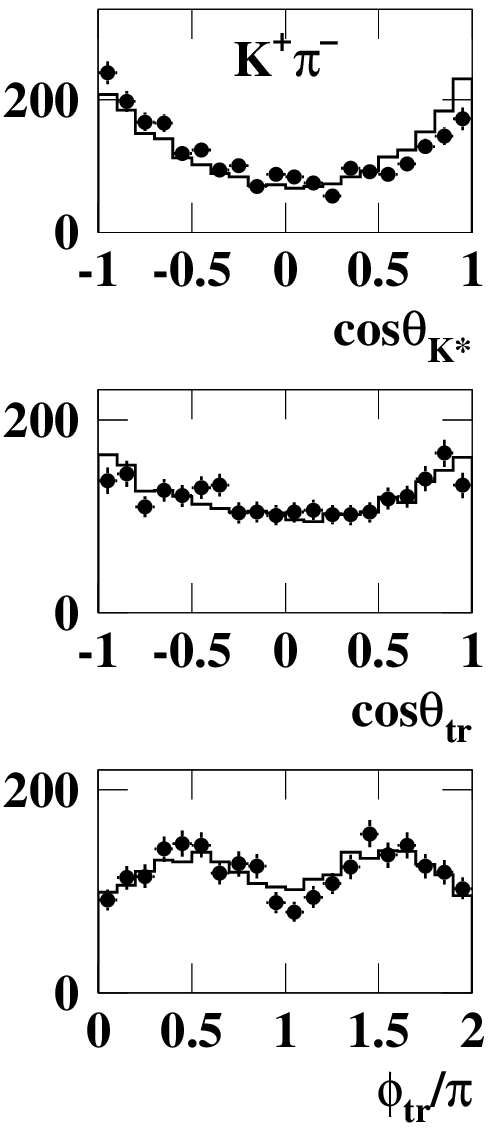}
\includegraphics[width=\thirdofsinglewidth]{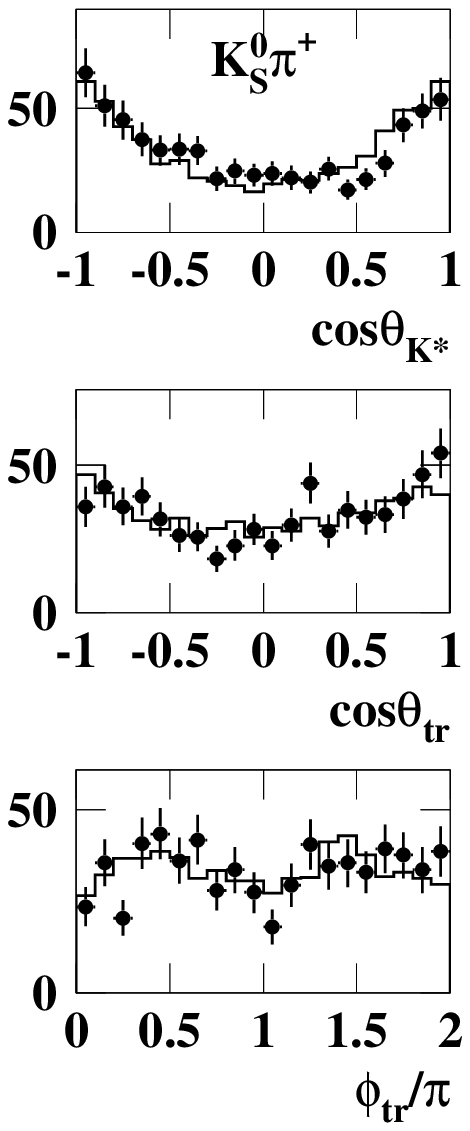}
\includegraphics[width=\thirdofsinglewidth]{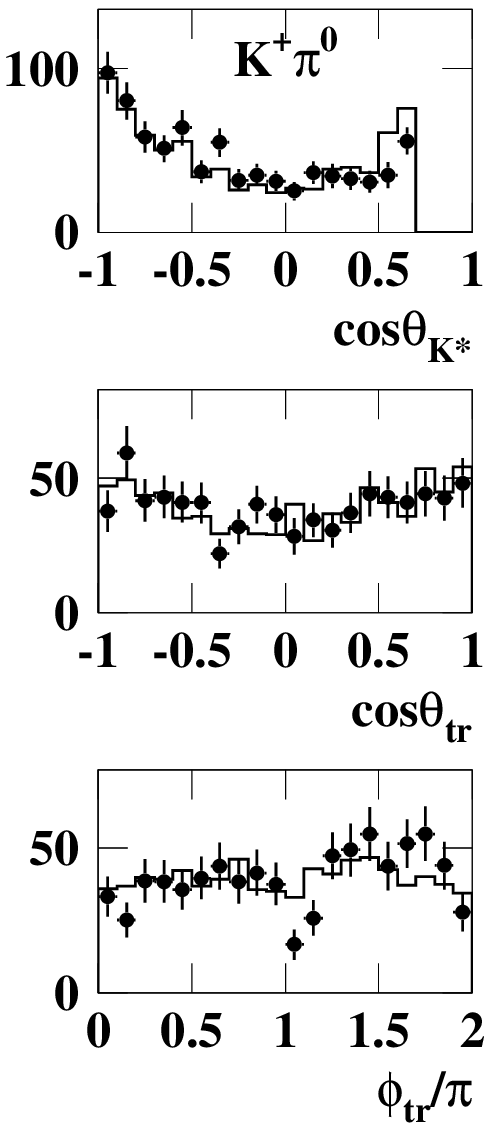}
\end{center}
\caption{Angular distributions.
Histogram: Inclusive \jpsi MC sample ($p^*_{\jpsi} > 1.3~\gevc$).
Points: Data.
The spectra are  acceptance-corrected, background-subtracted, and normalized
to the estimated yields (Table~\ref{tab:resmes}).
The visible forward-backward discrepancy in the  $\cos\theta_{K^*}$ distribution is 
due to the $K\pi$ $S$-wave amplitude present in the data, and absent in the MC sample.
The related systematic uncertainties in the measurements of the decay amplitudes 
are listed in  line 7 of Table
\protect\ref{tab:syst-ch234}.
 \label{fig:1Dang}
}
\end{figure}

Angular distributions for the three channels are shown in
Fig.~\ref{fig:1Dang}.  A forward-backward asymmetry is clearly
visible in the comparison of the distributions of $\cos\theta_{K^*}$
for (pure $P$-wave) MC, generated with the amplitudes found in the data, and for the data samples themselves. This is due to \SP interference.

In a series of 168 simulated experiments of the same size as the data sample,
we find that the probabilities for obtaining a larger likelihood
than that observed for the data are 11\%, 47\%, 58\%, and 25\% for the
$\Kpm\pimp, \KS\pipm, \Kpm\piz$, and combined samples, respectively.

The results for \jpsi \Kstarz and \jpsi \Kstarpm decays are found to be
compatible with each other (Table~\ref{tab:fitted-values});
this confirms the expectation of isospin symmetry.

  From Eq.~(\ref{eqn:angFitNum}), we note that $\delta_\| - \delta_0$ differs from $\pi$
by 3.6 standard deviations and that $\delta_\perp - \delta_0$ differs from
0 by 2.0 standard deviations.
In order to determine the uncertainty in  $\delta_\| - \delta_\perp$, the combined data sample is
refit using  $\delta_\|- \delta_\perp$
and $\delta_0 - \delta_\perp$ as phase parameters.
The resulting
amplitudes and the value of $\delta_0 - \delta_\perp$ were as before, and 
this refit yields
\begin{eqnarray}
\delta_\parallel-\delta_\perp &=& (\pi - (0.60 \pm 0.08 \pm 0.02)) \rad,
\label{eqn:deltapar_deltaperp}
\end{eqnarray}
where the systematic uncertainties have been estimated as in Sec.~\ref{subsec:Systematics}. The $\delta_\parallel-\delta_\perp$ statistical uncertainty agrees
with that expected from Eq.~(\ref{eqn:angFitNum})
taking into account the $65\%$ correlation between the $\delta_\parallel-\delta_0$ and $\delta_\perp-\delta_0$ parameters.
The departure from $\pi$ is $7.6$ standard deviations, and this demonstrates quite clearly the presence of final-state interactions between the $\jpsi$ and the $K^*$.

\section{Resolving the Strong Phase  Ambiguity}
\label{sec:SolvingtheStrongPhasesAmbiguity}

In our earlier publication \cite{BaBarFullAngularAnalysisPRL2001} we
presented evidence for the presence of a $K\pi$ $S$-wave amplitude 
in the  $1.1 < m_{K \pi} < 1.3 \gevcc$ range.
We study this $S$ wave in more detail here, in particular its
interference with the $P$ wave in the vicinity of the $\Kstar(892)$ resonance.
We then use this interference to resolve the strong phase ambiguity for the
$\B\to\jpsi\Kstar(892)$ decay amplitudes, using the observations of
Sec.~\ref{subsec:ProbingthepresenceofaKpiSwave}.

In the following we will denote the two strong phase solutions
obtained in the analysis of Sec. 
\ref{sec:AngularAnalysis} based on a purely  $P$-wave angular distribution,
 by:
\begin{eqnarray}
 {\rm Solution\ I:\ }
(\delta_\parallel-\delta_0 , \delta_\perp-\delta_0) &\simeq &(2.7,0.2), \label{eqn:sol1}
\\
{\rm Solution\ II:\ }
(\delta_\parallel-\delta_0 , \delta_\perp-\delta_0) &\simeq &(-2.7,\pi-0.2). \nonumber \\\label{eqn:sol2}
\end{eqnarray}

\begin{figure}[tbp]
\begin{center}
\includegraphics[width=\singlewidth]{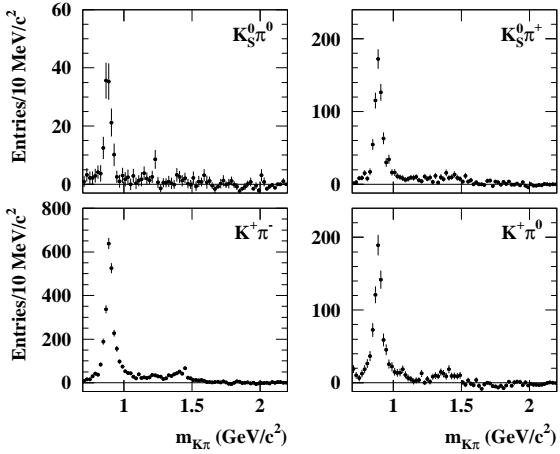}
\end{center}
\caption{The background-subtracted $K\pi$ invariant mass distributions for $\jpsi K\pi$ candidates in data. 
\label{fig:mkpi}}
\end{figure}

The $K\pi$ mass requirement mentioned in Sec.~\ref{sec:EventSelection} ($m_{K\pi}$ within 100\mevcc
of the nominal $\Kstar(892)$ mass) is now relaxed, and the whole kinematical domain for the $K\pi$ system
from $\B\to\jpsi K\pi$ decay is used.
The $m_{K \pi}$ spectra are shown in Fig.~\ref{fig:mkpi}.

\subsection{Probing the \SP interference}
\label{sec:probingTheSPInterference}
\begin{figure}[tbp]
\begin{center}
\includegraphics[width=\singlewidth]{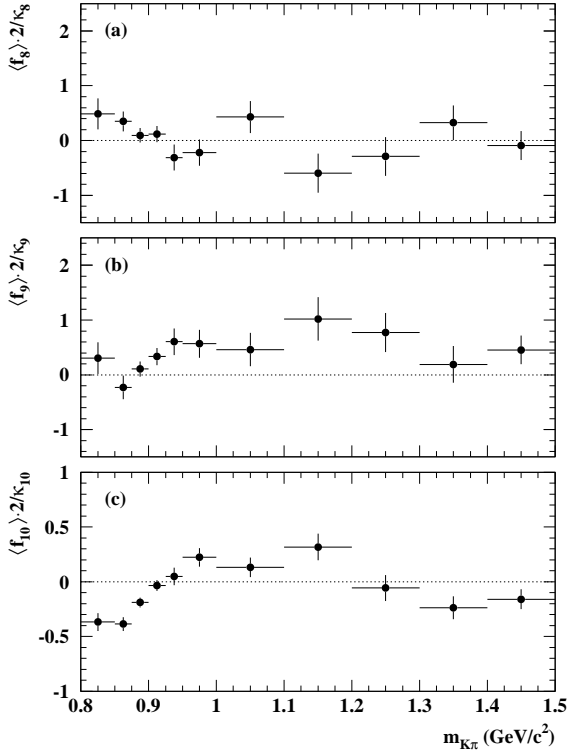}
\caption{\label{fig:moments_8_9_10}Measured values of (a) 
$\langle f_8\rangle \cdot 2/\kappa_8$, (b) $\langle f_9\rangle \cdot 2/\kappa_9$ and (c) $\langle f_{10}\rangle \cdot 2/\kappa_{10}$, defined in Eq.~(\ref{eqn:f_10_plotted}),
 as a function of  $m_{K\pi}$, for the $\jpsi\Kpm\pimp$ candidates in data. The three distributions show a clear variation near the $K^*(892)$ region.}
\end{center}
\end{figure}

We use the $\Kpm\pimp$ sample since it is the largest sample and has the lowest background level. We split this
sample into $K\pi$ mass intervals so that each interval has approximatly the same number of candidates. Equation~(\ref{eqn:g_S+P_reduite}) shows that
the presence of both $K\pi$ $P$-wave and $K\pi$ $S$-wave amplitudes (i.e. $\lambda\neq 0$ and $\lambda\neq\pi/2$)
implies the presence of \SP interference. Before fitting the data to the distribution of Eq.~(\ref{eqn:g_S+P_reduite}),
we check for the presence of such interference effects by evaluating the moments of the angular functions $f_{8,9,10}$. The orthogonality of these functions is expressed by
\begin{eqnarray}
\label{eqn:orthogonality-f-i}
\int f_i(\vomega) f_j(\vomega) {\rm d}{\vomega} =
 \delta_{ij} \kappa_i\;\;\;\; (i&=&8,9,10\;;\nonumber \\ j&=&1,\dots, 10),
\end{eqnarray}
with $\kappa_8 = \kappa_9 = 3/40\pi$ and $\kappa_{10} = 3/4\pi$.
The moments are defined by
\begin{equation}
\label{eqn:defMoments}
\langle f_{i} \rangle \equiv \int g_{S+P}(\vomega;m_{K\pi},\boldsymbol{A},\lambda)
f_i(\vomega) \; {\rm d}{\vomega},
\end{equation}
and are functions of $m_{K\pi}$. Using Eq.~(\ref{eqn:defMoments}), we obtain for $i=8,9,10$:
\begin{eqnarray}
\frac{2}{\kappa_8} \langle f_{8} \rangle &=& \sin 2\lambda \cos(\delta_\parallel -\delta_S)|A_\parallel |, \nonumber \\
\frac{2}{\kappa_9} \langle f_{9} \rangle &=& \sin 2\lambda \sin(\delta_\perp-\delta_S)|A_\perp|, \nonumber \\
\label{eqn:f_10_plotted}
\frac{2}{\kappa_{10}} \langle f_{10} \rangle &=& \sin 2\lambda \cos(\delta_S-\delta_0)|A_0|.
\end{eqnarray}
The behaviour with $m_{K\pi}$ of the right side of Eq.~(\ref{eqn:f_10_plotted}) terms in data can be displayed by evaluating, in each $K\pi$ mass interval, the related moments.
Their background-subtracted, acceptance-corrected distributions are shown in Fig.~\ref{fig:moments_8_9_10} for data.
They show rapid variation near the position of the $K^*(892)$,
where the phase of the $P$-wave changes most rapidly.
Similar distributions obtained from inclusive \jpsi MC samples, in which
no interference between $S$ and $P$ waves is simulated, show values of the moments
compatible with zero in the corresponding mass range.
In addition, the fact that the
moments $\langle f_8\rangle$, $\langle f_9\rangle$, $\langle f_{10}\rangle$ show
significant deviation
from zero in the $K\pi$ mass region
above 0.8 \gevcc is a clear indication of
the presence of an $S$-wave $K\pi$ amplitude in this region, interfering with the $P$-wave amplitudes.

We also note that the $\cos\theta_{K^*}$ forward-backward asymmetry
\begin{eqnarray}
A_{FB} &\equiv&  \frac{N(\cos\theta_{K^*}>0)-N(\cos\theta_{K^*}<0)}{N(\cos\theta_{K^*}>0)+N(\cos\theta_{K^*}<0)} \nonumber\\
&=&  \frac{\sqrt{3}}{2}\sin2\lambda\cos(\delta_S-\delta_0) |A_0|
\end{eqnarray}
is proportional to $\langle f_{10}\rangle$ (Eq.~(\ref{eqn:f_10_plotted})).
The distribution of $\langle f_{10}\rangle$ (Fig.~\ref{fig:moments_8_9_10}(c)) has a mean value of $-0.14\pm 0.03$ in the 0.8---1.0 \gevcc $K\pi$ mass range,
thus indicating a global $\cos\theta_{K^*}$ backward trend in the $K^*(892)$ region, as observed in Fig.~\ref{fig:1Dang}.

\subsection{Fitting for $\mathbf{\delta_S-\delta_0}$}
\label{subsec:Fittingforgamma}
\begin{figure}[tbp]
\begin{center}
\includegraphics[width=\singlewidth]{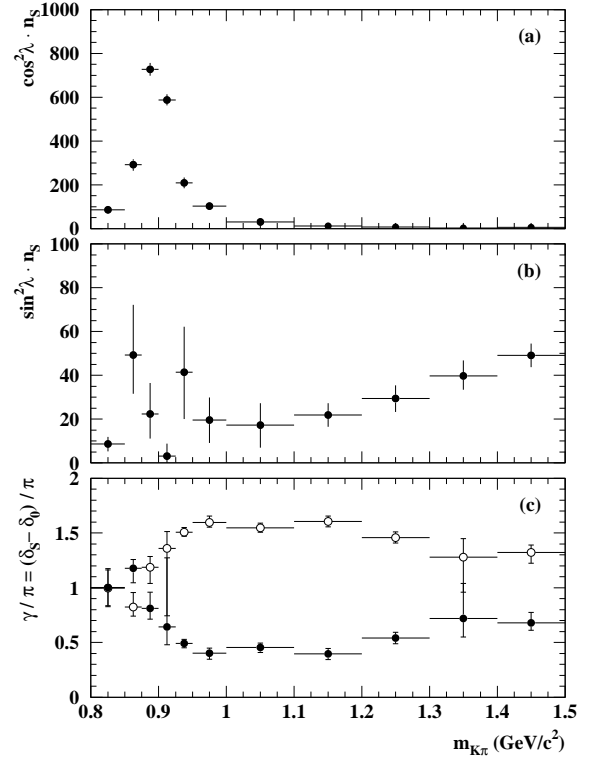}
\caption{\label{fig:break-amb-ch2}
(a) The $P$-wave intensity times number of events; i.e., $n_S \cos^2\lambda $;
(b) the $S$-wave intensity  times number of events; i.e., $ n_S\sin^2\lambda$.
These are the numbers of events that would be observed in each interval for the amplitude under consideration, if it were the only amplitude.
The  fraction of  $S$-wave intensity  integrated over the range  $0.8 < m_{K\pi} < 1.0 \gevcc$ is found to be $(7.3 \pm 1.8)\%$.
(c) The evolution of $\gamma/\pi $ with $m_{K\pi}$, for the two sets of strong phases.
The mirror symmetry described by 
Eq.~(\protect\ref{eqn:phase_S+P_ambiguity}) is clearly visible as 
$\gamma \leftrightarrow  2\pi-\gamma$.
The error bars represent the statistical uncertainty in the fit 
of ($\lambda, \gamma$) in the  $m_{K\pi}$ interval considered. All distributions are from fits to $\jpsi\Kpm\pimp$ candidates in data.}
\end{center}
\end{figure}

The $S+P$ angular distribution (Eq.~(\ref{eqn:g_S+P_reduite})) is fit to the data
in each $K\pi$ mass interval of Fig.~\ref{fig:moments_8_9_10} in order to obtain the
values of $\lambda$ and $\gamma= \delta_S - \delta_0$.
Separate fits are performed for the two possible strong phase solutions (Eqs.~(\ref{eqn:sol1},\ref{eqn:sol2})).
We fix the $P$-wave amplitudes to the values obtained previously (Eq.~(\ref{eqn:angFitNum})).
The methods for acceptance correction and background subtraction described in
Sec. \ref{sec:AngularAnalysis} are also applied here. Any variation of the acceptance with $m_{K\pi}$ is neglected.

The fit results for the $K^\pm\pi^\mp$ channel are shown in 
Fig.~\ref{fig:break-amb-ch2}. 
Figure~\ref{fig:break-amb-ch2}(a) shows the $P$-wave
intensity, namely $\cos^2\lambda\cdot n_S$, and
Fig.~\ref{fig:break-amb-ch2}(b) shows the $S$-wave intensity. 
 The quantity $n_S$ is the
estimated number of signal events in the given $m_{K\pi}$ mass interval and is obtained
from a fit similar to that in Fig.~\ref{fig:resmesde}, but with
the Gaussian parameters fixed to the values obtained there.

Figure~\ref{fig:break-amb-ch2}(c)
 shows the phase $\gamma$ as a function of $m_{K\pi}$ for the two
solutions for the strong phases (Eqs.~(\ref{eqn:sol1},\ref{eqn:sol2})). We see that the two solutions show opposite behavior in each mass interval,
as they must (see Eq.~(\ref{eqn:phase_S+P_ambiguity})).
The large excursion in the relative phase in passing through the
$K^*(892)$ region supports our assumption that the phases of the decay
amplitudes reflect the phases of the simple $K\pi$ system.

The full points  of Fig.~\ref{fig:break-amb-ch2}(c)
are  obtained with strong phases of ``Solution II'', for which
$\gamma$ is decreasing in the $K^*(892)$ region,
as required for the physical solution. A conservative estimate of the discrimination between the two solutions is made by fitting for the
slope $d\gamma/dm_{K\pi}$ in the range $0.8 <m_{K\pi}< 1.0$ \gevcc; we find
\begin{center}
\begin{tabular}{lr} 
Solution I: & $16.2 \pm 2.7$ rad/\gevcc\\ 
Solution II: & $-16.2 \pm 2.7$ rad/\gevcc\\  
\end{tabular}
\end{center}
As they must, these two slopes have opposite values.
The two fits have a $\chi^2$ per degree of freedom of 1.6.
Finally, interpreting Solution II as the
physical solution, we obtain the unique result
\begin{eqnarray}
\delta_\parallel-\delta_0 &=& (-2.73 \pm 0.10 \pm 0.05) \rad, \nonumber \\
\label{eqn:solutionIIvalues}
\delta_\perp-\delta_0 &=& (+2.96 \pm 0.07 \pm 0.05) \rad,
\end{eqnarray}
{\it i.e.}, the two relative phase  values are approximately equal in magnitude but with opposite sign.\\

It should be noted that this phase solution is not that selected in previous papers, nor in Eq.~(\ref{eqn:angFitNum}).\\

\subsection{Examining the Moments}
\label{subsec:Examiningthemoments}

The values of the moments $\langle f_{8}\rangle$, $\langle f_{9}\rangle$, and $\langle f_{10}\rangle$ (Eq.~(\ref{eqn:f_10_plotted}))
are unchanged under the strong phases transformation
Eq.~(\ref{eqn:phase_S+P_ambiguity}), and, as such,
do not allow us to distinguish between Solution I and II; but we show here that their variation, in particular that of $\langle f_9 \rangle$,
with $m_{K\pi}$,
together with the physical requirement that ${\rm d}\gamma/{\rm d} m_{K\pi}<0$,
allows us to resolve the ambiguity, without relying on the explicit solutions displayed in Fig.~\ref{fig:break-amb-ch2}(c).

Since $\lambda$ is small and positive, $\sin2\lambda>0$ and from Eq.~(\ref{eqn:f_10_plotted}) we can write
\begin{eqnarray}
\label{eqn:deriv_f8}
\frac{{\rm d} \langle f_{8}\rangle}{{\rm d} m_{K\pi}} &\sim& +\sin(\delta_\|-\delta_0-\gamma)\frac{{\rm d} \gamma}{{\rm d} m_{K\pi}},  \\
\label{eqn:deriv_f9}
\frac{{\rm d} \langle f_{9}\rangle}{{\rm d} m_{K\pi}} &\sim& -\cos(\delta_\perp-\delta_0-\gamma)\frac{{\rm d} \gamma}{{\rm d} m_{K\pi}}, \\
\label{eqn:deriv_f10}
\frac{{\rm d} \langle f_{10}\rangle}{{\rm d} m_{K\pi}} &\sim& -\sin\gamma\frac{{\rm d} \gamma}{{\rm d} m_{K\pi}}.
\end{eqnarray}
Given that the values for $\delta_\|$ and $\delta_\perp$ are close to $0$ or $\pi$ (Eqs.~(\ref{eqn:sol1},\ref{eqn:sol2})),
we can approximate Eqs.~(\ref{eqn:deriv_f8},\ref{eqn:deriv_f9}) by
\begin{eqnarray}
\frac{{\rm d} \langle f_{8}\rangle}{{\rm d} m_{K\pi}} &\sim& - \cos(\delta_\|-\delta_0)\sin\gamma \frac{{\rm d}\gamma}{{\rm d} m_{K\pi}},  \\
\frac{{\rm d} \langle f_{9}\rangle}{{\rm d} m_{K\pi}} &\sim& - \cos(\delta_\perp-\delta_0)\cos\gamma \frac{{\rm d}\gamma}{{\rm d} m_{K\pi}}.
\end{eqnarray}
On Fig.~\ref{fig:moments_8_9_10}(c) we observe, in the $\Kstar(892)$ region, that
\begin{eqnarray}
 \langle f_{10}\rangle &\sim& +\cos\gamma \ < 0, \\
\frac{{\rm d} \langle f_{10}\rangle}{{\rm d} m_{K\pi}} &\sim& -\sin\gamma\frac{{\rm d} \gamma}{{\rm d} m_{K\pi}} \ > 0,
\end{eqnarray}
meaning that
\begin{eqnarray}
\label{eqn:sign_from_f8}
\frac{{\rm d} \langle f_{8}\rangle}{{\rm d} m_{K\pi}} && {\rm has\ the\ sign\ of\ \ } \cos(\delta_\|-\delta_0),\\
\label{eqn:sign_from_f9}
\frac{{\rm d} \langle f_{9}\rangle}{{\rm d} m_{K\pi}} && {\rm has\ the\ sign\ of\ \ } \cos(\delta_\perp-\delta_0)\frac{{\rm d}\gamma}{{\rm d} m_{K\pi}}.
\end{eqnarray}
The variation of $\langle f_{8}\rangle$ observed on  Fig.~\ref{fig:moments_8_9_10}(a) is compatible with Eq.~(\ref{eqn:sign_from_f8}), whichever strong
phase solution is considered (Eqs.~(\ref{eqn:sol1},\ref{eqn:sol2}))
and thus cannot distinguish between the physical solution and the non-physical one. Figure~\ref{fig:moments_8_9_10}(b)
shows that ${\rm d} \langle f_{9}\rangle / {\rm d} m_{K\pi} > 0$, meaning that either
\begin{eqnarray}
\cos(\delta_\perp-\delta_0) > 0 &{\rm\ and\ }& \frac{{\rm d}\gamma}{{\rm d} m_{K\pi}} > 0,\\
\cos(\delta_\perp-\delta_0) < 0 &{\rm\ and\ }& \frac{{\rm d}\gamma}{{\rm d} m_{K\pi}} < 0.
\end{eqnarray}
We note that $\cos(\delta_\perp-\delta_0) > 0$ for Solution I (Eq.~(\ref{eqn:sol1})), and $\cos(\delta_\perp-\delta_0) < 0$ for Solution II (Eq.~(\ref{eqn:sol2})).
The variation of $\langle f_9 \rangle$ with $m_{K\pi}$ provides thus the association of Solution II with the physical requirement ${\rm d}\gamma/{\rm d} m_{K\pi} < 0$,
and of Solution I with the non-physical case ${\rm d}\gamma/{\rm d} m_{K\pi} > 0$. This leads to select Solution II as the physical solution,
consistently with the previous section.

\subsection{Checking the $(K\pi)_{P-{\rm wave}}$ lineshape}
\begin{figure}[tbp]
\begin{center}
\includegraphics[width=\singlewidth]{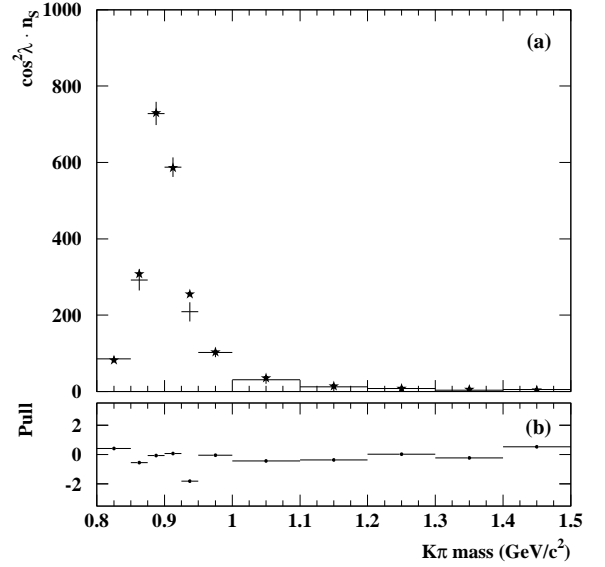}
\caption{\label{fig:kstshape-ch2}(a) Comparison of the $K^\pm\pi^\mp$ $P$-wave intensity with a Breit-Wigner lineshape,
including a centrifugal barrier factor, with world average parameters for $K^*(892)$~\cite{ref:pdg}.
The lineshape is integrated in each mass interval (star markers) and compared with the measured intensity in that interval, after
a minimun $\chi^2$ fit of the overall normalization to the data in the  0.8--1.3 \gevcc mass range. The $\chi^2$ per degree of freedom is 0.86.
(b) Pull (i.e. difference of measured and expected intensities, normalized to the uncertainty)
 in each mass interval.}
\end{center}
\end{figure}

Figure~\ref{fig:kstshape-ch2} compares the $P$-wave intensity (as already shown in Fig.~\ref{fig:break-amb-ch2}(a)) with a Breit-Wigner lineshape, 
including a centrifugal barrier factor,
using the world average~\cite{ref:pdg} parameter values for $K^*(892)$. (The mass resolution is about 3 MeV/c$^2$ and is negligible in its effect.)  The overall
normalization is fit in the 0.8--1.3 \gevcc mass range. The $\chi^2$ per degree of freedom is 0.86. The good agreement observed between the data and the Breit-Wigner lineshape
suggests that the final-state interactions observed at the
end of Sec.~\ref{subsec:Results},
though statistically significant, are not so great as to distort the lineshape.
This is consistent with
our hypothesis of small interaction  between the $\jpsi$ meson and the $K\pi$ system,
made at the end of Sec.~\ref{subsec:ProbingthepresenceofaKpiSwave}.

\subsection{Comparison with $\boldsymbol{K^- p\to K^- \pi^+ (n)}$ Scattering Results}
\label{subsec:Swave}

\begin{figure}[tbp]
\begin{center}
\includegraphics[width=\singlewidth]{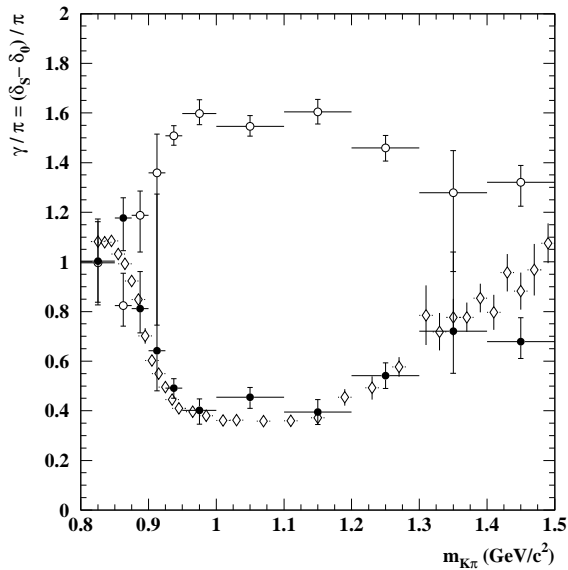}
\caption{\label{fig:gamma-babar-lass-ch2}
Comparison of the variation of $\gamma=\delta_S-\delta_0$ with $m_{K\pi}$
for the $\jpsi \Kpm\pimp$ events, for
 ``Solution I'' (open points, Eq.~(\ref{eqn:sol1})) and 
 ``Solution II'' (full points, Eq.~(\ref{eqn:sol2})), 
with that measured by the LASS experiment~\cite{AwajiHakase1986,Aston:1987ir,LASSreFit} (diamond markers).}
\end{center}
\end{figure}

In Fig.~\ref{fig:gamma-babar-lass-ch2} we compare the evolution of
$\gamma$ observed in Fig.~\ref{fig:break-amb-ch2}(c) with that obtained from the LASS
measurement of $\Km p\to \Km \pip (n)$ scattering.
The LASS points~\cite{LASSreFit} (based on  data from Refs.~\cite{AwajiHakase1986,Aston:1987ir}),
represented as diamonds, show the phase difference
\[
\delta_{S(I=1/2)} - \delta_{P(I=1/2)}
\]
as a function of  $m_{K\pi}$. 
Only the $I=1/2$ amplitude is retained
 since this is the only one produced 
by the $\B\to\jpsi K \pi$ process.
The LASS  analysis takes into account the $D$ wave, while the present
analysis does not, but the $D$ wave ($\Kstar_2(1430), \Gamma \sim 100$
MeV) has an effect only at high  $m_{K\pi}$. 
An overall shift of $\pi$ radian is added to the LASS phase difference
measurements
in order to match the sign of the
forward-backward asymmetry observed with the $\B\to \jpsi K\pi$ events.
This shift does not modify the slope and general shape that are of interest here.
The shift corresponds merely to changing the relative sign between
the $S$  and $P$ wave amplitudes.
The need for such a global shift  is not surprising since the production processes
are unrelated.
We can see that the agreement between ``Solution II'' and LASS is
striking (Fig.~\ref{fig:gamma-babar-lass-ch2}).

\section{Measurement of $\boldsymbol{\cos 2\beta}$ }
\label{sec:Measurementofcos2beta}

To measure $\ctwob$,
we perform an unbinned maximum likelihood fit to 
the differential decay rate as a function of proper time and the
three angular variables (Eqs.~(\ref{eqn:time_dependent_g},\ref{eqn:amplitude_de_t})) for the
$\Bz \to \jpsi \Kstarz (\KS \piz)$ sample. The amplitude
parameters  $\vert A_0 \vert$, $\vert A_{\perp} \vert$, $\vert A_{\parallel} \vert$,
$\delta_{\parallel}-\delta_{0}$, and $\delta_{\perp}-\delta_{0}$  in Eq.~(\ref{eqn:amplitude_de_t}) are fixed to those
obtained by the angular analysis of the three high-statistics flavor-specific
$B \to \jpsi \Kstar$ channels, presented in Secs.~\ref{sec:AngularAnalysis} and~\ref{sec:SolvingtheStrongPhasesAmbiguity}.
In particular, the strong phases $\delta_{\parallel}-\delta_{0}$ and $\delta_{\perp}-\delta_{0}$ are fixed to ``Solution II'' (Eq.~(\ref{eqn:sol2},\ref{eqn:solutionIIvalues})),
described in Sec.~\ref{sec:SolvingtheStrongPhasesAmbiguity}.

We examine each event with a 
$\Bz \to \jpsi \Kstarz (\KS \piz)$ candidate, indicated by $B_{CP}$,
 for evidence
that the other neutral \B meson, $B_{tag}$, decayed
as a \Bz\ or a \Bzb (flavor tag, as described below).  We also determine the decay
proper-time difference $\deltat \equiv t_{CP} - t_{tag}$,  which
corresponds to the variable $t$ in  Eqs.~(\ref{eqn:time_dependent_g})-(\ref{eqn:asymm_CP}).
To a good approximation
$c \cdot\deltat = \deltaz / (\gamma_B \cdot  \beta \cdot \gamma)$
where $\deltaz$ is the separation between the $B_{CP}$   
and $B_{tag}$ vertices along the \epem collision axis
and $\gamma_B \cdot\beta \cdot \gamma = (p_{e^-} - p_{e^+}) / (2\cdot m_B)$.

The background level in the $\Bz\to\jpsi\KS\piz$ sample is higher than in
the other $\B\to\jpsi K\pi$ channels. In particular, some \CP-violating backgrounds
 tend to peak in the signal region ($\mes > 5.27$~\gevcc),
making the $\mes < 5.27$~\gevcc region a poorer representation
of the background behaviour than in the other $\B\to\jpsi K\pi$ channels.
In contrast with the method used for the angular analysis 
described above, the \CP analysis is performed by the maximizing a
likelihood function that contains the PDFs of  both the signal and the background.
Only events from the signal region are used. Monte Carlo samples are
used to determine
the angular acceptance, and the background composition and angular dependence, as described
in the following Sections.

\subsection{Background Contributions}

The inclusive \jpsi Monte Carlo sample is used
to determine the composition of the $\Bz \to \jpsi \Kstarz (\KS \piz)$ candidate
sample. The results are shown in Table~\ref{tab:cp-backgrounds}.
 Aside from the signal and cross-feed contributions, the 
dominant contributions are from $\Bz \to \chic1 \KS$ and 
$\Bz \to \jpsi K^{**}$ decays, where $K^{**}$ denotes higher-mass $K^*$ resonances.
The angle- and \deltat -dependent PDF for each of these backgrounds
is described in the next Section.

\begin{table}
\caption{\label{tab:cp-backgrounds}
Composition of the reconstructed $\Bz \to \jpsi \Kstarz (\KS\piz)$ candidate sample
in the region $\mes > 5.27$ \gevcc.
The background fractions are estimated from
an inclusive \jpsi Monte Carlo 
sample with an equivalent integrated luminosity of
590 \invfb with the requirement $p^*_{\jpsi} > 1.3~\gevc$.
The first
uncertainty is statistical; the second is  systematic and is
based on the  uncertainty in the corresponding branching fraction \protect\cite{ref:pdg}
if it is known, and otherwise is based on a 50\% uncertainty on
the branching fraction used in the Monte Carlo generator.}
\begin{center}
\begin{ruledtabular}
\begin{tabular}{lcc}
 & Fraction(\%) & Effective \CP \\ \hline
Signal & $83.0\pm 1.2 \pm 5.7$ &  \\
Cross-feed & $3.2 \pm 0.5 \pm 0.3$ & $0$ \\
$\Bz \to \chi_{c1} \KS$ & $4.0 \pm 0.6 \pm 1.1$ & $-1$ \\
Higher-mass \Kstar resonances & $3.5 \pm 0.6 \pm 1.8$ & $0\pm1$ \\
Non-resonant $\Bz \to \jpsi \KS\piz$ & $2.8 \pm 0.5 \pm 1.4$ & $0\pm1$ \\
Other $\B \to \jpsi X$ & $3.5 \pm 0.6 \pm 1.8$ & $0\pm1$ \\
\end{tabular} \end{ruledtabular}
\end{center}
\end{table}

\subsection{Acceptance Description}

The  acceptance and the combinatorial background PDF are described by
expansions in terms of products of spherical harmonics.
The orthonormal basis functions used are
\begin{eqnarray}
\label{eqn:expansion-basis}
{\cal Y}_{LRM}(\vomega) &\equiv& \sqrt{2\pi} Y_{LM}(\theta_\psi,\chi) Y_{RM}(\theta_{K^*},0),
\end{eqnarray}
where $\theta_\psi,\chi$ are the helicity angles 
corresponding to the transversity angles $\theta_{tr}, \phi_{tr}$ \cite{StephThese} and the $Y_{lm}$ are spherical harmonic functions.
These ${\cal Y}_{LRM}$ functions describe the partial waves involved in a ``Scalar $\to$ Vector $X$'' decay, where $X$ can be of arbitrary spin~\cite{DunietzPRD1990}.
A function of $\vomega$, say $\epsilon(\vomega)$, is expanded as:
\begin{eqnarray}
\label{eqn:spherical-harmonics-expansion}
\epsilon(\vomega) &=& \sum_{LRM} T^\epsilon_{LRM} {\cal Y}_{LRM}(\vomega),
\end{eqnarray}
where the sums over $L$ and $R$ run, in principle, from 0 to infinity,
and the sum over $M$ from $-\min(L,R)$ to $+\min(L,R)$.

The moments of the acceptance are estimated from Monte Carlo simulation,
with $N_{gen}$ events generated with PDF $g$, $N_{obs}$ events being
finally selected:
\begin{eqnarray}
\label{eqn:moments-efficiency}
T^\epsilon_{LRM} &\equiv& \int \epsilon(\vomega) {\cal Y}^*_{LRM}(\vomega) d\vomega \nonumber \\
&\approx& \frac{1}{N_{gen}} \sum_{i=1}^{N_{obs}}\frac{1}{g(\vomega_i)} {\cal Y}^*_{LRM}(\vomega_i).
\end{eqnarray}
The moments for the background  PDFs are computed using the
reconstructed events of the MC background sample distributed as $b(\vomega)$:
\begin{eqnarray}
\label{eqn:moments-background}
T^b_{LRM} &=& \int b(\vomega) {\cal Y}^*_{LRM}(\vomega) d\vomega  \nonumber \\
&\approx& \frac{1}{N_b}\sum_{i=1}^{N_b} {\cal Y}^*_{LRM}(\vomega_i).
\end{eqnarray}

We note that the analytical expressions for the background 
PDF and for the efficiency are not needed to compute these moments.
The expansion is done up to rank $L=R= 4$ for the acceptance (signal distribution is of rank 2) and 
up to rank $L=R= 6$ for the background. These ranks are chosen to be large enough so that no significant
deviation of the fitted parameters $\sin2\beta$ and $\cos2\beta$ is observed in high-statistics Monte Carlo samples when compared to the generated values.

\subsection{The $\mathbf{B_{flav}}$  Sample}

The fit is additionally performed on  a large sample of
fully reconstructed $\Bz$ decays to flavor 
eigenstates ($B_{flav}$) with decays 
$\Bz \to D^{(*)\pm}h^{\mp}$, 
where $h=\pi, \rho, a_{1}$ and $\Bz \to \jpsi \Kstarz$
$(\Kstarz\to \Kpm \pimp)$.
These events are used to measure the parameters of the
 flavor-tagging algorithm and of the  \deltat-resolution
functions.
Flavor tagging performance is shown to be independent of the fully-reconstructed \B meson and the $\Delta t$
resolution is dominated by the vertex resolution of the incompletely reconstructed tagging \B meson.
Thus both tagging and $\Delta t$ resolution can be studied with these large, well understood samples and the results applied to the channels of interest.

The fully reconstructed \B meson, i.e., 
$B_{flav}$ or $B_{CP}$, is denoted by $B_{rec}$.

\subsection{Flavor Tag Determination}

We use a multivariate technique \cite{BaBarSin2BetaPRL2002}
to determine the flavor of the $\B_{tag}$ meson.
Separate neural networks are trained to identify primary leptons,
kaons, soft pions from $D^*$ decays, and high-momentum charged particles 
from \B decays.
Events are assigned to one of five mutually exclusive tagging
categories based on the estimated mistag probability and the source of
the tagging information:
{\tt Lepton, Kaon I, Kaon II, Inclusive} and 
{\tt Untagged}.  The {\tt Untagged} events are not
used in this analysis.

We determine the average dilution $\langle D \rangle$ 
and dilution  difference $\Delta D$, defined as 
\begin{eqnarray}
\overline{D} &\equiv& 1-2 \overline{w}, \nonumber \\
D &\equiv& 1-2 w, \nonumber \\
\langle D \rangle &\equiv& \frac{D+\overline{D}}{2},\nonumber \\
\Delta D &\equiv& D-\overline{D},
 \label{eqn:tagging-dilution-def}
\end{eqnarray}
for each tagging category, where $w$ ($\overline{w}$) is the probability that
a flavor tag determination is  incorrect when the true tag 
is a $\Bz$ ($\Bzb$).  
The quality of the tagging is expressed in terms of the effective
efficiency $Q \equiv \sum_{k}\epsilon_{k}(1-2w_{k})^2$, where
$\epsilon_{k}$ and $w_{k}$ are the efficiencies and mistag
probabilities, respectively, for events tagged in category $k$.
The tagging performance is measured in a large data sample of fully
reconstructed \B decays.
The effective tagging efficiency is $(28.1 \pm 0.7)\%$
\cite{BaBarSin2BetaPRL2002}.
The tagging efficiency asymmetry between \Bz\ and \Bzb\ has been
studied \cite{StephThese,625} using the full simulation of the
experiment and has been found to be negligible for this analysis.

\subsection{Determination of \deltat\ and \deltat\ Resolution}
\label{subsec:DeltatResolution}

The proper-time difference \deltat between the decays of the two \B
mesons in the event ($\B_{rec},\B_{tag}$) is  determined from the 
measured separation along the collision axis, \deltaz,
between the $\B_{rec}$ and the $\B_{tag}$ vertices
(Eq.~(22) of Ref.~\cite{BaBarTheBigPRD_CP_AsymFlavorOsc}.)
The $\B_{tag}$ decay vertex is obtained by fitting tracks that  do not
belong to the $\B_{rec}$ candidate, imposing constraints from the 
 $\B_{rec}$ momentum and the beam spot location.
The average \deltat resolution is approximately 1.1 ps.
We require that the measured proper-time difference between the 
 $\B_{rec}$ and the $\B_{tag}$ decays satisfies $|\deltat|<20~\text{ps}$ 
and that the estimated uncertainty in \deltat, $\sigma_{\deltat}$,
which is derived from the vertex fit for the event,
be less than 2.5 ps.

The \deltat-resolution function ${\cal R}$ is represented 
by a sum of three
Gaussian distributions (called the core, tail, and outlier components):
\begin{eqnarray}
{\cal R}(\delta(\deltat) &\equiv& f_{core}{\rm G}(\delta(\deltat);\mu_{core},\sigma_{core}) + \nonumber\\
                                           &~& f_{tail}{\rm G}(\delta(\deltat);\mu_{tail},\sigma_{tail}) + \nonumber\\
\label{eqn:dt_res_function}
                                           &~&  f_{outlier}{\rm G}(\delta(\deltat);\mu_{outlier},\sigma_{outlier}), \\ \nonumber
\end{eqnarray}
where $G$ is the Gaussian function, $\delta(\deltat) \equiv \deltat-\deltat_{\rm true}$,
$\deltat_{\rm true}$ is the actual decay time difference, and $f_{core}$, $f_{tail}$, and $f_{outlier}$ the fractions of each component.

For the width of the core and tail Gaussians ($\sigma_{core}$, $\sigma_{tail}$), we use the measurement 
uncertainty $\sigma_{\deltat}$ and
allow separate scale factors $S_{core}$ and $S_{tail}$ to
accommodate an overall underestimate ($S_k > 1$) or overestimate 
($S_k < 1$) of the errors for all events, so that $\sigma_{core}=S_{core}\sigma_{\deltat}$ and
$\sigma_{tail}=S_{tail}\sigma_{\deltat}$.

The core and tail Gaussian distributions are allowed to have a nonzero
mean ($\mu_{core}$, $\mu_{tail}$)
to account for charm decay products possibly included in the $B_{\rm tag}$ vertex.
In the resolution function, these mean offsets
are scaled by the event-by-event measurement error $\sigma_{\deltat}$
to account for an observed correlation
\cite{BaBarTheBigPRD_CP_AsymFlavorOsc} between the mean of the
$\delta(\deltat)$ distribution and the measurement error
$\sigma_{\deltat}$ in Monte Carlo simulation.
For the core we allow different means for each flavor-tagging category.
 One common mean is used for the tail components.
The third Gaussian has a fixed width $\sigma_{outlier}=8$\ps\ and no offset ($\mu_{outlier}=0$); it
accounts for fewer than 1\% 
of events, typically due to incorrectly
reconstructed vertices.

\subsection{Likelihood Function}
\label{subsec:Likelihood}

We maximize the log-likelihood given by
\begin{eqnarray}
\label{eqn:log-likelihood-CP-total}
L_{total} = L_{CP}+ L_{flav},
\end{eqnarray}
\begin{eqnarray}
\label{eqn:log-likelihood-CP}
 L_{CP} &=&
\sum_{tag,c} \sum_{i=1}^{N_{obs}}\ln \bigg[ f_{0}b_{0} + f_{\pm}b_{\pm}\nonumber \\
        &~&  \hspace{1.0cm}+\left(1 - ( f_{0} + f_{\pm})\right) g^{\KS\piz,\ {\rm obs}}\bigg], 
\end{eqnarray}
where $f_{0}$ and $f_{\pm}$ are the fractions of ``neutral'' and
``charged'' background in the \CP sample, and $b_0$ and $b_{\pm}$ are
the corresponding background PDFs, described below.  The signal
PDF $g^{\KS\piz,\ {\rm obs}}$ is

\begin{widetext}
\begin{eqnarray}
\label{eqn:obs_pdf_summary_CP}
g_{\zeta}^{\KS\piz,\ {\rm obs}}(\vomega,\deltat;\boldsymbol{A},\sin 2\beta,\cos 2\beta) 
&=&
\epsilon(\vomega) \frac{\Gamma_0}{2}{\cal A}(\vomega;\boldsymbol{A})\times  \nonumber \\
~ & ~ & \hspace{-3cm}
\Bigg\{
\left[1+\zeta\left(\frac{\Delta D}{2}\right)\right]
\ER(\deltat)
-\zeta\langle D \rangle
\bigg[
\CR(\deltat)
 \frac{{\cal P}(\vomega;\boldsymbol{A})}{{\cal A}(\vomega;\boldsymbol{A})}+ 
 \SR(\deltat)\left(
 \frac{{\cal S}(\vomega;\boldsymbol{A})}{{\cal A}(\vomega;\boldsymbol{A})} \sin 2\beta + 
 \frac{{\cal C}(\vomega;\boldsymbol{A})}{{\cal A}(\vomega;\boldsymbol{A})} \cos 2\beta
 \right)
 \bigg]
 \Bigg\}\nonumber\\
~ & ~ & \hspace{-3.4cm}
\left/ 
\left\{\left[1+\zeta\frac{\Delta D}{2}\right]
\sum_{k=1,2,3,5}\!\!\!{\cal A}_k \Phi_k 
-\zeta\langle D \rangle \frac{1}{1+x^2_d}
\sum_{k=4,6}\!{\cal A}_k \Phi_k\right\} \right.
,
\end{eqnarray}
\end{widetext}
where $\zeta$ labels the flavor of the tagging \B meson
($\zeta=1$) or \Bzb meson ($\zeta=-1$), and $x_d={\Delta m}/{\Gamma_0}$.
${\cal A}_k$ is defined by Eq.~(\ref{eqn:numerics}), and 
 $\Phi_k$  is the diagonal part of $\Phi^{\jchan=\KS\piz}_k$
defined in  Eq.~(\ref{eqn:coefs:Eff}).
Only $\varepsilon^{\jchan \to \jchan}, \jchan=\KS\piz$ (Eq.~(\ref{eqn:coefs:Eff})) is considered because the cross-feed is treated separately here, in the background contribution,
as it does not contribute to \CP violation.
The  \deltat-resolution function ${\cal R}$ (Eq.~(\ref{eqn:dt_res_function})) appears in  the following convolutions:
\begin{eqnarray}
\ER(\deltat)
&\equiv& e^{-\Gamma_0|\deltat_{true}|}\otimes {\cal R}(\delta(\deltat)), \nonumber \\
\SR(\deltat)
&\equiv& e^{-\Gamma_0|\deltat_{true}|}\sin(\Delta m \deltat_{true})\otimes {\cal R}(\delta(\deltat)), \nonumber \\
\CR(\deltat) 
&\equiv& e^{-\Gamma_0|\deltat_{true}|}\cos(\Delta m \deltat_{true})\otimes {\cal R}(\delta(\deltat)). \nonumber \\
\label{eqn:time-resolution-convolutions}
\end{eqnarray}

Most of the background is due to inclusive decays of \B mesons to \jpsi (see Table~\ref{tab:backgrounds}).
We account for backgrounds with the following PDF's:
\begin{itemize}
\item
Backgrounds from neutral-\B decays (see Table~\ref{tab:cp-backgrounds}) are parametrized  with a form
analogous to the one that describes $\jpsi\KS$, but with an effective \CP eigenvalue,
$\eta_{CP}$, and angle dependence $b(\vomega)$:
\begin{eqnarray}
\label{eqn:pdf-BG-CP-content}
b_{0,\zeta}(\deltat,\vomega;\stwob, \eta_{CP}) &\equiv& \frac{\Gamma_0}{2}e^{-\Gamma_0 |\deltat|} \times
 \nonumber \\
& &
\hspace{-4.cm}
\left[
\left(1+\zeta \frac{\Delta D}{2}
\right)
 -\zeta \langle D \rangle \eta_{CP} \sin2\beta \sin(\Delta m \deltat)
\right] b(\vomega). 
 \nonumber \\
\end{eqnarray}
For  $\Bz \to \chi_{c1} K^0_S$ the angular dependence $b(\vomega)$ is estimated with the  Monte Carlo,
and parametrized using an  expansion in  ${\cal Y}^*_{LRM}$.
For higher-mass \Kstar resonances, non-resonant $\Bz \to \jpsi \KS\piz$,
 and  other $\B \to \jpsi X$ sources, a flat angular dependence is used.
\item 
Backgrounds from charged \B decays (see Table~\ref{tab:cp-backgrounds}) are dominantly due to cross-feed from $\Bpm\to\jpsi\Kstarpm$. They
have a $\Delta t$ distribution characterized by the decay rate $\Gamma_+$.
They are represented by
\begin{eqnarray}
\label{eqn:pdf-BG-charged}
b_{\pm,\zeta}(\deltat,\vomega) &\equiv& \frac{\Gamma_+}{2}e^{-\Gamma_+ |\deltat|} 
\left(1 + \zeta \frac{\Delta D}{2}\right) b_{\pm}(\vomega).
 \nonumber \\
\end{eqnarray}

\end{itemize}

The $B_{flav}$ sample, which is used to determine the tagging features, enters
the log-likelihood through the $L_{flav}$ term (Eq.(\ref{eqn:log-likelihood-CP-total})), which is based on PDFs for ``mixed'' and ``unmixed'' events as is
appropriate for these neutral \B decays.
The background PDFs include a zero-lifetime contribution, a contribution
with an effective lifetime, and a contribution with an effective
lifetime and an oscillating factor.
The signal PDFs are
\begin{eqnarray}
h_{u,\zeta}(\Delta t) & \propto & \left[ \left(1 + \zeta  \frac{\Delta D}{2}\right) +u \langle D \rangle\cos(\Delta m \Delta t)\right],
 \nonumber \\
\end{eqnarray}
where $u=1$ and  $u=-1$ for unmixed and mixed events, respectively.
A complete description of the log-likelihood $L_{flav}$ term is provided in Ref.~\cite{BaBarTheBigPRD_CP_AsymFlavorOsc}
($\ln{\cal L}_{\rm mix}$ term in Eq.(6) of Ref.~\cite{BaBarTheBigPRD_CP_AsymFlavorOsc}).\\

Finally  the free parameters  in the fit are (see Table~\ref{tab:CP_fit_result})
\begin{itemize}
\item  $\sin 2\beta$ and $\cos 2\beta$ (2),
\item the parameters for the signal \deltat-resolution function (8),
\item the  tagging  parameters  for signal (8),
\item  the  parameters for the  background  $B_{flav}$ \deltat resolution function  (3),
\item  the  parameters describing the composition of the  background PDF for the $B_{flav}$ sample (13).
\end{itemize}
In total there are 34 parameters. We fix $\Gamma_0$ and $\Delta m$ to
their world average values~\cite{ref:pdg}.

\subsection{Validation}
\label{subsec:Validation:CP}

The fitting scheme has been validated using the full simulation and
the large parametrized MC
samples mentioned above. No statistically significant bias is observed
(Table~\ref{tab:Valid_CP_fit}).
\begin{table} [htb]
\caption{\label{tab:Valid_CP_fit} 
Validation on full MC simulation
[inclusive \jpsi MC sample ($p^*_{\jpsi} > 1.3~\textrm{GeV}/c$) and  
$B_{flav}$ samples] 
and large parametrized MC samples.
The generated values of $\sin 2\beta$ and $\cos 2\beta$ are $0.700$
and $0.714$, respectively.}
\begin{center}
\begin{tabular}{llll}\hline\hline
$CP$ Sample  & & $\sin2\beta$ & $\cos2\beta$ \\
\hline
\rule{0mm}{4mm} 
Full MC   & (0.6 \invab) & $0.61\pm 0.16$   & $0.20\pm 0.32$\\
Parametrized MC & (16 \invab)  & $0.709\pm0.017$ & $0.705\pm0.036$ \\
\hline \hline
\end{tabular}
\end{center}
\end{table}

As a further cross-check, the data samples for the $\Bpm \to\jpsi\Kstarpm$,
channels, which are not expected to show any sizeable
\CP violation in the SM, are examined.
For these channels, the differential decay rate does not have a $\sin\Delta m\deltat$ contribution,
so that the coefficients analogous to $\sin2\beta$ and $\cos 2\beta$ should vanish.
No significant deviation this expectation is observed (Table \ref{tab:check_charged_B}).
\begin{table}[htbp]
\caption{\label{tab:check_charged_B}Fit results for the $\Bpm \to \jpsi \Kstarpm$ data control samples.}
\begin{center}
\begin{ruledtabular}
\begin{tabular}{lcc} 
Sample & $\sin2\beta$ & $\cos2\beta$ \\ \hline
\rule{0mm}{4mm} 
$\Bpm \to \jpsi (\KS\pipm)$ & $0.21\pm0.20$ & $-0.21\pm0.47$ \\
$\Bpm \to \jpsi (\Kpm\piz)$ & $0.20\pm0.20$ & $-0.26\pm0.46$ \\
\end{tabular} \end{ruledtabular}
\end{center}
\end{table}

\subsection{Systematic Uncertainties}
\label{subsec:SystematicUncertainties}

The contributions to the systematic uncertainty are summarized in Table~\ref{tab:syst_CP}. 
Systematic uncertainties (a) -- (j) and (q) -- (t) are in common 
with the $\stwob$ analysis \cite{BaBarTheBigPRD_CP_AsymFlavorOsc}
 and are estimated in the same way.
Systematic uncertainties (k) -- (p) are specific to this $\jpsi (\KS\piz)$ analysis
and are elaborated in the following:
\begin{enumerate}
\item[(k)]
The systematic uncertainty due to imperfect knowledge of the fractions and \CP values of the background sources
is obtained by varying the fractions (see
Table~\ref{tab:cp-backgrounds}) by one standard deviation, if the
background is measured, or by 50\% of the branching fraction used in
the  Monte Carlo  otherwise. The effective \CP values (see
Table~\ref{tab:cp-backgrounds}) of unmeasured background is set to $-1$
and then to $+1$ to evaluate the effect on the measured  parameters.
\item[(m)]
Backgrounds are assumed to have the same dilutions  as the signal.
To evaluate the related uncertainty, the dilutions obtained from
the \Bpm sample are used  and the difference in the results is 
taken as the systematic uncertainty.
\item[(n)] 
Random sets of amplitude moduli and strong phases are generated, according to a
multi-Gaussian distribution,
based on the covariance matrix obtained in the fit for the amplitudes and on the sytematic uncertainties 
in the amplitudes.  These amplitudes are used in place of the nominal amplitudes to evaluate 
the variation in the CP parameters.  This procedure incorporates the uncertainties in the
$S$-wave amplitude as well.
\item[(o)] 
The limited size of the Monte Carlo sample induces an uncertainty in
the moments used to determine the acceptance and the
background distribution. This is evaluated by splitting the Monte
Carlo into ten samples, leading to ten \CP measurements, and taking as an
estimate of the uncertainty the RMS divided by $\sqrt{10}$.
\item[(p)] 
A flat angular distribution has been assumed for some of the background components
(Sec.~\ref{subsec:Likelihood}).  We
estimate the magnitude of the related bias by computing the background
moments from low-$m_{ES}$ events.
\end{enumerate}
\begin{table}[!htb]
\caption{\label{tab:syst_CP}
Summary of systematic and statistical uncertainties on $\sin2\beta$ and $\cos2\beta$.}
\begin{center} 
\begin{ruledtabular}
\begin{tabular}{lcc} 
 Source & $\sin2\beta$ & $\cos2\beta$ \\\hline
 \multicolumn{3}{c}{Signal Properties} \\ \hline 
(a) \deltat-resolution function              & $\pm0.002$ & $\pm0.002$ \\ 
(b) signal dilution $B_{CP}$ vs $B_{flav}$   & $\pm0.012$ & $\pm0.013$ \\ 
(c) Gaussian model for \textit{outliers}  & $\pm0.001$& $\pm0.000$\\ 
(d) $f_{tail}$ parameter                     & $\pm0.002$ & $\pm0.003$ \\ 
(e) resolution/tagging correlation           & $\pm0.001$ & $\pm0.001$ \\ 
(f) SVT alignment                            & $\pm 0.010$ & $\pm0.030$ \\ \hline
 \multicolumn{3}{c}{Background properties: $B_{flav}$} \\ \hline 
(g) signal probability                       & $\pm0.001$ & $\pm0.001$ \\ 
(h)  ARGUS  $m_0$ parameter                    & $\pm0.002$ & $\pm0.010$ \\ 
(i) oscillating contribution                 & $\pm0.001$ & $\pm0.022$ \\ 
(j) $\delta_{peak}$ contribution             & $\pm0.001$ & $\pm0.003$ \\ \hline
 \multicolumn{3}{c}{\jpsi (\KS\piz) specific} \\ \hline
(k) background fraction and \CP parity       & $\pm0.032$ & $\pm0.142$ \\ 
(m) background dilutions                     & $\pm0.002$ & $\pm0.006$ \\ 
(n) amplitude uncertainties                  & $\pm0.016$ & $\pm0.154$ \\ 
(o) statistics used for moments              & $\pm0.030$ & $\pm0.030$ \\ 
(p) angular background distribution          & $\pm0.024$ & $\pm0.064$ \\ \hline
 \multicolumn{3}{c}{External parameters} \\ \hline
(q) $z$ scale and ``\textit{boost}''       & $\pm0.001$ & $\pm0.001$ \\ 
(r) beam spot                                & $\pm0.010$ & $\pm0.040$ \\ 
(s) \Bz lifetime                             & $\pm0.014$ & $\pm0.040$ \\ 
(t) $\Delta m$                           & $\pm0.018$ & $\pm0.032$ \\ \hline
 \multicolumn{3}{c}{Monte Carlo} \\ \hline
(u) Monte Carlo statistics                 & $\pm0.130$ & $\pm0.140$ \\ 
\hline
Total systematic uncertainty               &$\pm0.14$ & $\pm0.27$ \\ 
Statistical uncertainty                    &$\pm0.57$ & $^{+0.76}_{-0.96}$ \\ 
\end{tabular} \end{ruledtabular}
\end{center}
\end{table}

\subsection{Results}
\label{subsec:ResultsCP}

\begin{figure*}[tbp]
\begin{center}
\includegraphics[width=\linewidth]{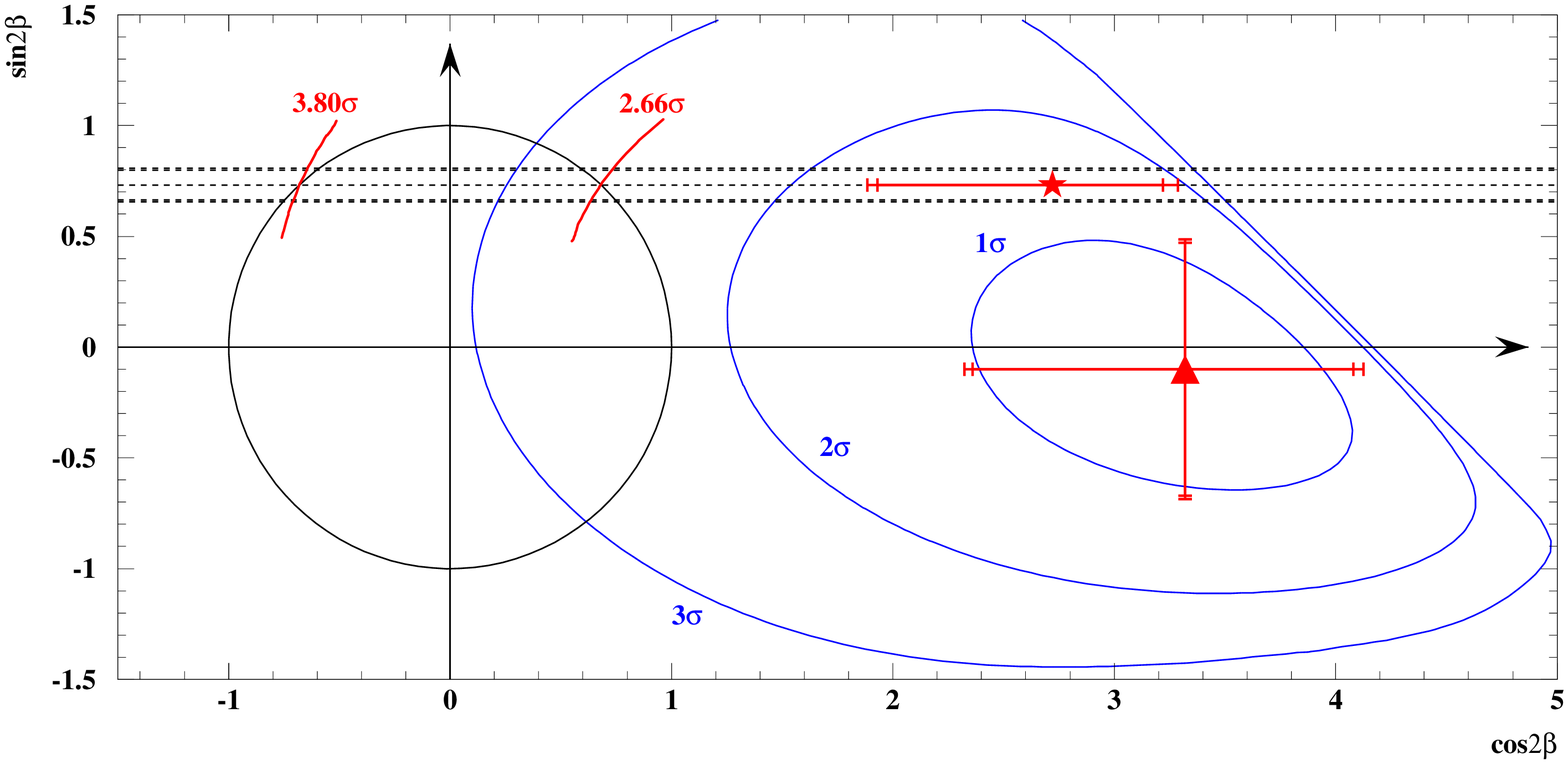}
\caption{\label{fig:iso-lk}
Contour plots in the $\ctwob$, $\stwob$ plane.
The triangle denotes the result of the fit. 
The error bars show the statistical uncertainty and the quadratic sum
of the statistical and  systematic uncertainties.
The star denotes the result of the fit with $\stwob$ fixed at $\stwob =0.731$ 
\protect\cite{PDG2004}.
The value of $\stwob$ of reference \protect\cite{PDG2004} and its uncertainties are represented
as dashed horizontal lines.
The $n\sigma (n=1,2,3)$ contour corresponds to a decrease of $0.5 n^2$  in the log-likelihood  with respect to the maximum value.
The unit  circle ($\cos^2 2\beta + \sin^2 2\beta =1$) on which the true values must lie  is also shown.
}
\end{center}
\end{figure*}

The results of the fit are given in Table \ref{tab:CP_fit_result}.
\begin{table*}
\caption{\label{tab:CP_fit_result} 
Global \CP fit  of the \jpsi \Kstarz(\KS\piz) events 
together with the $B_{flav}$ sample. 
The transversity amplitudes used are those
measured in the angular analysis. 
The $\ctwob$ value shown is the one corresponding to ``Solution II''
for the strong phases. 
The $b$'s are the coefficients of the linear dependence of the \deltat
off-set on \deltat uncertainty : $\langle \deltat \rangle = b \times
\sigma_{\deltat}$. }
\begin{center}
\begin{ruledtabular}
\begin{tabular}{lccc} 
Parameter & Value & Correlation with $\sin2\beta$ & Correlation with $\cos2\beta$\\ \hline
$\sin2\beta$ & $ -0.10 \pm 0.57 $ & $+1.000$ & $-0.368$\\ 
$\cos2\beta$ & $ 3.32^{+0.76}_{-0.96} $ & $-0.368$ & $+1.000$ \\
\hline
\multicolumn{4}{c}{Signal resolution function}\\ \hline
$S_{core}$ & $\phm 1.093 \pm 0.048 $ & $-0.020$ & $+0.028$\\
$S_{tail}$ & $\!\!3.0$ (fixed) & & \\ 
$b_{core}$ {\tt Lepton}    & $\phm 0.012 \pm 0.063$ & $+0.017$ & $-0.010$\\
$b_{core}$ {\tt Kaon~I}    & $-0.226 \pm 0.052$ & $+0.008$ & $-0.050$\\
$b_{core}$ {\tt Kaon~II}   & $-0.248 \pm 0.046$ & $+0.013$ & $-0.023$\\
$b_{core}$ {\tt Inclusive} & $-0.212 \pm 0.047$ & $+0.020$ & $-0.020$\\
$b_{tail}$                 & $ -1.01\hphantom{0} \pm 0.29 \hphantom{0}$ & $-0.021$ & $+0.029$\\
$f_{tail}$                 & $\phm 0.109 \pm 0.020$ & $+0.022$ & $-0.030$\\
$f_{out}$                  & $\phm 0.002 \pm 0.001$ & $-0.004$ & $+0.006$\\ \hline
\multicolumn{4}{c}{Signal dilutions}\\ \hline
$\langle D \rangle$, {\tt Lepton }    & $\phm 0.933 \pm 0.013 $ & $-0.002$ & $-0.004$\\
$\langle D \rangle$, {\tt Kaon~I }    & $\phm 0.799 \pm 0.014 $ & $-0.009$ & $+0.051$\\
$\langle D \rangle$, {\tt Kaon~II}    & $\phm 0.582 \pm 0.016 $ & $-0.001$ & $-0.006$\\
$\langle D \rangle$, {\tt Inclusive } & $\phm 0.368 \pm 0.017 $ & $+0.009$ & $+0.024$\\

$\Delta D$, {\tt Lepton } & $\phm 0.031 \pm 0.022 $ & $-0.003$ & $-0.001$\\
$\Delta D$, {\tt Kaon~I } & $\phm 0.023 \pm 0.022 $ & $-0.010$ & $+0.039$\\
$\Delta D$, {\tt Kaon~II} & $\phm 0.090 \pm 0.024 $ & $-0.008$ & $+0.002$\\
$\Delta D$, {\tt Inclusive } & $\phm 0.050 \pm 0.026 $ & $+0.002$ & $+0.004$\\ \hline
\multicolumn{4}{c}{Background properties ($B_{flav}$)} \\ \hline
$\tau$ [ps] & $\phm 1.335 \pm 0.064 $ & $+0.001$ & $+0.000$\\
$f(\tau=0 )$ {\tt Lepton } & $\phm 0.29\hphantom{0} \pm 0.17\hphantom{0} $ & $+0.000$ & $-0.001$\\
$f(\tau=0 )$ {\tt Kaon~I } & $\phm 0.631 \pm 0.027 $ & $+0.000$ & $-0.001$\\
$f(\tau=0 )$ {\tt Kaon~II} & $\phm 0.659 \pm 0.024 $ & $+0.000$ & $-0.001$\\
$f(\tau=0 )$ {\tt Inclusive } & $\phm 0.684 \pm 0.023 $ & $+0.000$ & $+0.000$\\ \hline
\multicolumn{4}{c}{Background resolution function} \\ \hline
$S_{core}$ & $\phm 1.398 \pm 0.019 $ & $+0.001$ & $-0.002$\\
$b_{core}$ & $-0.043 \pm 0.013 $ & $-0.001$ & $+0.002$\\
$f_{out}$ & $\phm 0.015 \pm 0.002 $ & $+0.000$ & $+0.001$\\ \hline
\multicolumn{4}{c}{Background dilutions} \\ \hline
$\langle D \rangle$, {\tt Lepton} , $\tau = 0$ & $\phm 1.36\hphantom{0} \pm 0.69\hphantom{0}$ & $+0.000$ & $+0.001$\\
$\langle D \rangle$, {\tt Kaon~I} , $\tau = 0$ & $\phm 0.648 \pm 0.030$ & $+0.001$ & $-0.004$\\
$\langle D \rangle$, {\tt Kaon~II} , $\tau = 0$ & $\phm 0.393 \pm 0.023$ & $+0.000$ & $+0.000$\\
$\langle D \rangle$, {\tt Inclusive} , $\tau = 0$ & $\phm 0.158 \pm 0.024$ & $-0.001$ & $-0.002$\\

$\langle D \rangle$, {\tt Lepton} , $\tau > 0$ & $\phm 0.17\hphantom{0} \pm 0.11\hphantom{0}$ & $+0.000$ & $+0.000$\\
$\langle D \rangle$, {\tt Kaon~I} , $\tau > 0$ & $\phm 0.251 \pm 0.048$ & $+0.000$ & $+0.000$\\
$\langle D \rangle$, {\tt Kaon~II} , $\tau > 0$ & $\phm 0.278 \pm 0.042$ & $+0.000$ & $+0.000$\\
$\langle D \rangle$, {\tt Inclusive} , $\tau > 0$ & $\phm 0.031 \pm 0.046$ & $+0.000$ & $+0.000$\\ 
\end{tabular} \end{ruledtabular}
\end{center}
\end{table*}
Figure~\ref{fig:iso-lk} shows the contour plots in the $\ctwob$, $\stwob$ plane.
We obtain 
\begin{eqnarray} 
\ctwob &=& +3.32 ~^{+0.76}_{-0.96} \stat \pm
0.27 \syst, \nonumber \\
\stwob &=& -0.10 \pm 0.57 \stat \pm 0.14 \syst.
\end{eqnarray}
The quality of the fit is estimated  by generating 2000 experiments 
using the parametrized MC, and with the same sample size that is observed
for the data.
The probability to obtain a likelihood lower than that 
obtained from the data is found to be $(22 \pm 1)$\%.

When $\stwob$ is fixed to the value measured in \B decays
to $\jpsi \KS$ and related modes, $\stwob = \stwobz \equiv 0.731$
\cite{PDG2004}, we find 
\begin{eqnarray}
\ctwob=+2.72_{-0.79}^{+0.50}\stat \pm 0.27\syst .
\end{eqnarray}

\subsection{Graphical Representation} 
\label{subsec:GraphicalRepresentation} 
\begin{figure}[tbp]
\begin{center}
\includegraphics[width=\singlewidth]{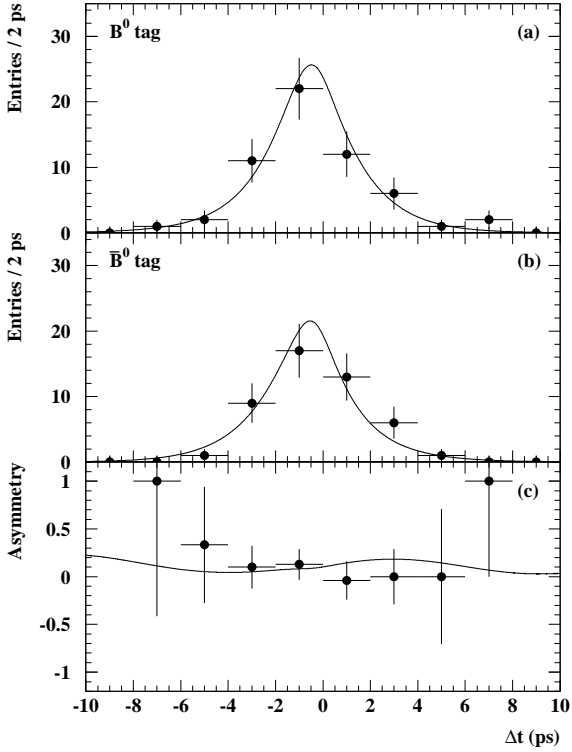}
\caption{\label{fig:delta-t}
The distribution of $\Delta t$ for events in the signal region, for (a) \Bz and (b) \Bzb tags with the fit result (full curve)
overlaid.
In (c) we show the raw asymmetry in the number of \Bz\ and \Bzb\ tags in the signal region,
      $(N_{\Bz}-N_{\Bzb})/(N_{\Bz}+N_{\Bzb})$,
 for data, with the fit result (full curve)
overlaid. Note that above distributions are not sensitive to  \ctwob since this dependence vanishes
when integrated over the angular variables.
}
\end{center}
\end{figure}

The distribution of the time difference \deltat is shown in
Figs.~\ref{fig:delta-t}(a) and~\ref{fig:delta-t}(b), and
the  time-dependent asymmetry is shown in Fig.~\ref{fig:delta-t}(c).
Note that in the case of perfect acceptance this asymmetry is
not sensitive to \ctwob~\cite{StephThese}.

A graphical representation of the sensitivity of the data to $\ctwob$
is obtained  from the time dependence of the moment of $\cal C$.
Since $\cal C$ is orthogonal to both
 $\cal A$ and $\cal S$, we obtain, using Eq.~(\ref{eqn:asymm_CP_1})
\begin{eqnarray}
 \langle {\cal C} \rangle_{\pm}(\deltat) &\equiv& \int g_{\pm}(\vomega,\deltat;\boldsymbol{A},\sin 2\beta,\cos 2\beta) 
{\cal C}(\vomega;\boldsymbol{A}) \; {\rm d}\vomega \nonumber \\
&=& \pm\frac{\Gamma_0}{2} e^{-\Gamma_0 |\deltat|}\times \nonumber \\
&~& \left\{\cos(\Delta m \deltat) \int {\cal P}(\vomega;\boldsymbol{A}) {\cal C}(\vomega;\boldsymbol{A}) \; {\rm d}\vomega \right. + \nonumber\\
&~& \left. \sin(\Delta m \deltat) \cos 2\beta
 \int {\cal C}^2(\vomega;\boldsymbol{A}) \; {\rm d}\vomega \right\}. \nonumber \\
\label{eqn:moment_C}
\end{eqnarray}
We see that the magnitude of the $\sin(\Delta m \deltat)$ oscillation
is proportional to $\cos 2\beta$.
The introduction of the angular acceptance $\epsilon(\vomega)$ in principle breaks the
above orthogonality, causing $\int{\cal A C}\epsilon {\rm d}\vomega$ and $\int{\cal S C}\epsilon {\rm d}\vomega$
terms to appear in the $\langle {\cal C} \rangle_{\pm}(\deltat)$ expression, in addition to the
$\int{\cal P C}\epsilon {\rm d}\vomega$ and $\int{\cal C}^2\epsilon {\rm d}\vomega$ terms.
These quantities are estimated using Monte Carlo and are found to be at the percent level
of $\int{\cal C}^2\epsilon {\rm d}\vomega$ for $\int{\cal A C}\epsilon {\rm d}\vomega$ and $\int{\cal S C}\epsilon {\rm d}\vomega$,
and 14\% of $\int{\cal C}^2\epsilon {\rm d}\vomega$ for $\int{\cal P C}\epsilon {\rm d}\vomega$.

\begin{figure}[tbp]
\begin{center}
\includegraphics[width=\singlewidth]{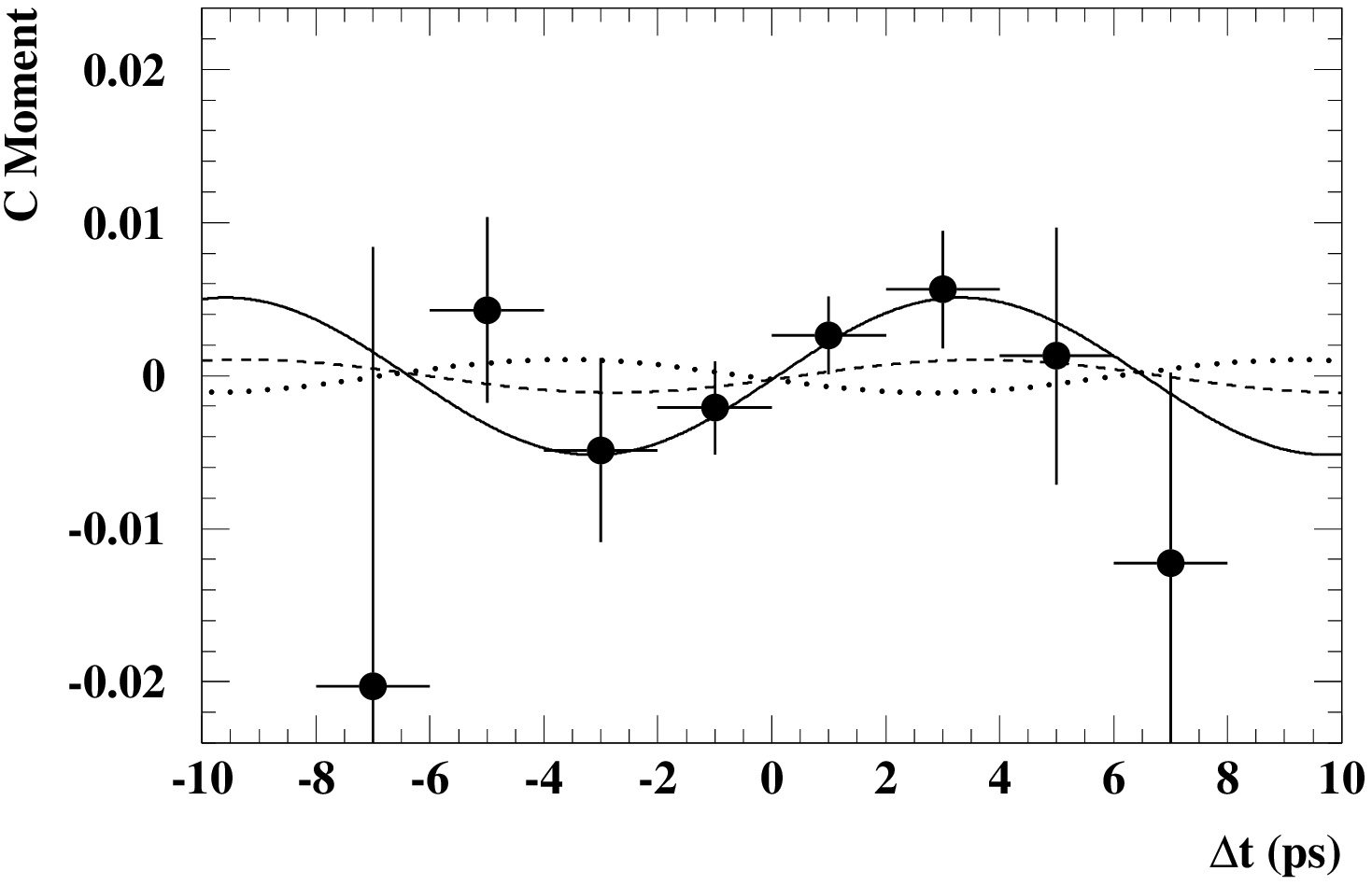}
\caption{\label{fig:moment_C_data}The moment of $\cal C$ as a function of 
\deltat. The overlaid curve corresponds to 
the fit results (Table~\ref{tab:CP_fit_result}). The dashed curve corresponds to 
$\ctwob = +\sqrt{1-\sin^22\beta_0} = +0.68 $ 
\protect\cite{PDG2004}, the dotted one to the non-standard solution $\ctwob = -0.68 $.}
\end{center}
\end{figure}
Figure~\ref{fig:moment_C_data} shows the moment of $\cal C$ as a
function of $\Delta t$, overlaid with a function obtained from
Eq.~(\ref{eqn:moment_C}) that takes the acceptance into account.

\subsection{Confidence Level for Positive $\boldsymbol{\cos2\beta}$ Solution}
\label{sec:confidenceLevelcos2beta}
The value $\stwob=0.731\pm0.056$~\cite{PDG2004} measured in the charmonium-$\Kz$ channel is in good agreement
with expectations from the measurements of the sides of the Unitarity Triangle if the choice 
$\beta\approx 0.41\equiv \beta_0$ is made. However, the alternative solutions $\beta\approx \pi/2-0.41, 0.41+\pi,$ and $3\pi/2 -0.41$
could turn out to be correct if there is a significant contribution from outside the Standard Model.
We show here that we can exclude at a significant level
of confidence the possibilities $\pi/2-0.41$ and $3\pi/2-0.41$,  assuming that the value of $\stwob$ that would be 
inferred from a high-statistics measurement of the $\jpsi\Kstar$ channel
would conform to the measurement of Ref.~\cite{PDG2004}.
We therefore constrain \stwob to  \stwobz.
The systematic uncertainty on $\beta$, induced by the uncertainty
in $\stwob$, ($\pm   0.056$)
\cite{PDG2004} is  $\pm 0.043$, which is negligible here.

We define $\ctwobz \equiv +\sqrt{1- \sin^2 2\beta_0} \approx
+0.68$. In the following, we estimate the confidence level at which
the $-\ctwobz$ hypothesis can be excluded against the $+\ctwobz$ solution.

\subsubsection{Assuming Gaussian Statistics}
\label{sec:cl_ctwob_gaussian}
\begin{figure}[tbp]
\begin{center}
\includegraphics[width=\singlewidth]{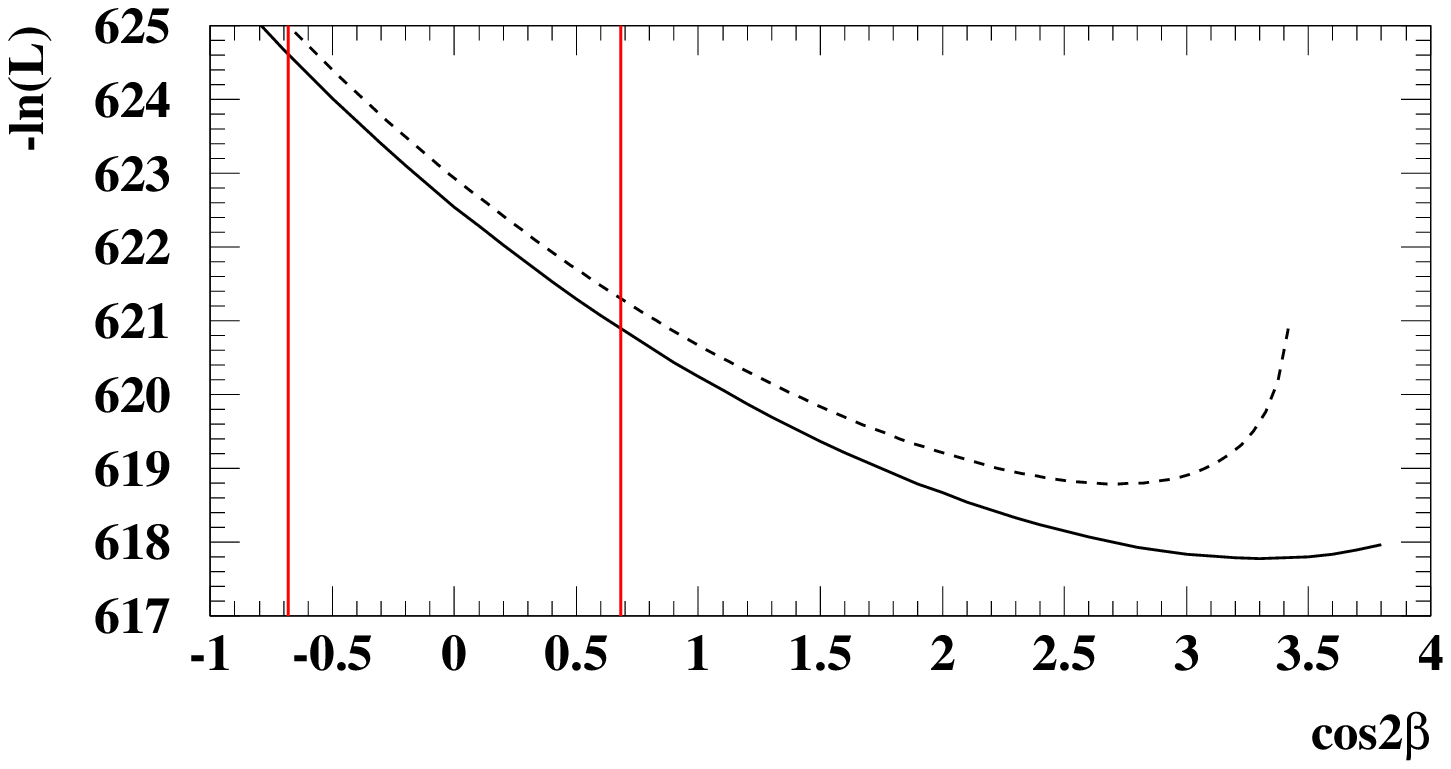}
\caption{\label{fig:sinfize}
The negative logarithm of the likelihood as a function of
$\ctwob$.
Continuous line: $\stwob$ is a free parameter.  Dashed line: $\stwob$ is fixed at $\stwob_0 =0.731$ \protect\cite{PDG2004}. }
\end{center}
\end{figure}

Figure~\ref{fig:sinfize} shows the variation of the likelihood 
as a function of $\ctwob$. 
In the case of fixed $\stwob$, the optimum is obtained at $\ctwob
=+2.72$, $2.2\sigma$ from $+\ctwobz$ and $3.5\sigma$ from $-\ctwobz$.
For a Gaussian distribution, the probabilities to observe values $2.2$ and $3.5$ $\sigma$ from the mean value are, respectively, 3.25\% and 0.08\%.
In a Bayesian approach, assuming equal {\it a priori} probabilities for the
$\pm\ctwobz$ hypotheses, the probability
that the $+\ctwobz$ choice is wrong would be $0.08/(3.25+0.08)=2.4\%$.
\begin{figure}[tbp]
\begin{center}
\includegraphics[width=\singlewidth]{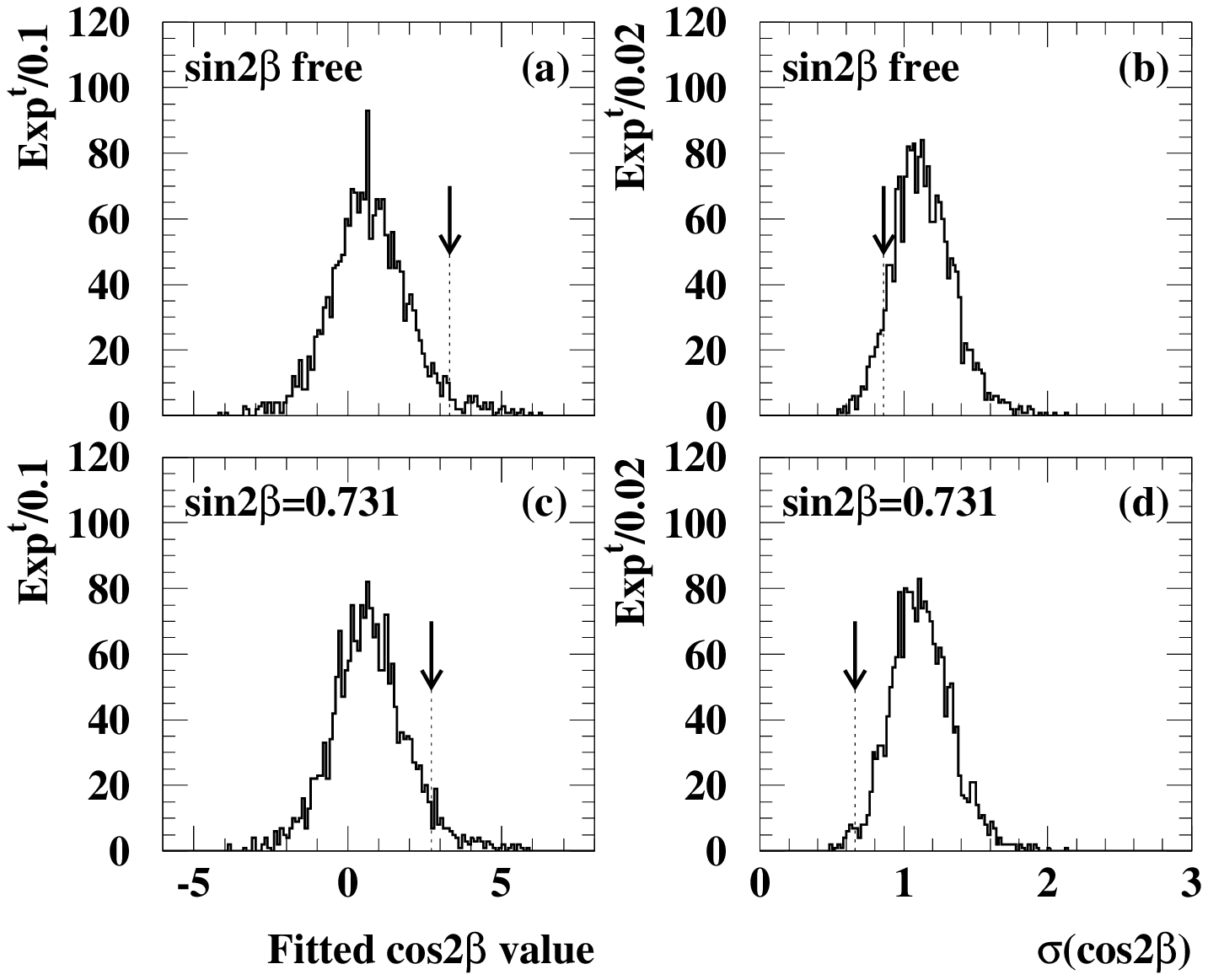}
\caption{\label{fig:toys}
The distribution of the values of \ctwob ((a) and (c)), 
and of the statistical uncertainties ((b) and (d)), 
obtained in 2000 simulated experiments, each based on a sample of the same size as the data;
i.e., 104 events.
These are taken from the parametrized MC sample mentioned above, with the generated \ctwob value $+0.68$.
In (a) and (b) \stwob is also free in the fit.
In (c) and (d), \stwob is fixed to the world average.
The vertical arrows show the positions of the values obtained from the data.
  }
\end{center}
\end{figure}

\subsubsection{Using the Distribution of \ctwob Values Obtained from Simulated Experiments}
\label{sec:cl_ctwob_toymc}
\begin{figure}[tbp]
\begin{center}
\includegraphics[width=\singlewidth]{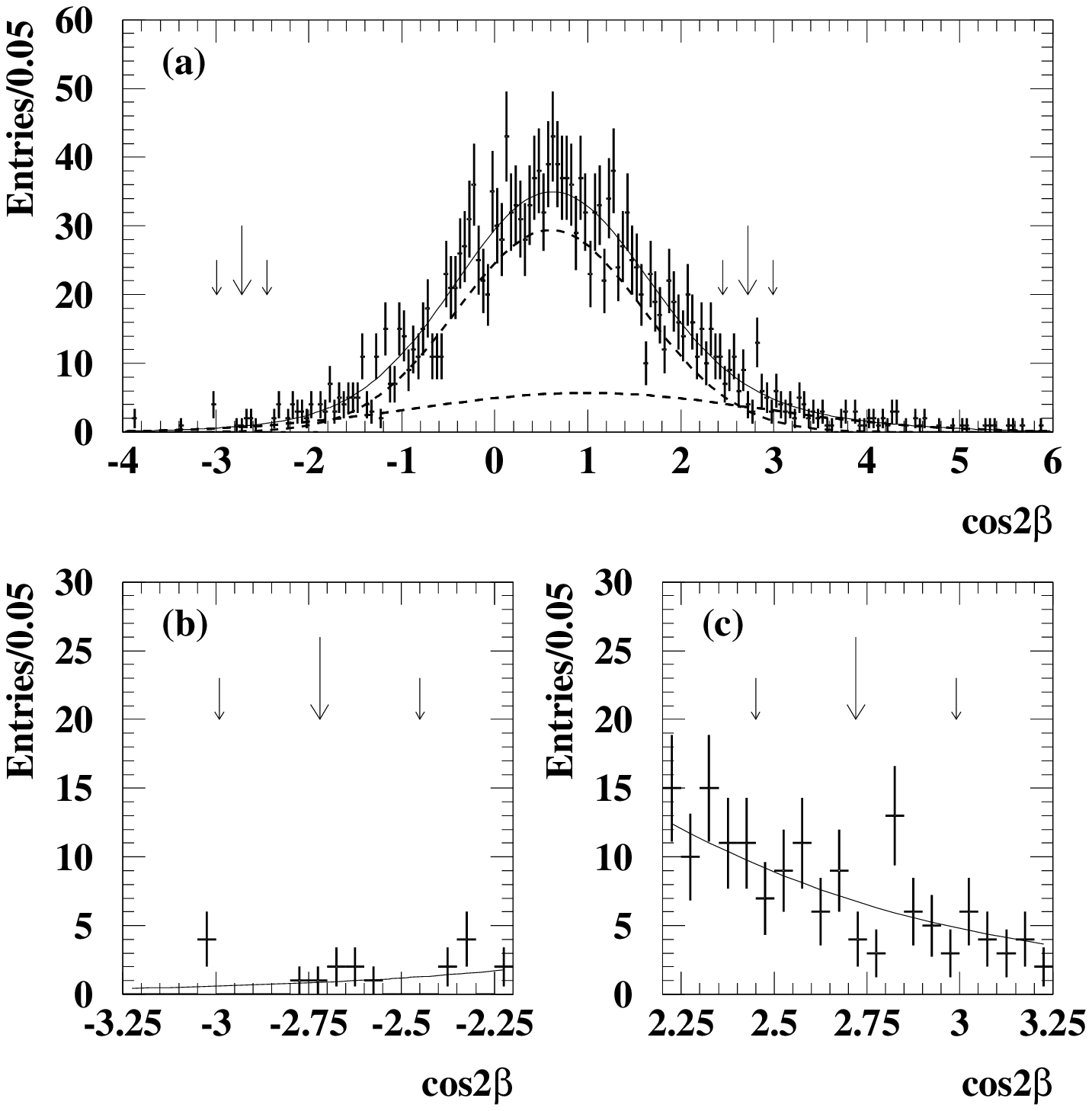}
\caption{\label{fig:CL}
(a) The $\ctwob$ distribution for 2000 
simulated experiments of the same size as the data sample, where the generated values for \stwob and \ctwob are $+0.731$ and $+0.68$, respectively (same as Fig.~\ref{fig:toys}(c)).
An unbinned likelihood fit is performed to this distribution, using the sum of two Gaussian functions.
The fit result is shown on the plot (full line) with the individual Gaussian contributions (dashed lines). The long vertical arrows show the \ctwob values $\pm 2.72$.
The small arrows indicate the extent of systematic uncertainties. (b) and (c) zoom on the  $\ctwob = \pm 2.72$ regions of Fig.~\ref{fig:CL}(a). The densities of points
at $+2.72$ and $-2.72$ are used to
discriminate between the $\ctwob=\pm\sqrt{1-\sin^22\beta_0}=\pm0.68$ hypotheses, as explained in the text.}
\end{center}
\end{figure}

To take into account the  nonparabolic shape of the log-likelihood
as a function of \ctwob,
\ctwob values are measured with 2000 simulated samples, each the same
size as the data sample (104 events) (Fig.~\ref{fig:toys}).
For the $+\cos2\beta_0$ hypothesis, the distribution
$\frac{dN^+}{d\ctwob}$
of \ctwob
values is that shown in Fig.~\ref{fig:toys}(c).
An unbinned likelihood fit is performed to the sample of the 2000 \ctwob values, with a sum of two Gaussian functions, $h(\ctwob)$.
The fit result is shown in Fig.~\ref{fig:CL}(a) (where $h(\ctwob)$ is scaled by 100, i.e., 2000 times the bin size).

The distribution 
$\frac{dN^-}{d\ctwob}$
is obtained by the transformation $\ctwob\to-\ctwob$: i.e., we have
\[
\frac{dN^+}{d\ctwob}(\ctwob) = \frac{dN^-}{d\ctwob}(-\ctwob).
\]

In a frequentist approach, we consider the probability that a result would disfavor, by at least as much as ours,
the $-\ctwobz$ hypothesis against the $+\ctwobz$ one, by computing the probability to observe a ratio
$r(\ctwob) \equiv \frac{dN^-}{d\ctwob}(\ctwob) / \frac{dN^+}{d\ctwob}(\ctwob)$ smaller than or equal to the one we obtain, $r(+2.72)$~\cite{statistics}.
Because this ratio $r(\ctwob)$ has a monotonic decreasing behaviour with $\ctwob$, the probability to obtain $r(\ctwob) \leq r(+2.72)$, if the
true $\ctwob$ value is indeed $-\ctwobz$, is
\begin{eqnarray}
\alpha &\equiv& \int_{+2.72}^{+\infty} \frac{dN^-}{d\ctwob}(\ctwob)\; d\!\ctwob \nonumber \\
&=& 0.6\%,
\end{eqnarray}
leading to the confidence level at which the $-\ctwobz$ hypothesis is excluded:
\begin{eqnarray}
CL^{\rm Freq.}(-\cos2\beta_0\;{\rm excluded}) &\equiv& 1-\alpha \nonumber \\
&=& 99.4\%.
\end{eqnarray}
If we ask how likely it is to obtain a result in the above $(+2.72,+\infty)$ range, if the true value of \ctwob is +\ctwobz, we find:
\begin{eqnarray}
\int_{+2.72}^{+\infty}\frac{dN^+}{d\ctwob}(\ctwob)\; d\!\ctwob &=& 5.7\%.
\end{eqnarray}
In a frequentist interpretation, a high value for this last quantity would have indicated, together with the
high $CL^{\rm Freq.}(-\cos2\beta_0\;{\rm excluded})$ value obtained,
that the $(+2.72,+\infty)$ domain would have allowed a sharp distinction between the two $\pm\ctwob$ hypotheses.
The rather low value observed here ($5.7\%$) expresses that, at the present level of statistics,
the discrimination between the $\pm \ctwobz$ hypotheses is rather modest.
We can conclude however that our result would be somewhat more
improbable ($0.6\%$) if the true value of \ctwob were $-\ctwobz$ than it would be ($5.7\%$) if the true value were $+\ctwobz$.\\

In a Bayesian approach, assuming that the two $\pm \cos2\beta_0$ hypotheses have {\it a priori} equal probabilities, 
the confidence level at which the $-\cos2\beta_0$ solution is excluded, $CL(-\cos2\beta_0\;{\rm excluded})$, is obtained from
$\frac{dN^+}{d\ctwob}(+2.72)$ and $\frac{dN^-}{d\ctwob}(+2.72)$ as follows:
\begin{eqnarray}
CL(-\cos2\beta_0\;{\rm excluded}) &=& \frac{h(+2.72)}{h(+2.72)+h(-2.72)} \nonumber\\
&=& \frac{6.64\pm0.38}{(6.64\pm0.38)+(0.86\pm0.15)} \nonumber\\
\label{eqn:CL}
&=&(88.6\pm 2.0)\%.
\end{eqnarray}
The probability to select incorrectly the $+\ctwobz$ solution is significantly larger than for the previous Bayesian estimate based on Gaussian statistics (Sec.~\ref{sec:cl_ctwob_gaussian}).
The uncertainty in Eq.~(\ref{eqn:CL}) comes from the statistical uncertainties
on $h(+2.72)$, $h(-2.72)$  (limited by the 2000 simulated experiments used),
and their correlation ($-6\%$).
The systematic effects on the $\cos2\beta$ measurement contribute to a $\pm0.4\%$ variation of $CL(-\cos2\beta_0\;{\rm excluded})$ and are included in quadrature in the above uncertainty.
We include a $-1\sigma$ safety margin on $CL(-\cos2\beta_0\;{\rm excluded})$, and thus report
\begin{equation}
\label{eqn:CL_report}
CL(-\cos2\beta_0\;{\rm excluded})=86.6\%.
\end{equation}

\section{Conclusion}
\label{sec:Conclusion}

We measure the transversity amplitudes of the decay to flavor
eigenstates, $B\to \jpsi K^{*0}(\Kpm\pimp$) and $B\to \jpsi K^{*\pm}$
($\Kpm\piz$ and $\KS\pipm$), with improved precision with respect to 
existing measurements.
We determine
\begin{eqnarray}
\delta_\parallel-\delta_0 &=& (-2.73 \pm 0.10 \pm 0.05) \rad, \nonumber \\
\delta_\perp-\delta_0 &=& (+2.96 \pm 0.07 \pm 0.05) \rad, \nonumber \\
\vert A_0 \vert^2 &=&  0.566 \pm 0.012 \pm 0.005,  \nonumber \\ 
\vert A_\parallel \vert^2 &=&  0.204 \pm 0.015 \pm 0.005,  \nonumber \\ 
\vert A_\perp\vert^2 &=&  0.230 \pm 0.015 \pm 0.004,
\label{eqn:trans_final_result}
\end{eqnarray}
and
\begin{eqnarray}
\delta_\parallel-\delta_\perp &=& (0.60 \pm 0.08 \pm 0.02) \rad.
\end{eqnarray}

We observe the presence of a significant $S$-wave amplitude interfering
with the $P$-wave amplitude in the region of the $\Kstar(892)$.
Using a novel method based on the dependence on the $K\pi$ invariant
mass of the interference between the $S$- and $P$-waves, we resolve
the ambiguity in the determination of the strong phases involved in \B decays
to $\jpsi\Kstar(892)$.

\begin{figure}[htbp]
\begin{center}
\includegraphics[width=\singlewidth]{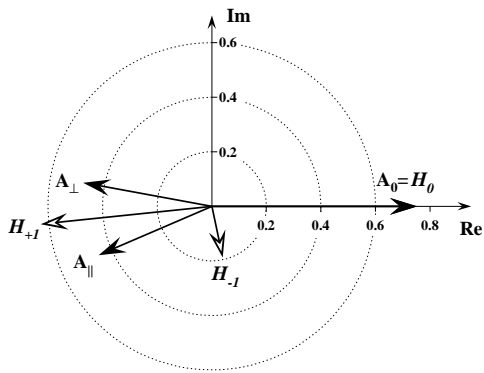}
\caption{\label{fig:amp_draw}Representation in the complex plane of the measured transversity amplitudes $(A_0, A_\|, A_\perp)$ (Eq.~(\ref{eqn:trans_final_result}))
and the equivalent helicity amplitudes $(H_0, H_{-1}, H_{+1})$ obtained using Eq.~(\ref{eqn:helicity-transversity-amplitudes}). The values for $H_+$ and $H_-$ are
quoted in Table~\ref{tab:heli_amp}.}
\end{center}
\end{figure}

The values obtained for $|A_\parallel |^2$ and
$|A_\perp|^2$ are  consistent with being equal.
The additional unambiguous determination of the phases relative to that of $A_0$ indicates that they have similar size but opposite sign,
with a difference,
$\delta_\|-\delta_\perp$, of  $34\pm 5$ degrees.
Using the relations between the helicity amplitudes and the transversity
amplitudes, 
\begin{eqnarray}
H_{+1} & \equiv & (A_\parallel + A_\perp)/\sqrt{2} \nonumber\\
       &\equiv& |H_+|e^{i\delta_+}, \nonumber\\
H_{-1} & \equiv & (A_\parallel - A_\perp)/\sqrt{2}  \nonumber\\
       & \equiv& |H_-|e^{i\delta_-}, \nonumber\\
H_0    & \equiv & A_0,
\label{eqn:helicity-transversity-amplitudes}
\end{eqnarray}
we obtain the moduli and phases given in Table~\ref{tab:heli_amp}.
\begin{table}
\caption{\label{tab:heli_amp}Helicity-amplitude moduli and phases for $H_+$ and $H_-$
obtained from the measured transversity amplitudes (Eq.~(\ref{eqn:trans_final_result}))
using Eq.~(\ref{eqn:helicity-transversity-amplitudes}). The corresponding configuration is shown in Fig.~\ref{fig:amp_draw}. The uncertainties are statistical only.}
\begin{center}
\begin{ruledtabular}
\begin{tabular}{cccc}
$|H_+|^2$ & $|H_-|^2$ & $\delta_+$ (\rad) & $\delta_-$ (\rad) \\ \hline
$0.396 \pm 0.015$ & $0.0379 \pm 0.009$ &  $-3.04 \pm 0.08$ &  $-1.36 \pm 0.12$
\end{tabular} \end{ruledtabular}
\end{center}
\end{table}
This determines the hierarchy of the helicity amplitudes in the decay to be
$\vert H_{0}\vert : \vert H_{+1}\vert : \vert H_{-1}\vert \sim 0.75 : 0.63 : 0.19$.
The corresponding configurations  of
the helicity and transversity amplitudes
in the complex plane are illustrated in Fig.~\ref{fig:amp_draw}.

We confirm the presence of nonzero relative strong phases, with the
difference between the phases of $A_0$ and $A_\parallel$ deviating from
$\pi$ with a significance of 3.6$\sigma$, and the phase difference between the
two transverse amplitudes being 7.6$\sigma$ from zero.

Treating $\sin 2\beta$ and $\cos 2\beta$ as independent quantities
in the fit to the data, 
we obtain $\cos2\beta = +3.32 ~^{+0.76}_{-0.96} (\stat) \pm 0.27 (\syst)$.
When $\stwob$ is fixed to
the value measured in the charmonium-$K^0$ modes, 
$\stwob =0.731$ \cite{PDG2004}, we find 
\begin{eqnarray}
\ctwob =+2.72_{-0.79}^{+0.50} \pm 0.27.
\end{eqnarray}
The sign of $\cos2\beta$ is found to be positive at the 86\% C.L.
This is compatible with the sign inferred from the
Standard-Model-based fits of the Cabibbo-Kobayashi-Maskawa triangle, thus
limiting the possible presence of unknown physics beyond the Standard
Model.

We are grateful for the 
extraordinary contributions of our \pep2\ colleagues in
achieving the excellent luminosity and machine conditions
that have made this work possible.
The success of this project also relies critically on the 
expertise and dedication of the computing organizations that 
support \babar.
The collaborating institutions wish to thank 
SLAC for its support and the kind hospitality extended to them. 
This work is supported by the
US Department of Energy
and National Science Foundation, the
Natural Sciences and Engineering Research Council (Canada),
Institute of High Energy Physics (China), the
Commissariat \`a l'Energie Atomique and
Institut National de Physique Nucl\'eaire et de Physique des Particules
(France), the
Bundesministerium f\"ur Bildung und Forschung and
Deutsche Forschungsgemeinschaft
(Germany), the
Istituto Nazionale di Fisica Nucleare (Italy),
the Foundation for Fundamental Research on Matter (The Netherlands),
the Research Council of Norway, the
Ministry of Science and Technology of the Russian Federation, and the
Particle Physics and Astronomy Research Council (United Kingdom). 
Individuals have received support from 
CONACyT (Mexico),
the A. P. Sloan Foundation, 
the Research Corporation,
and the Alexander von Humboldt Foundation.\\

\appendix

\section{Uncertainties with a Pseudo-log-likelihood, and Validations}
\label{appendx:pseudoLK}

In the angular analysis, the background correction is performed using a pseudo-log-likelihood 
 $L'$, defined in Eq.~(\ref{eqn:pseudo-log}).
As $L'$ is not a log-likelihood, the uncertainties yielded by the minimization 
program (\minuit)~\cite{minuit} that is used are  incorrect estimates of the actual uncertainties.
The  correct estimate is given by
\begin{widetext}
\begin{eqnarray}
{\rm\bf Cov}[\boldsymbol{A}] &=& 
{\rm\bf Cov^{H}}[\boldsymbol{A}]
\Bigg\{ {\bf 1} \,+\, 
\Bigg[n_B (1+k) \int {b(\vomega) \left(\frac{\vec{\nabla} g^{\jchan,obs}(\vomega)}{g^{\jchan,obs}(\vomega)}\right)^{2} \dd \vomega}
\label{eqn:single_experiment_covariance_1}
\\
 & ~ &\hspace{2.7cm}
+ N^2_{B} \sigma^2_k \left( \int {b(\vomega) \frac{\vec{\nabla} g^{\jchan,obs}(\vomega)}{g^{\jchan,obs}(\vomega)}\dd \vomega}\right)^{2}
\;\;\;\;\;\;\Bigg] {\rm\bf Cov^{H}}[\boldsymbol{A}]
\Bigg\}, \nonumber
\end{eqnarray}
\end{widetext}
where
\begin{itemize}
\item ${\rm\bf Cov^{H}}[\boldsymbol{A}]$ is the covariance matrix of $\boldsymbol{A}$ at the maximum
of $L'$, estimated by the {\tt HESSE} routine of \minuit \cite{minuit} after the fit has converged.
\item In the expressions for the  bilinear forms $\int {b(\vomega) (\vec{\nabla} g^{\jchan,obs}(\vomega)/g^{\jchan,obs}(\vomega))^{2} \dd \vomega}$
and 
\\
$ \left(\int {b(\vomega) \vec{\nabla} g^{\jchan,obs}(\vomega)/g^{\jchan,obs}(\vomega)\dd \vomega}\right)^{2}$,
$\vec{\nabla}$ denotes the gradient, i.e., differentiation with respect to the fit parameters $\boldsymbol{A}$.
The ``square'' is not to be understood as a ``scalar product'', but as a ``direct product'', i.e.
$\vec{v}^{\,2}=\vec{v}^{\,\dag}\vec{v}$, so that the resulting quantity is a square matrix.
\item $g^{\jchan,obs}(\vomega)$ and $b(\vomega)$ are the PDFs for the signal and the background.
Note that in practice the knowledge of the PDF of the background is not needed 
for the computation of Eq.~(\ref{eqn:single_experiment_covariance_1}) because
for any function $h(\vomega)$, $\int {b(\vomega) h(\vomega)\dd \vomega}$
is estimated by the average of $h$ over the \mes sideband background sample:
\begin{eqnarray}
\int {b(\vomega) h(\vomega)\dd \vomega} \approx 
\frac{1}{N_{B}}\sum_{i=1}^{N_{B}} h(\vomega_i).
\end{eqnarray}

\item $k$ is the scaling parameter $k = \tilde{n}_B/N_{B}$, and $\sigma_k$ is its uncertainty.
\end{itemize}

The estimated number of background events in the signal region
$\tilde{n}_B$ is obtained from an  ARGUS plus Gaussian fit to the \mes
spectrum.

The validation of the pseudo-log-likelihood method (i.e., the unbiased
nature of the fit parameters, which is not shown here, and of their uncertainties) comes from MC-based studies.
 We have simulated $10^{3}$ experiments with $10^{4}$ events each \cite{Evtgen},
using a signal PDF with 
$\theta_A = \phi_A = \delta_\parallel = \delta_\perp = 1$ rad
(as defined in Eq.~(\ref{eqn:polar_description_of_amplitudes})).

\begin{figure}[h]
\begin{center}
\includegraphics[width=\singlewidth]{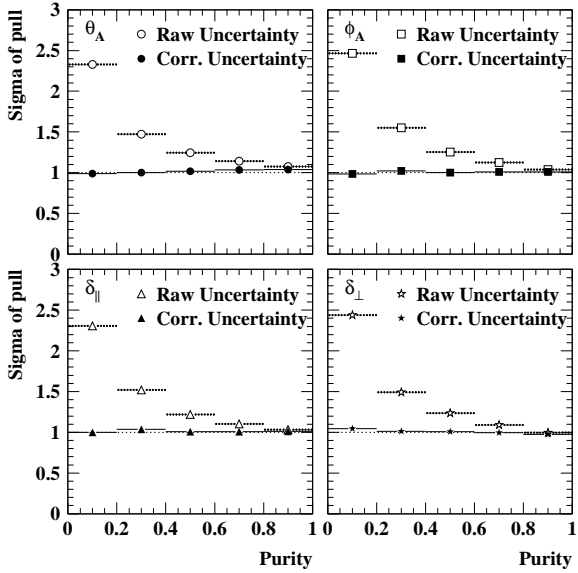}
\caption{\label{fig:pseuso-log-LK:validation}
Root mean square (RMS) of the pull distributions as a function of signal purity
(defined in the 5.2 -- 5.3 \gevcc \mes range), for the fitted
parameters $\theta_A, \phi_A, \delta_\parallel$, and $\delta_\perp $.
Open symbols denote the RMS of the pulls computed with
the uncertainties
 taken directly from \minuit. Closed symbols denote the uncertainties computed according to Eq.~\ref{eqn:single_experiment_covariance_1}.}
\end{center}
\end{figure}

We study the behavior of the fit  for various values of the purity,
adding the appropriate number of background events.
A variety of background shapes have been used.
Figure \ref{fig:pseuso-log-LK:validation} presents results using an
ARGUS \mes distribution with an angular distribution of
$a + b\cos^2(\theta_{K^*}) + c \cos^4(\theta_{K^*})$
\footnote{More precisely: $\frac{1}{\sqrt{8\pi}} {\cal Y}_{000} + \frac{1}{15} {\cal Y}_{040}$.}.

Figure~\ref{fig:pseuso-log-LK:validation} shows the results from the
Monte Carlo study.  As the purity decreases, the \minuit-reported uncertainties diverge more and more from the actual spread in the results.  The uncertainties
calculated from Eq.~(\ref{eqn:single_experiment_covariance_1}) correctly 
predict the behavior of the spread, even at low purity.


\end{document}